\documentclass[12pt]{article}
\usepackage{amssymb,amsmath,amsfonts}
\usepackage{graphicx}
\graphicspath{{Figures/}}
\usepackage{color}
\numberwithin{equation}{section}

\newcommand{\be}{\begin{equation}}
\newcommand{\ee}{\end{equation}}
\newcommand{\bea}{\begin{eqnarray}}
\newcommand{\eea}{\end{eqnarray}}

\renewcommand{\c}{{\rm col}}
\newcommand{\e}{{\rm e}}

\renewcommand{\l}{{\rm loc}}
\renewcommand{\d}{{\rm d}}

\newcommand{\av}[1]{\left\langle{#1}\right\rangle}

\newcommand{\LR}{L_{\vec{R}}}

\newcommand{\wL}{{\widetilde L}}
\newcommand{\wI}{{\widetilde I}}
\newcommand{\vlambda}{\vec{\lambda}}

\newcommand{\grintl}{[\kern-.18em [}
\newcommand{\grintr}{]\kern-.18em ]}

\newcounter{resultcounter}[section]

\newtheorem{thm}[resultcounter]{Theorem}
\newtheorem{lem}[resultcounter]{Lemma}
\newtheorem{prop}[resultcounter]{Proposition}

\newtheorem{definition}[resultcounter]{Definition}

\def\bed{\begin{definition}}
\def\eed{\end{definition}}
\def\one{{\mathchoice {\rm 1\mskip-4mu l} {\rm 1\mskip-4mu l} {\rm 1\mskip-4.5mu l} {\rm 1\mskip-5mu l}}}

\newcommand{\R}{{\mathbb R}}

\renewcommand{\r}{{\rm R}}
\newcommand{\s}{{\rm S}}

\renewcommand{\i}{{\rm i}}

\def\qed{\hfill $\Box$\medskip}

\newcommand{\scalprod}[2]{\left\langle {#1}, {#2}\right\rangle}
\newcommand{\bbbone}{\mathchoice {\rm 1\mskip-4mu l} {\rm 1\mskip-4mu l}
{\rm 1\mskip-4.5mu l} {\rm 1\mskip-5mu l}}

\renewcommand{\S}{\s}

\renewcommand{\R}{{\rm R}}

\begin{document}

\title{Dynamics of a Chlorophyll Dimer in \\
	Collective and Local Thermal Environments}

\author{ M. Merkli\footnote{Department of Mathematics and Statistics, Memorial University of Newfoundland, St. John's, NL, Canada A1C 5S7; merkli@mun.ca}\and G.P. Berman\footnote{Theoretical Division, Los Alamos National Laboratory, and the New Mexico Consortium,  Los Alamos, NM, 87544, USA; gpb@lanl.gov}  \and R.T. Sayre\footnote{Biological Division, B-11, Los Alamos National Laboratory and the New Mexico Consortium, 100 Entrada Dr., Los Alamos, NM 87544, USA; rsayre@newmexicoconsortium.org} \and S. Gnanakaran\footnote{Theoretical Division, Los Alamos National Laboratory, Los Alamos, NM, 87545, USA; gnana@lanl.gov} \and M. K\"onenberg\footnote{Department of Mathematics and Statistics, Memorial University of Newfoundland, St. John's, NL, Canada A1C 5S7; present address: Fachbereich Mathematik, Universit\"at Stuttgart,  Germany, 		martin.koenenberg@mathematik.uni-stuttgart.de}\and A.I. Nesterov\footnote{Departamento de Fisica, CUCEI, Universidad de Guadalajara, Av. Revoluci\'on 1500, Guadalajara, CP 44420, Jalisco, M\'exico; nesterov@cencar.udg.mx} \and H. Song\footnote{Tianjin University of Technology, Tianjin, China; song\_haifeng@126.com}}

\maketitle

\hfill \small LA-UR-15-29509

\begin{abstract}
	We present a theoretical analysis of exciton transfer and decoherence effects in a photosynthetic dimer interacting with collective (correlated) and local (uncorrelated) protein-solvent environments. Our approach is based on the framework of the spin-boson model. We derive explicitly the thermal relaxation and decoherence rates of the exciton transfer process, valid for arbitrary temperatures and for arbitrary (in particular, large) interaction constants between the dimer and the environments. We establish a generalization of the Marcus formula, giving reaction rates for dimer levels possibly individually and asymmetrically coupled to environments.  We identify rigorously parameter regimes for the validity of the generalized Marcus formula.
	The existence of long living quantum coherences at ambient temperatures emerges naturally from our approach.
\end{abstract}

\section{Introduction}

When a sunlight photon is absorbed by a light-sensitive molecule (such as chlorophyll or carotenoid) in a light-harvesting photosynthetic complex (LHC), the photon energy is stored in the molecule in the form of an exciton, an excited electron state of the molecule. The exciton then travels very quickly (some picoseconds) inside the LHC and reaches the reaction center (RC), where charge separation, and afterwards relatively slow chemical reactions  take place \cite{QEB}. Both the primary processes of exciton dynamics and charge separation occur in the presence of a protein (and solvent) environment at ambient temperature. In the framework of F\"{o}rster's resonance excitation transfer theory \cite{Forster}, the energy transfer is so fast that both fluorescence and recombination (due to the environment) can be neglected.

When modeling these primary exciton transfer (ET) processes, the light-sensitive molecule is usually associated with a geometrically localized site, $n$, having excited electron energy  $E_n$ \cite{QEB,Lloyd1}. The total number of sites, $N$, depends on the photosynthetic system. For example, $N=15$ in the CP29 LHC which is closely associated with photosystem II (PSII) \cite{CP29}.  The sites interact via dipole-dipole (or exchange) interaction, which is described by matrix elements $V_{nm}$. Similar to many molecular systems, the simplest unit of this picture is a dimer, which describes the interaction between two (not necessarily neighboring) sites. The dimer is characterized by a donor (site $n$) and an acceptor (site $m$). Denote by $\Delta E_{nm}=E_n-E_m$ the difference between the two excited electron energy levels. The donor and acceptor can be weakly coupled ($|V_{nm}/\Delta E_{nm}|\ll 1$) or they can be strongly coupled ($|V_{nm}/\Delta E_{nm}| \gtrsim 1$) \cite{Dimer}. The dimer is embedded in, and interacting with, a protein-solvent environment. One then introduces the constants of interaction $\lambda_{n}$ and $\lambda_m$ characterizing the strength of the interaction between the dimer sites and the environment. The protein-solvent environment is characterized by its correlation function, which depends on the temperature and some parameters which describe the spectral density (the `spectral function') of protein-solvent fluctuations at low and high frequencies. In order to describe different environmental effects, this correlation function can be taken to vary from a quite standard form  \cite{CF1} to a rather complicated one \cite{CF2}.

The quantum dynamics of the dimer is characterized by its $2\times 2$ time-dependent reduced density matrix, which is obtained from the `total' dimer-environment density matrix by averaging over (tracing out) the environmental degrees of freedom. The reduced dimer dynamics is then formally equivalent to that of an effective spin $1/2$. The protein-solvent environment is often modeled by a set of linear quantum oscillators (bosonic degrees of freedom) which characterize the dynamics of the protein and solvent atoms. In this ``spin-boson" model, the interaction between the dimer and the environment is usually reduced to the self-consistent (adiabatic) renormalization of the dimer ``donor" and ``acceptor" energy levels \cite{XuSch,MBSa}. See also \cite{T1,T2} for different approaches. This so-called ``diagonal interaction" \cite{XuSch} can be generalized, if needed, to a ``non-diagonal interaction" \cite{MBSa}.

The dynamics of the diagonal components of the reduced dimer density matrix (relative to the energy basis) is characterized by the ET rate.  The dynamics of the non-diagonal components of the reduced density matrix is characterized by the decoherence rate. It describes the evolution of quantum coherent effects which are currently the focus of intense theoretical and experimental studies \cite{QEB,Vegte} (see also references therein).    

Even the relatively simple spin-boson model has surprisingly many unresolved and rather subtle theoretical issues. The first one is related to the structure of the dimer electron energy spectrum. A real dimer based, for example, on two chlorophyll molecules, has many vibrational degrees of freedom \cite{Vib1,Vib2,Vib3,Vib4}. This means that the Hilbert space of the electron energy spectrum of a dimer is in actual fact very complicated (not two-dimensional!), with the two-level effective spin $1/2$ system only being a reasonable approximation to reality. The second issue is related to the protein-solvent environment. There are many models which describe this environment, but there is no consensus or clear understanding as to which one is the right one \cite{QEB}. 
The third issue is related to the dimer-environment interaction. The problem is that the interaction constants $\lambda_{n}$ and $\lambda_m$ are not small (compared to $V_{nm}$). A standard perturbation theory can therefore not be used. Indeed, the famous Marcus formula for the electron transfer rate, $\gamma$, from an initially populated donor has the form,  $\gamma \propto ({1/\lambda})\exp(-f(\lambda)/\lambda^2)$ \cite{XuSch}, where the interaction constant, $\lambda$, appears in the denominators of both the prefactor and in the exponent. In the standard perturbation theory approach (such as  the Bloch-Redfield theory \cite{MBSa}), the rate is proportional to $\lambda^2$. It is therefore clear that the Marcus rate cannot be derived by using a standard perturbative approach.

In this paper, we resolve the third issue mentioned above, i.e., we develop a theory which allows for strong (indeed, arbitrary) values of the interactions. The donor and the acceptor of the dimer can be localized close to each other ($\lesssim 1nm$). For a protein-solvent environment whose correlation length exceeds this distance, it is reasonable to consider the donor and acceptor to be coupled to a single reservoir, a situation which we call the collective reservoir model. The opposite case, when each dimer site is coupled to its own, independent environment, is called the local reservoirs model. Collective and local interactions were discussed previously in \cite{Vegte} (see also references therein) and in \cite{NB} for stochastic environments. The advantage of the current work is that our mathematical analysis is rigorous, meaning that any approximation made can be controlled, and conditions of validity of these approximations can be given.  This is especially important because, as mentioned above, standard perturbation approaches can not be used.

The main result of this work is a controlled expansion, for small $V_{nm}$, of the reduced dimer density matrix, valid for all times $t\ge 0$ and for arbitrary coupling constants $\lambda_1,\lambda_2$ and temperatures $T$. From this expansion, we derive the rates of the processes of electron (excitation) transfer and decoherence. We show that in certain regimes, those rates coincide with the expressions obtained previously from the usual Marcus formula.

We discuss the conditions on the spectral function of the reservoir at low frequencies which lead to long-time quantum coherences. This issue is of significant interest in recent research on the ET dynamics in LHCs, see e.g.  \cite{Vib4} and references therein.

The paper is organized as follows. In Section \ref{descrmodel}, we describe the model and give an outline of our main results. In Section \ref{mainresdetsect}, we give the results in more mathematical detail and present the results of our numerical simulations. Section \ref{sectdynres} is devoted to the mathematical approach
of the dynamical resonance theory. We summarize our results in a Conclusion section. Finally, in Appendices \ref{LSOsect}-\ref{reservoirssect} we present the detailed derivation of a few intermediate results.

\section{Description of the model}
\label{descrmodel}
\begin{figure}
\begin{center}
\scalebox{0.4}{\includegraphics{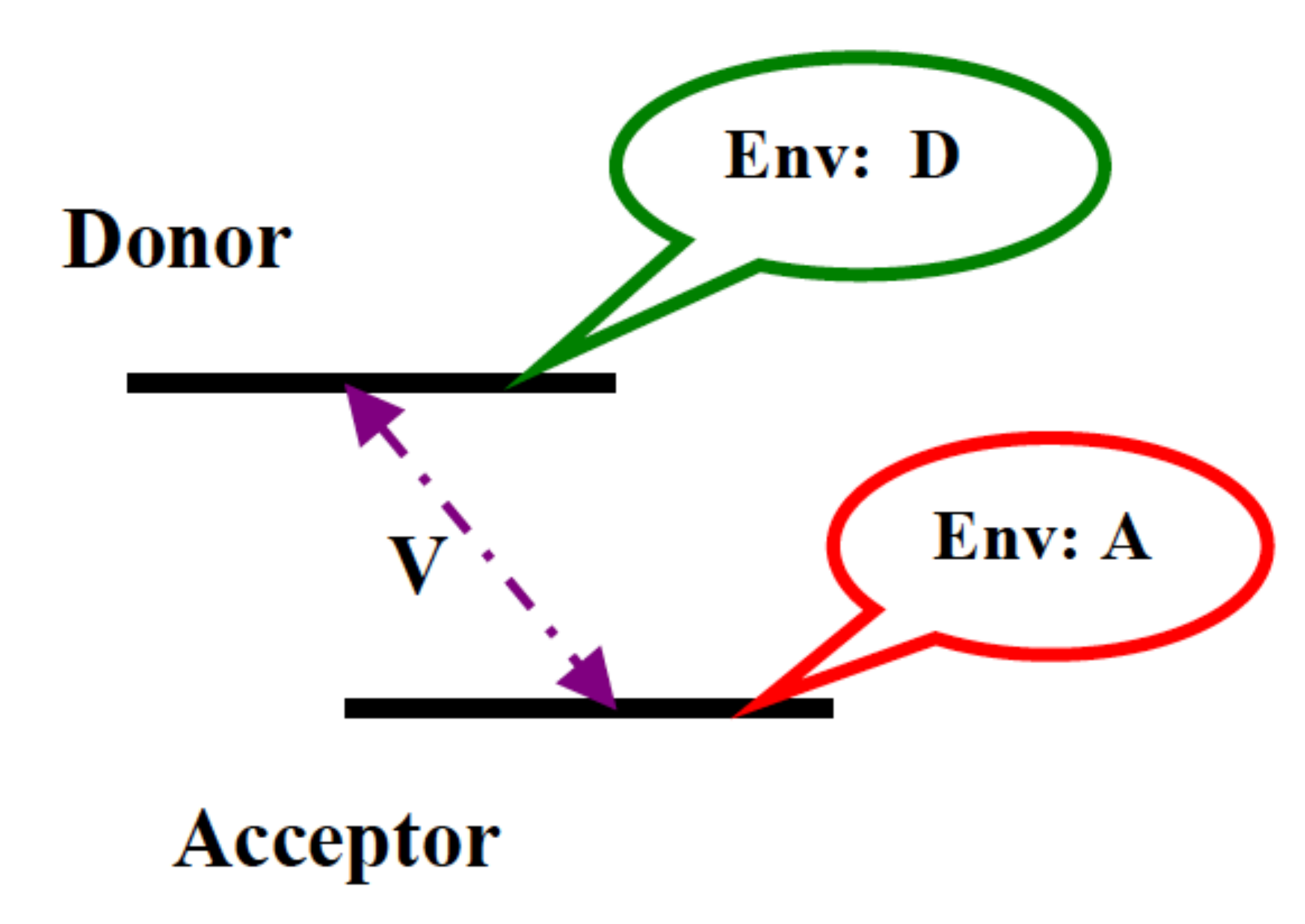}}	
\end{center}
\caption{(Color online) The schematic of the chlorophyll dimer (D-donor, A-acceptor) interacting with the protein-solvent environments. 
\label{D1}}
\end{figure}

We consider two types of dimer-reservoir interactions. In one, both dimer sites  are coupled to the {\em same}, collective, heat bath. In the other one, each dimer site is coupled to an {\em independent}, local, heat bath. The Hamiltonian of the {\bf collective reservoir model} is
\begin{equation}
	\label{1'}
	H_\c=\frac12 \begin{pmatrix}
		\epsilon & V\\
		V & -\epsilon
	\end{pmatrix}
	+H_\r + \begin{pmatrix}
		\lambda_1 & 0\\
		0& \lambda_2
	\end{pmatrix}\otimes\phi (g),
\end{equation}
where
\begin{equation}
	\label{2'}
	H_\r = \sum_k \omega_{k} a^\dagger_k a_k,\ \  \mbox{and} \ \ \phi(g) = \frac{1}{\sqrt2}\sum_k g_k a^\dagger_k +{\rm h.c.}
\end{equation}
Here, $\omega_k>0$ is the frequency of mode $k$ and $g_k\in\mathbb C$ is the `form factor', determining how strongly the mode $k$ is coupled to the dimer. 
While the quantities in \eqref{2'} are written down for discrete modes indexed by $k$, we consider a reservoir with {\em continuous modes}, where $k$ becomes a continuous variable  (e.g. $k\in{\mathbb R}$ or a finite interval in $\mathbb R$, or $k\in{\mathbb R}^3$). 
 Protein-solvent environments are usually described by an ensemble of atoms, called a ``molecular reservoir model",  in contrast to an ensemble of modes like phonons (see also \cite{PSE}). In Sections \ref{sectdynres} and \ref{reservoirssect}, we discuss the relation between these two environment models and we show that they both lead to the same results for the exciton transfer dynamics discussed in this paper.  
 
 In the continuous mode limit, the form factor $g_k$ and the creation and annihilation operators $a^\dagger_k$, $a_k$ become functions of the continuous variable $k$, which we denote by $g(k)$ and $a^*(k)$, $a(k)$, respectively.

The Hamiltonian of the {\bf local reservoirs model} is
\begin{equation}
	\label{1}
	H_\l=\frac12 \begin{pmatrix}
		\epsilon & V\\
		V & -\epsilon
	\end{pmatrix}
	+H_{\r_1} +H_{\r_2} +\lambda_1 \begin{pmatrix}
		1 & 0\\
		0& 0
	\end{pmatrix}\otimes\phi_1(g_1) +\lambda_2 \begin{pmatrix}
	0 & 0\\
	0 & 1
\end{pmatrix}\otimes\phi_2(g_2).
\end{equation}
Now we have {\em two} form factors $g_1(k)$, $g_2(k)$ and {\em two} sets of creation and annihilation operators, $a^*_j(k)$ and $a_j(k)$, $j=1,2$. The values $j=1,2$ label the reservoirs. All operators of one reservoir commute with all those of the other.

We consider {\bf initial states} of the form
\begin{equation}
	\rho_{\rm in} = \rho_\s\otimes \rho_{\r} \quad\mbox{or}\quad 
	\rho_{\rm in} = \rho_\s\otimes \rho_{\r_1}\otimes\rho_{\r_2}
\end{equation}
for the collective or local models, respectively. Here, $\rho_\s$ is an arbitrary initial density matrix of the dimer and $\rho_{\r}$, $\rho_{\r_1}$, $\rho_{\r_2}$ are the reservoir equilibrium states at a temperature $T=1/\beta>0$. The {\bf reduced dimer density matrix} at time $t$ is obtained by evolving the whole dimer-reservoir density matrix and then tracing out the reservoirs, 
\begin{equation}
	\label{a2}
	\rho_\s(t) = {\rm Tr}_{\rm Reservoir(s)} \left(\e^{-\i t H} \rho_{{\rm in}} \e^{\i t H}\right).
\end{equation}
Here, $H=H_\c$ or $H=H_\l$, depending on the model considered. 

\medskip

The chlorophyll molecules in the dimer have the optical transitional frequencies, with excited energy levels $E_{\rm exc}\approx 1.5-2.5\rm eV$. The thermal bath usually has much lower excited modes energy, $E_T\approx 25\rm meV$. 

\subsection{Outline of results}

Our main goal is to describe the dynamics of the reduced dimer density matrix.  In Theorems \ref{thm1} and \ref{thm2} below, we give the law of evolution of the populations (diagonal elements of the density matrix in the energy basis) and of decoherence (off-diagonal). These results exhibit a main term in the dynamics and a remainder which is {\em controlled for all times $t\ge0$}. In order to derive the dynamical laws mathematically, we need to impose a regularity condition on the form factors $g$, $g_{1,2}$ in \eqref{1'}, \eqref{1}. It is best expressed as a condition on the {\bf spectral function of the reservoir}, which is defined as \cite{MBSa,Leggett} 
\begin{equation}
\label{mr39}
J(\omega) =\sqrt{2\pi}\, \tanh(\beta\omega/2)\,\widehat{\cal C}(\omega),\qquad \omega\ge 0,
\end{equation}
where $\widehat{\cal C}(v)=\frac{1}{\sqrt{2\pi}}\,\int_{-\infty}^\infty \e^{-\i vt}{\cal C}(t) \d t$, $v\in{\mathbb R}$, is the Fourier transform of the 
symmetrized correlation function
\begin{equation}
\label{mr37}
{\cal C}(t) =\tfrac12\left( \av{\phi_t(g)\phi(g)}_\beta + \av{\phi(g)\phi_t(g)}_\beta\right).
\end{equation}
Here, $\av{\cdot}_\beta$ is the average of a reservoir observable in the reservoir equilibrium at temperature $T=1/\beta$, $g$ is the form factor appearing in \eqref{1'}, \eqref{1} and $\phi_t(g)=\e^{\i tH_\r}\phi(g)\e^{-\i tH_\r}$, $H_\r$ the uncoupled Hamiltonian of the reservoir(s). Of course, in the model where we have two reservoirs, there are two corresponding spectral functions $J_1(\omega)$, $J_2(\omega)$. We point out that the product $\tanh(\beta\omega/2)\widehat{\cal C}(\omega)$ appearing in the definition \eqref{mr39} is actually {\em independent} of $\beta$, expressible purely in terms of the form factor (c.f. \cite{MBSa,Leggett} and also Section \ref{reservoirssect}). The mathematical regularity condition we impose (for each reservoir) is 
\begin{equation}
\label{regj}
J(\omega) = \frac{\omega^{2p+2}}{(1+\omega)^\sigma} \widetilde J(\omega),
\end{equation}
where 
\begin{equation}
\label{1.7j}
p=\tfrac12,\tfrac32,\tfrac52,\tfrac72,\quad\mbox{or}\quad p>4 \quad \mbox{and}\quad \sigma >3/2,
\end{equation}
and where $\widetilde J(\omega)$ is a bounded function of $\omega\ge 0$. We remark that in principle, a `minimal condition' on the low-modes behavior is $p>-1/2$, but so far our mathematical techniques require the more restrictive values in \eqref{1.7j}.\footnote{While our current rigorous mathematical approach requires the restriction \eqref{1.7j} on $p$, we believe  that the only `rigid' requirement is $p>-1/2$. This same restriction occurs in Leggett's work \cite{Leggett} and is needed in the application of a unitary transformation (polaron transformation, which is defined only if $p>-1/2$) to obtain a Hamiltonian in which $\lambda_{1,2}$ do not have to be considered to be small parameters, but where $V$ is the small perturbation parameter. An improvement of our current mathematical approach is likely to remove some of the restriction on $p$.  See Sections \ref{regularitysect} and \ref{reservoirssect} for further detail on this point.}

\subsubsection{Relaxation of the dimer}

We denote the population of site (level) $1$ by
\begin{equation}
	\label{a50}
	p(t) = \scalprod{\varphi_1}{\rho_\s(t)\varphi_1} = [\rho_\s(t)]_{11},
\end{equation}
where the dimer site basis for $V=0$ (or, energy basis) is 
\begin{equation}
	\label{eigenbasis}
	\varphi_1 =
	\begin{pmatrix}
		1\\
		0
	\end{pmatrix}
	\qquad\mbox{and}\qquad
	\varphi_2 =
	\begin{pmatrix}
		0\\
		1
	\end{pmatrix}.
\end{equation}
The initial population is $p(0)\in [0,1]$. We show in Theorem \ref{thm1} that, for arbitrary values of $\lambda_1$ and $\lambda_2$ and sufficiently small values of $V$, and for {\em all times $t\ge 0$},
\begin{equation}
	\label{introa51}
	p(t) = p_\infty +\e^{-\gamma t}\left( p(0)-p_\infty\right) +R(t).
\end{equation}
The relation \eqref{introa51} is valid for both the local and collective reservoirs models, with different expressions for the relaxation rate $\gamma$. Namely, $\gamma$ takes specific values $\gamma_\c$ and $\gamma_\l$ (c.f. \eqref{b12}) for the collective and local reservoirs models, which we discuss below in Section \ref{ratessection}. It satisfies $\gamma\propto V^2$ and $\gamma>0$ for $V>0$. The remainder term has the bound $|R(t)|\le C/t$ for some constant $C$. The final value $p_\infty$ is the population of the dimer at equilibrium with the reservoir(s),
\begin{equation}
	\label{introa52}
	p_\infty=  \frac{\e^{-\frac{\beta}{2}(\epsilon-\alpha_1)}}{\e^{-\frac{\beta}{2}(\epsilon-\alpha_1)}+\e^{-\frac{\beta}{2}(-\epsilon-\alpha_2)}}+O(V) =\frac{1}{1+\e^{\beta\hat\epsilon}} +O(V),
\end{equation}
where
\begin{equation}
\label{epsilonhat}
\hat\epsilon =\hat\epsilon(\lambda_1,\lambda_2)= \epsilon -\frac{\alpha_1-\alpha_2}{2}.
\end{equation}
Here, $\alpha_{1,2}\ge 0$ are {\em renormalizations} of the dimer donor and acceptor energies (see Section \ref{renormsect}), which are caused by the interaction with the reservoir(s). Their precise expressions are given in \eqref{a57} and they satisfy $\alpha_j\propto\lambda^2_j$.

\pagebreak[1]
\medskip

{\bf Discussion}
\begin{itemize}
	
\item {\em Properties of the final population $p_\infty$.}

If $\alpha_1=\alpha_2$ then we say that the reservoirs are {\em coupled in a symmetric way} to the dimer, and we have $\hat \epsilon =\epsilon>0$. This happens for instance if $\lambda^2_1=\lambda^2_2$ and, for the local reservoirs case, if additionally $g_1=g_2$. 

However, the sign of $\hat\epsilon$ can be positive or negative depending on which reservoir coupling constant is stronger. From \eqref{epsilonhat} and $\alpha_j\propto\lambda^2_j$  we see readily that if $\lambda^2_1>\!\!> \max\{\lambda^2_2, \epsilon\}$ then $\hat\epsilon \propto -\lambda^2_1$ and if $\lambda^2_2>\!\!> \max\{\lambda^2_1,\epsilon\}$ then $\hat\epsilon\propto \lambda^2_2$. For high temperatures we have
\begin{equation}
	\label{pinfty}
p_\infty \approx \frac12-\frac{\hat\epsilon}{T},\qquad T>\!\!>|\hat\epsilon|.
\end{equation}
The small correction to the value $1/2$ is {\em negative} for the symmetrically coupled case ($\hat\epsilon=\epsilon$) and if level two is coupled to the reservoir(s) much more strongly than level one. If the level one is coupled more strongly, then that correction is {\em positive}. This shows that if one level is coupled more strongly to the reservoir(s) than the other, the final population of the more strongly coupled level is increased. While this effect is small for high temperatures (see \eqref{pinfty}), it is large for low temperatures:
\begin{equation}
	\label{pinftylow}
	p_\infty \approx
	\left\{
	\begin{array}{cl}
		1, & \mbox{if} \quad \lambda^2_1>\!\!>\max\{\lambda_2^2,\epsilon\} \\
		0, & \mbox{if} \quad \lambda^2_2>\!\!>\lambda_1^2
	\end{array}
	\right.
	\quad \mbox{and}\quad T<\!\!<|\hat\epsilon|
\end{equation}
We conclude that if $T<\!\!<|\hat\epsilon|$, then one can entirely populate level one by coupling it very strongly to the reservoir(s), or entirely depopulate it by coupling the other level very strongly to the reservoir(s).
	
\item {\em Domain of usefulness of expansion \eqref{introa51}.} 

The expansion is meaningful if 
\begin{equation}
	\label{goodbound}
	|R(t)| \le  | p_\infty+\e^{-\gamma t}(p(0)-p_\infty)|.
\end{equation}
If $\gamma t<\!\!<1$ then the right side is just $p(0)$ and so the bound is satisfied if in addition $t\ge C/p(0)$, where $C$ is the constant in the upper bound of $R(t)$ (see before \eqref{introa52}). Hence \eqref{goodbound} holds for 
\begin{equation}
	\label{bound1}
C/p(0)<\!\!< t<\!\!<1/\gamma \propto V^{-2}.
\end{equation}
This is useful if $p(0)$ is not very small. Other domains can be found as follows. If $p(0)-p_\infty\ge 0$ then \eqref{goodbound} is satisfied for $t\ge C/p_\infty$. Similarly, if $p(0)-p_\infty<0$, then we have $q(0)-q_\infty>0$, where $q(t)=1-p(t)$ and $q_\infty=1-p_\infty$ are the level 2 populations. Now  \eqref{introa51} gives $q(t)=q_\infty +\e^{-\gamma t}(q(0)-q_\infty)-R(t)$, where the remainder is the same as in \eqref{introa51}. Hence, this remainder is small compared to the main term for $t\ge C/q_\infty$.
At high temperatures, $p_\infty\approx 1/2\approx q_\infty$ and the remainder is small for times $t\ge 2C$, regardless of the initial condition. 

\item{\em A different remainder estimate. } The dynamical resonance theory of \cite{KoMe} which we use to prove the expansion \eqref{introa51} is designed to give a remainder that decays as $t\rightarrow\infty$. Instead, one can modify this theory to yield a remainder which is $O(V)$ for all times, but may not vanish at $t=\infty$.

\end{itemize}

\subsubsection{Decoherence of the dimer}

Let 
$$
[\rho_\s(t)]_{12} = \scalprod{\varphi_1}{\rho_\s(t)\varphi_2}
$$
be the off-diagonal density matrix of the dimer in the basis \eqref{eigenbasis}. We show in Theorem \ref{thm2} that for arbitrary $\lambda_1, \lambda_2$, for small enough $V$, and {\em all times $t\ge0$}, 
\begin{equation}
\label{introa64}
{}[\rho_\s(t)]_{12} = \e^{-\gamma t/2}\, \e^{-\i t(\hat\epsilon+x_{\rm LS})}\  \e^{-\Gamma_\infty} \,  [\rho_\s(0)]_{12} +O(V) +R(t),
\end{equation}
where $\gamma$ is the relaxation rate (the same as in \eqref{introa51}, see \eqref{b12} for the explicit expression) and  $x_{\rm LS}\in\mathbb R$ is the Lamb shift (c.f. \eqref{lambshift}). Relation \eqref{introa64} is valid for both the local and collective reservoirs models (the $\gamma$ having different expressions). The term $O(V)$ is independent of time $t$, and $R(t)$ satisfies $|R(t)|\le C/t$, for some constant $C$. The constant $\Gamma_\infty\ge 0$ is given explicitly in \eqref{gammainfty}. It describes the large time decoherence of the dimer under the dynamics with $V=0$.   Namely, for $V=0$ the dimer dynamics can be solved exactly (see Section \ref{subsectB2}),
\begin{equation}
\label{introa65}
{}[\rho_\s(t)]_{12} = \e^{-\i t\hat\epsilon} \, {\cal D}(t) \, [\rho_\s(0)]_{12},\qquad (V=0)
\end{equation}
where ${\cal D}(t)$ (which can be expressed in terms of a reservoir correlation function, see Section \ref{decoorigin}) satisfies
\begin{equation}
\label{introa66}
\lim_{t\rightarrow\infty} {\cal D}(t) =\e^{-\Gamma_\infty}.
\end{equation}

{\bf Discussion}

\begin{itemize}
	
	\item {\it Two regimes: full and partial phase decoherence ($V=0$)}

	The dynamics for $V=0$ leaves the populations invariant, and the off-diagonal density matrix element of the dimer (energy basis) satisfies
	\begin{equation}
	\label{e2}
	\big| [\rho_\s(t)]_{12}\big|  =\big| [\rho_\s(0)]_{12}\big|
	\left\{
	\begin{array}{ll}
	\e^{-\frac1\pi(\lambda_1-\lambda_2)^2Q_2(t)} & \mbox{collective reservoir}\\
	\e^{-\frac1\pi\lambda_1^2Q_2^{(1)}(t)}\e^{-\frac1\pi\lambda_2^2Q_2^{(2)}(t)} & \mbox{local reservoirs}
	\end{array}
	\right.
	  \quad (V=0),
	\end{equation}
	for some functions $Q_2(t), Q_2^{(j)}(t)\ge 0$ (see \eqref{b4}, \eqref{b3}). The process described by \eqref{e2} is called phase decoherence. We say that {\em full} phase decoherence takes place if $[\rho_S(t)]_{12}\rightarrow 0$ as $t\rightarrow \infty$. Otherwise we call the phase decoherence {\em partial}. Full phase decoherence happens if and only if $\lim_{t\rightarrow\infty} Q_2(t)=\infty$ and $\lambda_1\neq\lambda_2$ (collective) or at least one $\lim_{t\rightarrow\infty}Q_2^{(j)}(t)=\infty$ and $\lambda_j\neq 0$ (local). We show in \eqref{iff} that 
	\begin{equation}
	\label{e3}
	\mbox{Full phase decoherence}\qquad \Longleftrightarrow\qquad p\le 0,
	\end{equation} 
	where $p$ is the parameter in \eqref{regj} (see also Section \ref{decoorigin}). 
	
	In the situation of partial phase decoherence (for $V=0$), i.e., when $p>0$,  the spectral function of the reservoir, \eqref{regj}, satisfies  $J(\omega)\rightarrow 0$ as $\omega\rightarrow 0$. This low-mode behavior of $J$ is not always satisfied in noisy protein environments. In particular, it is not true in the presence of the so-called $1/f$ noise \cite{1f}. At the same time, a vanishing spectral density in the limit of low frequencies can be realized in many LHCs, see \cite{Vib4} (as well  references therein).

	Our technical condition \eqref{1.7j}, used to derive the evolution of the dimer mathematically rigorously, places us in the regime of partial phase decoherence. But as mentioned above, we expect that the mathematical analysis can be pushed to cover the range $p>-1/2$. It is then reasonable to include a discussion of the relaxation rates in the situation of full phase decoherence ($-1/2<p\le 0$) as well, in what follows. 
	
	\item {\it Domain of usefulness of expansion \eqref{introa64}.}
	
	Analogously to the discussion of \eqref{goodbound}, one sees that the remainder in \eqref{introa64} is small if 
	$$
	C\e^{\Gamma_\infty} / [\rho_\s(0)]_{12} <\!\!< t<\!\!<  1/\gamma \propto V^{-2}.
	$$
	We may view $\e^{-\Gamma_\infty}[\rho_\s(0)]_{12}$ in \eqref{introa64}  as a {\em shifted initial condition}, whose absolute value (but not phase) is that of the asymptotic dynamics with $V=0$. Under the dynamics with $V=0$, the absolute value of the off-diagonal matrix element reaches its final value $\e^{-\Gamma_\infty}\big| [\rho_\s(0)]_{12}\big|$ at the reservoir correlation time independent of $V$ (c.f. \eqref{introa66}, \eqref{be6}, Lemma \ref{lemma1}). For $V$ small, the factor $\e^{-\gamma t/2}\approx 1$ in \eqref{introa64} has not yet started to decay at that point in time.

	\item {\it Possible improvement of the resonance expansion for small times.}
	
	We expect that an expansion \eqref{introa64} holds with $\e^{-\Gamma_\infty}$ replaced by $1$ and a remainder $R(t)=O(V)$ for all times. However, the existing dynamical resonance theory must be modified to reach that result. In its present form, it is not accurate to describe decoherence for small times. Indeed, it follows from \eqref{introa64} that  $R(0)=[\rho_\s(0)]_{12}\cdot  O(1-\e^{-\Gamma_\infty})+O(V)$, which may not be small. We explain how the factor $\e^{-\Gamma_\infty}$ appears naturally within the resonance description of the dynamics, and why it is not present (rather, $\Gamma_\infty=0$) in the dynamics of the populations. See Section \ref{explansect}, equation \eqref{-04}.
	
\item{{\em Relation between decoherence and relaxation rates for arbitrary interaction strength.}} 

It is well known from the Bloch-Redfield weak coupling theory that if $\lambda$ is small and $J(0)=0$,  then the relaxation rate $\gamma_{\rm relax}$ (decay of diagonal) and the decoherence rate $\gamma_{\rm deco}$ (decay of off-diagonal) satisfy the relation 
\begin{equation}
	\label{therelation}
	\gamma_{\rm deco}=\gamma_{\rm relax}/2.
\end{equation} 
Our results \eqref{introa51} and \eqref{introa64} show that the relation \eqref{therelation} is {\em valid for arbitrary $\lambda$}.

\end{itemize}

 \subsubsection{Relaxation rates}
 \label{ratessection}
 
 The relaxation rates of the collective and local reservoirs models are given by 
 \begin{eqnarray}
 \begin{split}
 \label{b12}
 \gamma_\c &= V^2\lim_{r\rightarrow 0_+}\int_0^\infty \!\!\e^{-r t} \cos(\hat\epsilon t) \ \cos\left[\frac{(\lambda_1-\lambda_2)^2}{\pi}Q_1(t)\right] \exp\left[ -\frac{(\lambda_1-\lambda_2)^2}{\pi}Q_2(t)\right] \d t,\\
 \gamma_\l &= V^2 \lim_{r\rightarrow 0_+}\int_0^\infty\!\!\e^{-r t} \cos(\hat\epsilon t)\, \cos\left[ \frac{\lambda_1^2}{\pi}Q_1^{(1)}(t) +\frac{\lambda_2^2}{\pi}Q_1^{(2)}(t)\right] \exp \left[-\frac{\lambda_1^2}{\pi}Q_2^{(1)}(t) -  \frac{\lambda_2^2}{\pi}Q_2^{(2)}(t)\right] \d t,
 \end{split}
 \end{eqnarray}
 where $\hat\epsilon$ is defined in \eqref{epsilonhat} and where (collective reservoir)
 \begin{equation}
 \begin{split}
 Q_1(t) &= \int_0^\infty \frac{J(\omega)}{\omega^2}\sin(\omega t) \d\omega,\\ 
 Q_2(t) &= \int_0^\infty \frac{J(\omega)(1-\cos(\omega t))}{\omega^2}\coth(\beta\omega/2) \d\omega
 \label{b4}
 \end{split}
 \end{equation}
 and (local reservoirs)
 \begin{equation}
 \begin{split}	
 Q_1^{(j)}(t) &=\int_0^\infty \frac{J_j(\omega)}{\omega^2} \sin(\omega t) \d\omega,\\
 Q_2^{(j)}(t) &=\int_0^\infty \frac{J_j(\omega) (1-\cos(\omega t))}{\omega^2} \coth(\beta\omega/2) \,  \d \omega.
 \label{b3}
 \end{split}
 \end{equation}
 The limit $r\rightarrow 0_+$ appears in \eqref{b12} because the analysis involves Green's functions close to the real axis. The correctness of the formula \eqref{b12} depends crucially on the presence of this limit when the function to be integrated does not decay to zero as  $t\rightarrow\infty$. This happens in the case of partial phase decoherence, where $Q_2(t)\not\rightarrow\infty$ for large times. In the contrary case, when the integrand in \eqref{b12} for $r=0$ is integrable, we can leave out the limit in the expression for $\gamma_\c$ and $\gamma_\l$ (just set $r=0$). Both situations are reasonable, but they need to be discussed separately. Note that for $\lambda_1=\lambda_2$, we obtain from \eqref{b12} that $\gamma_\c=0$ (for $\epsilon\neq 0$).
 
 \medskip
 
 {\em Remark.\ } The expansion \eqref{introa51} is valid for small values of $V$. In particular (see \eqref{need2}), $V\le {\rm const.}\widetilde\gamma$, where $\widetilde\gamma$ is $\gamma_\c/V^2$ or $\gamma_\l/V^2$ (which is independent of $V$). One can expand $\gamma$, \eqref{b12}, for small $\lambda$. To lowest order ($\lambda^2$), one then recovers the expression predicted by the Bloch-Redfield theory (Fermi golden rule) in the weak coupling limit (small $\lambda$). This has been shown in \cite{MBSa}, Section 5.

\medskip 
 
 We now give more manageable expressions of $\gamma_\c, \gamma_\l$ than \eqref{b12}, in both regimes of full and of partial phase decoherence. These easier (approximate) expressions are given in \eqref{e13}, \eqref{e13'} and \eqref{e18}, \eqref{MRL} and are derived from \eqref{b12} in Section \ref{derivationsubsect}. Let us consider spectral functions of the reservoir of the form \eqref{1.7j} with $p>-1/2$ and with an exponential high-mode cutoff $\omega_c>0$,
 \begin{equation}	
 \label{e1}
 J(\omega) =A_p\,\omega^{2p+2} \e^{-\omega/\omega_c},
 \end{equation}
 where $A_p>0$. In the local reservoirs model, each reservoir's spectral function $J_j(\omega)$ is of the form \eqref{e1} with possibly different $p$ and $\omega_c$. Set
 \begin{equation}
 \label{e15}
 \nu =\int_0^\infty \frac{J(\omega)}{\omega}\d\omega = A_p\,\omega_c^{2p+2} \int_0^\infty y^{2p+1}\e^{-y}\d y \qquad\mbox{and}\qquad
 \nu_j =\int_0^\infty \frac{J_j(\omega)}{\omega}\d\omega.
 \end{equation}
 
 \medskip
 \noindent
{\bf Dimensionalities.\ } We adopt units in which $\hbar=1$. The dimensions are as follows.

\smallskip

-- Dimensionless quantities: $J(\omega)$, $\nu$, $\nu_j$, $A_p\,\omega^{2p+2}$

-- Dimension energy: $\epsilon$, $\omega$, $\omega_c$, $T$,  $\lambda_j^2$

-- Dimension 1/energy:  $Q_j$, $Q_j^{(k)}$, $g_j(k)$ (c.f. \eqref{1})

\smallskip

Since the spectral function \eqref{e1} is dimensionless, the coefficient $A_p$ has the same dimension as $\omega^{-2p-2}$, which is $({\rm energy})^{-2p-2}$ and depends on $p$. $J(\omega)$ is quadratic in the form factor(s) $g$ ($g_j$) (c.f. \eqref{mr39}, \eqref{mr37}) and the form factor(s) are defined only up to multiplication with the coupling constants $\lambda_1,\lambda_2$, see \eqref{1'} and \eqref{1}. Therefore, only the combination $\lambda_j \sqrt{A_p}$ ever appears. The expressions for the relaxation rates do not depend on $A_p$.
 
 \medskip
 
 {\bf Relaxation rate in the regime of partial phase decoherence, $p>0$.}
 
\noindent
 Consider first the {\bf collective reservoir} model and the regime
 \begin{equation}
 \label{regime1}
 \omega_c <\!\!<T, \quad (\lambda_1-\lambda_2)^2\nu  <\!\!< \omega_c^2/T  \quad \mbox{and}\quad   |\epsilon-\tfrac\nu\pi (\lambda_1^2-\lambda_2^2)| <\!\!< \omega_c.
 \end{equation}
 The first inequality in (\ref{regime1}), $\beta\omega_c\ll 1$, is called the high temperature regime \cite{XuSch} and usually covers room temperatures.
 The dimer relaxation rate \eqref{b12} has the (approximate) expression
 \begin{equation}
 \label{e13}
 \gamma_\c  = V^2 \omega_c^{-1} \big( 1- \e^{-\frac{2TB}{\pi} (\lambda_1-\lambda_2)^2}\big),
 \end{equation}
 where $B=B(p,\omega_c)=\int_0^\infty \tfrac{J(\omega)}{\omega^3}\d\omega$. For instance, $B(1/2,\omega_c)= A_{1/2}\,\omega_c$.
 
 Similarly, for the {\bf local reservoirs} model, the spectral functions of the reservoirs are given by \eqref{e1} with (possibly different) $p_j>0$ and $\omega_{j,c}$. For ease of notation, we take $\omega_{1,c}\approx\omega_{2,c}\approx\omega_c$ and $p_1\approx p_2\approx p$ (the following bounds can be easily derived also when this approximate symmetry does not hold). In the high-temperature regime
 \begin{equation}
 \label{regime1'}
 \omega_c <\!\!<T, \qquad  \lambda_j^2\nu_j  <\!\!< \omega_c^2/T,\  j=1,2 \qquad \mbox{and}\qquad   |\epsilon-\tfrac1\pi (\nu_1\lambda_1^2-\nu_2\lambda_2^2)| <\!\!< \omega_c,
 \end{equation}
 the relaxation rate \eqref{b12} is given by
 \begin{equation}
 \label{e13'}
 \gamma_\l  = V^2 \omega_c^{-1} \big( 1- \e^{-\frac{2T}{\pi} (B_1\lambda_1^2+B_2\lambda_2^2)}\, \big),
 \end{equation}
where $B_j=B_j(p_j,\omega_{j,c}) =\int_0^\infty \frac{J_j(\omega)}{\omega^3} \d\omega$. 
 
 \bigskip
 
 {\bf Dimer relaxation in the regime of full phase decoherence, $-1/2<p\le0$.} 
  
\noindent
 Consider first the {\bf collective reservoir} model in the parameter region
 \begin{equation}
 \label{e20}
 \omega_c<\!\!< T\quad \mbox{and} \quad 	\omega_c^2 <\!\!< (\lambda_1-\lambda_2)^2 T\nu, 
 \end{equation}
 which also corresponds to the high-temperature regime. The dimer relaxation rate \eqref{b12} has the (approximate) expression
 \begin{equation}
 \label{e18}
 \gamma_\c = \left( \frac V2\right)^2
 \sqrt{\frac{2\pi}{T(\epsilon_{\c,1}+\epsilon_{\c,2})}}\ 
 \Big\{ \exp\Big[-\frac{(\epsilon-\epsilon_{\c,1})^2}{2T(\epsilon_{\c,1}+\epsilon_{\c,2})} \Big]  +\exp\Big[-\frac{(\epsilon+\epsilon_{\c,2})^2}{2T(\epsilon_{\c,1}+\epsilon_{\c,2})} \Big] \Big\}
 \end{equation}
 where two {\em reconstruction energies} are introduced,
 \begin{equation}
 \label{e19}
 \epsilon_{\c,j} = 2\frac\nu\pi(\lambda^2_j-\lambda_1\lambda_2),\qquad j=1,2.
 \end{equation}
 We call \eqref{e18} the {\bf Generalized Marcus Formula} (see the discussion after \eqref{Marcus}). As $\epsilon_{\c,1}+\epsilon_{\c,2}=2\frac\nu\pi(\lambda_1-\lambda_2)^2$, the second condition in \eqref{e20} is equivalent to 
 \begin{equation}
 \label{e21}
 \omega_c^2<\!\!< T(\epsilon_{\c,1}+\epsilon_{\c,2}).
 \end{equation}
 For symmetric coupling, $\lambda_1=-\lambda_2=\lambda$, we have $\epsilon_{\c,j}= \epsilon_\c=4\nu\lambda^2/\pi$ (for $j=1,2$) and \eqref{e19} takes the form
 \begin{equation}
 \label{equallambdasMarcus}
 \gamma_\c = \left(\frac V 2\right)^2 \sqrt{\frac\pi{T\epsilon_\c}}\Big( \e^{-\frac{(\epsilon-\epsilon_\c)^2}{4T\epsilon_\c}} + \e^{-\frac{(\epsilon+\epsilon_\c)^2}{4T\epsilon_\c}}\Big),
 \end{equation}
which coincides with the heuristically derived rate given in \cite{XuSch}. Typically, the second exponential is much smaller than the first one ($\epsilon,\epsilon_\c>0$) and is therefore neglected. The resulting formula,
\begin{equation}
\label{Marcus}
\gamma_\c = \left(\frac V 2\right)^2 \sqrt{\frac\pi{T\epsilon_\c}} \e^{-\frac{(\epsilon-\epsilon_\c)^2}{4T\epsilon_\c}},
\end{equation}
is the famous {\bf Marcus Formula} of electron transfer \cite{Marcus,MarcusNobel}. Hence the name Generalized Marcus Formula for \eqref{e18}, which extends the original one to the situation where the donor and acceptor may be coupled in an asymmetric way to a common reservoir. The energies $\epsilon_{\c,j}$, \eqref{e19}, play the role of generalized reorganization (reconstruction) energies. 

Similarly, for the {\bf local reservoirs} model, in the high-temperature regime
\begin{equation}
\label{e20'}
\omega_{j,c}<\!\!< T\quad \mbox{and} \quad \omega^2_{j,c}<\!\!<\lambda_j^2 T\nu_j, \quad j=1,2,
\end{equation}
the dimer relaxation rate \eqref{b12} has the expression,
\begin{equation}
\label{MRL}
\gamma_\l = \left(\frac{V}{2}\right)^2\sqrt{\frac{2\pi}{T(\epsilon_{\l,1}+\epsilon_{\l,2})}}\left\{ \exp\left[-\frac{(\epsilon-\epsilon_{\l,1})^2}{2T(\epsilon_{\l,1}+\epsilon_{\l,2})}\right] +  \exp\left[-\frac{(\epsilon+\epsilon_{\l,2})^2}{2T(\epsilon_{\l,1}+\epsilon_{\l,2})}\right]\right\},
\end{equation}
where the generalized reorganization (reconstruction) energies are
\begin{equation}
\label{elocj}
\epsilon_{\l,j} = 2\frac{\nu_j}{\pi}\lambda^2_j,\qquad j=1,2.
\end{equation}

\medskip

{\bf Discussion}

\begin{itemize}
	\item For $\lambda_1=-\lambda_2$, $\gamma_\c$ \eqref{e18}  reduces to the form derived in \cite{XuSch}. If in addition $\nu_1=\nu_2$, then $\gamma_\c=\gamma_\l$. The rate $\gamma_\l$ is a symmetric function of the coupling constants $\lambda_1$ and $\lambda_2$ (each reservoir acts independently in the same way), but $\gamma_\c$ is not. Indeed, $\gamma_\l$ only depends on the absolute values of the $\lambda_j$, while $\gamma_\c$ depends also on the sign of the product $\lambda_1\lambda_2$. If $\lambda_1=0$ or $\lambda_2=0$, then $\gamma_\c=\gamma_\l$.

	\item The {\em reconstruction energies} $\epsilon_{\c,j}$ of the collective reservoir model can be positive or negative, depending on the relative values and signs of the coupling constants. However, in the local reservoirs model, $\epsilon_{\l,j}\ge 0$ always.
	
	\item {\em Upper bound on $\gamma_\c$ in the high-temperature regime.} According to \eqref{e18} and \eqref{e21} we have 
 \begin{equation}
 	\label{upbnd2}
	\gamma_\c <2\sqrt{2\pi}(V/2)^2/\omega_c.
\end{equation}	
	
\end{itemize}
 
 \subsubsection{Derivation of \eqref{e13}, \eqref{e13'} and \eqref{e18}, \eqref{MRL} from \eqref{b12}}
 \label{derivationsubsect}
 
 {\em Derivation of \eqref{e13}.} The process is governed by two parameters: the (finite) asymptotic value
 \begin{equation}
 \label{e6}
 \lim_{t\rightarrow\infty} Q_2(t) \equiv Q_2(\infty)= \int_0^\infty \frac{J(\omega)}{\omega^2}\coth(\beta\omega/2) \d \omega
 \end{equation}
 and the characteristic time 
 \begin{equation}
 \label{e8}
 t_* = 1/\omega_c \quad \qquad \mbox{provided $\beta\omega_c <\!\!<1$}
 \end{equation}
 at which $Q_2(t)$ approaches the value $Q_2(\infty)$.\footnote{ To find the convergence speed of the limit \eqref{e6}, we calculate
 	\begin{equation}
 	\label{e7}
 	\int_0^\infty \frac{J(\omega)\cos(\omega t)}{\omega^2} \coth(\beta\omega/2)\d\omega \propto  \frac{T\omega_c}{(\omega_c t)^2+1} \qquad\mbox{for $\beta\omega_c <\!\!<1$.}
 	\end{equation}
 	Here, we have used $p=1/2$ (other values $p>0$ give the same conclusion) and that, due to the high frequency cutoff $\omega_c$ in $J$ (see \eqref{e1}), the integration is essentially over $\omega\le \omega_c$, and therefore $\coth(\beta\omega/2)\approx 2/\beta\omega$ provided $\beta\omega_c <\!\!<1$.} 
 To analyze $\gamma_\c$, \eqref{b12}, we split the integration in two domains. For $t>t_*$, we replace $\e^{ -\frac1\pi(\lambda_1-\lambda_2)^2 Q_2(t)}\approx \e^{-\frac1\pi(\lambda_1-\lambda_2)^2Q_2(\infty)}$. From\footnote{We have $Q_1(t)=A_p  \int_0^\infty\omega^{2p}\e^{-\omega/\omega_c} \sin(\omega t)\d\omega$ and upon making a change of variables $\omega'=\omega/\omega_c$, one immediately finds $Q_1(t)\le C A_p\omega_c^{2p+1}$, where $C$ is a dimensionless constant (depending on $p$).} 
 $Q_1(t)\le CA_p\omega_c^{2p+1}$ it follows that $\cos[\frac1\pi(\lambda_1-\lambda_2)^2Q_1(t)]\approx 1$ if $(\lambda_1-\lambda_2)^2 A_p\,\omega_c^{2p+1} <\!\!<1$. The last inequality is satisfied for (see \eqref{e15})
 \begin{equation}
 \label{e9}
 (\lambda_1-\lambda_2)^2\nu   <\!\!<\omega_c.
 \end{equation}
 Thus, under the conditions \eqref{e8} and \eqref{e9}, the contribution to $\gamma_\c$ from the integration over $t\ge t_*$ is
 \begin{eqnarray}
 V^2 \e^{-\frac1\pi(\lambda_1-\lambda_2)^2Q_2(\infty)}\lim_{r\rightarrow 0_+} \int_{t_*}^\infty  \e^{-r t }\cos(\hat\epsilon t)\d t &=& -V^2 \e^{-\frac1\pi(\lambda_1-\lambda_2)^2Q_2(\infty)}\  \tfrac{\sin(\hat\epsilon  t_*)}{\hat\epsilon}\nonumber\\
 & \approx&  -V^2 t_*\, \e^{-\frac1\pi(\lambda_1-\lambda_2)^2Q_2(\infty)},
 \label{e12}
 \end{eqnarray}
 where we use in the last step $|\hat\epsilon t_*|<\!\!<1$, or
 \begin{equation}
 \label{e10}
 |\hat\epsilon| <\!\!<\omega_c.
 \end{equation}
 Now we estimate the contribution to \eqref{b12} for small times $t\le t_*$. In this integral, we can set $r=0$ to begin with and we expand the integrand in \eqref{b12} as ($t$ small)
 \begin{eqnarray}
 	\lefteqn{
 \big(1-\tfrac12 \hat\epsilon^2 t^2\big) \left[ 1- \frac{(\lambda_1-\lambda_2)^4}{\pi^2}\nu^2 t^2\right] \ \left[1 -\frac{(\lambda_1-\lambda_2)^2}{\pi}T\nu t^2\right]}\nonumber\\
&& \approx 1-t^2\Big(\hat\epsilon^2/2 +(\lambda_1-\lambda_2)^4\nu^2/\pi^2 +(\lambda_1-\lambda_2)^2 T\nu/\pi\Big).
\label{est''}
\end{eqnarray}
The integral $\int_0^{t_*}...\d t$ of \eqref{est''} is then $t_*-\frac{t^3_*}{3}(\cdots)$ which we approximate by $t_*$ under the condition that $\frac{t^3_*}{3}(\cdots)<\!\!<t_*$, i.e., that 
$\hat\epsilon^2 +(\lambda_1-\lambda_2)^4\nu^2 +(\lambda_1-\lambda_2)^2 T\nu <\!\!< t_*^{-2}=\omega_c^2$. The latter inequality is implied by \eqref{regime1}. Combining this with \eqref{e12} and \eqref{e8} gives
 \begin{equation}
 \label{e11}
 \gamma_\c = V^2 \omega_c^{-1} \big( 1- \e^{-\frac1\pi(\lambda_1-\lambda_2)^2Q_2(\infty)}\big).
 \end{equation}
 Further, again for $\beta\omega_c<\!\!<1$,  we can replace $\coth(\beta\omega/2)$ in \eqref{e6} by $2/\beta\omega$. Then $Q_2(\infty) \approx 2T\int_0^\infty\frac{J(\omega)}{\omega^3}\d\omega$ and so \eqref{e13} follows from \eqref{e11}. Equation \eqref{e13'} is derived in the same way.
 
 \bigskip
 
 {\em Derivation of \eqref{e18}.}
 {}For $-1/2<p\le 0$ we have $Q_2(t)\rightarrow\infty$. The region of integration in \eqref{b4} is essentially $\omega\le\omega_c$, due to the high energy mode cutoff \eqref{e1}. Therefore, for $\beta\omega_c<\!\!<1$, we replace $\coth(\beta\omega/2)\approx 2/(\beta\omega)$ in \eqref{b4}, so that 
 \begin{equation}
\label{e14}
 Q_2(t) \approx 2T\int_0^\infty \frac{J(\omega)(1-\cos(\omega t))}{\omega^3}\d\omega\approx T\nu t^2,
\end{equation}
 where the last approximation is valid for times $t$ satisfying $\omega_c t<\!\!<1$ (in order that $1-\cos(\omega t)\approx\frac12\omega^2t^2$). The approximation $Q_2(t) \approx  T\nu t^2$ has also been used in \cite{XuSch}. As $Q_2$ is increasing, the exponential factor in \eqref{b12} for $t\ge\omega_c^{-1}$ is estimated as
 $$
 \e^{-\frac{(\lambda_1-\lambda_2)^2}{\pi} Q_2(t)}\approx \e^{-\frac{(\lambda_1-\lambda_2)^2}{\pi} T\nu t^2}\lessapprox \e^{-\frac{(\lambda_1-\lambda_2)^2}{\pi} T\nu/\omega_c^2} <\!\!<1,
 $$
 provided
 \begin{equation}
 	\label{e16}
 	(\lambda_1-\lambda_2)^2 T\nu >\!\!>\omega_c^2.
 \end{equation} 
The region $t\ge\omega_c^{-1}$ can thus be neglected in the integral and \eqref{b12} and we can also approximate $Q_1(t)$ in \eqref{b4} by its behaviour for small times (see also \cite{XuSch}),
 \begin{equation}
 \label{e14'}
 Q_1(t) \approx \nu t.
 \end{equation}
 Consequently,
 \begin{equation}
 \label{e17}
 \gamma_\c = V^2\int_0^\infty \cos(\hat\epsilon t) \cos[(\lambda_1-\lambda_2)^2\nu t/\pi] \ \e^{-\frac1\pi (\lambda_1-\lambda_2)^2 T\nu t^2}\  \d t.
 \end{equation}
 Using $\cos(x)\cos(y) = \frac12\{\cos(x+y)+\cos(x-y)\}$ and $\int_0^\infty \cos(bx) \e^{-ax^2}\d x =\sqrt{\pi/4a}\ \e^{-b^2/4a}$, we evaluate \eqref{e17} explicitly to obtain \eqref{e18}. Relation \eqref{MRL} is obtained in the same way.

\section{Main results: details}
\label{mainresdetsect}

\subsection{Energy renormalization}
\label{renormsect}

{}For the collective reservoir model and the local reservoirs model, repectively, set
 \begin{equation}
 	\label{a57}
 	\alpha_j = \frac{2\lambda^2_j\nu}{\pi},\ \ j=1,2\qquad \mbox{and}\qquad  \alpha_j = \frac{2\lambda^2_j\nu_j}{\pi},\ \ j=1,2,
 \end{equation}
 where the $\nu$, $\nu_j$ are given in \eqref{e15}. 
Consider $H_\c$, \eqref{1'} and $H_\l$, \eqref{1} with $V=0$ and  $\lambda_1, \lambda_2$ arbitrary.  We show in Appendix \ref{subsectB1} that the equilibrium state of the interacting dimer-reservoir system at temperature $T=1/\beta>0$ for the model with the collective reservoir is
\begin{eqnarray}
\rho_{\beta,\vlambda,0} &=& \frac{\e^{-\frac{\beta}{2}(\epsilon-\alpha_1)}}{\e^{-\frac{\beta}{2}(\epsilon-\alpha_1)}+\e^{-\frac{\beta}{2}(-\epsilon-\alpha_2)}} \  |\varphi_1\rangle\langle\varphi_1|\otimes W^*\big(\lambda_1g/(\i\omega)\big)\rho_{\r} W\big(\lambda_1g/(\i\omega)\big)\nonumber\\
&& +\frac{\e^{-\frac{\beta}{2}(-\epsilon-\alpha_2)}}{\e^{-\frac{\beta}{2}(\epsilon-\alpha_1)}+\e^{-\frac{\beta}{2}(-\epsilon-\alpha_2)}} \ |\varphi_2\rangle\langle\varphi_2| \otimes W^*\big(\lambda_2g/(\i\omega)\big)\rho_\r W\big(\lambda_2g/(\i\omega)\big),\quad
\label{01'}
\end{eqnarray}
where 
\begin{equation}
\label{weylop}
W(h) = \e^{\i \phi(h)}
\end{equation}
is the unitary Weyl operator.  By tracing out the reservoir in \eqref{01'} we find
\begin{equation}
\label{a58}
{\rm Tr}_\r\ \rho_{\beta,\vlambda,0} = \frac{\e^{-\beta H_\s^{\rm ren}}}{Z_\s^{\rm ren}},\quad \mbox{with}\quad H^{\rm ren}_S=\frac12
\begin{pmatrix}
\epsilon-\alpha_1  & 0\\
0 & -\epsilon-\alpha_2
\end{pmatrix}.
\end{equation}
In this sense, the system energies $\pm\frac12\epsilon$ are {\em renormalized} by $-\frac12\alpha_j\propto-\lambda_j^2$, due to the interaction with the reservoir. Note though, that {\em the system is entangled with the reservoir in the equilibrium state} $\rho_{\beta,\vlambda,0}$.

The situation for the model with the two local reservoirs is analogous. The interacting dimer-reservoirs equilibrium state ($V=0$) is given by
\begin{eqnarray}
\rho_{\beta,\vlambda,0} &=& \frac{\e^{-\frac{\beta}{2}(\epsilon-\alpha_1)}}{\e^{-\frac{\beta}{2}(\epsilon-\alpha_1)}+\e^{-\frac{\beta}{2}(-\epsilon-\alpha_2)}} \  |\varphi_1\rangle\langle\varphi_1|\otimes W^*\big(\lambda_1g_1/(\i\omega)\big)\rho_{\r_1} W\big(\lambda_1g_1/(\i\omega)\big)\otimes\rho_{\r_2}\nonumber\\
&& +\frac{\e^{-\frac{\beta}{2}(-\epsilon-\alpha_2)}}{\e^{-\frac{\beta}{2}(\epsilon-\alpha_1)}+\e^{-\frac{\beta}{2}(-\epsilon-\alpha_2)}}\  |\varphi_2\rangle\langle\varphi_2|\otimes\rho_{\r_1} \otimes W^*\big(\lambda_2g_2/(\i\omega)\big)\rho_{\r_2} W\big(\lambda_2g_2/(\i\omega)\big).\nonumber\\
& 
\label{01}
\end{eqnarray}
By tracing out the reservoir in \eqref{01} we find again a dimer state of the form \eqref{a58}, with $\alpha_{1,2}$ given by \eqref{a57}. 

We point out that if $\alpha_1=\alpha_2$ (e.g. if $\lambda_1=\lambda_2=\lambda$ and $g_1=g_2=g$), then $H_S^{\rm ren}$ differs from $H_\s$ (with $V=0$) just by a constant and the reduced dimer equilibrium state is the same for the coupled and the uncoupled system.

\subsection{Relaxation and decoherence}

In Section \ref{sectdynres}, we prove the following result.

\begin{thm}[Population dynamics, relaxation]
\label{thm1}
Let $\lambda_1$, $\lambda_2$ be arbitrary. There is a $V_0>0$ such that if  $0<|V|<V_0$, then 
\begin{equation}
\label{a51}
p(t) = p_\infty +\e^{-\gamma t}\left( p(0)-p_\infty\right) +R(t),
\end{equation}
where 
\begin{equation}
\label{a52}
p_\infty=  \frac{\e^{-\frac{\beta}{2}(\epsilon-\alpha_1)}}{\e^{-\frac{\beta}{2}(\epsilon-\alpha_1)}+\e^{-\frac{\beta}{2}(-\epsilon-\alpha_2)}} +O(V)
\end{equation}
is the level 1 population of the  dimer density matrix at equilibrium coupled with the reservoir(s). Here $\alpha_{1,2}$ are the renormalization energies given in \eqref{a57}, and  the relaxation rate $\gamma$ is given by \eqref{b12}, for the model with the collective reservoir and that with the local ones, respectively. The remainder term $O(V)$ is independent of time $t$ and $R(t)$ satisfies
\begin{equation}
\label{a63}
R(0)=0 \quad \mbox{and}\quad  |R(t)|\le C_1/t, 
\end{equation}
for some constant $C_1$.
\end{thm}

{\em Remark.\ } The dynamical resonance theory of \cite{KoMe} which we use to prove the expansion \eqref{a51} is set up so as to give a remainder that decays as $t\rightarrow\infty$. Instead, one could modify this theory to yield a remainder which is $O(V)$ for all times, but may not vanish at $t=\infty$. We do not explore this modification here.

\bigskip
We now discuss the decoherence properties of the dimer. The models with $V=0$ can be solved exactly, see Section \ref{subsectB2}. Namely,
\begin{equation}
\label{a65}
{}[\rho_\s(t)]_{12} = \e^{-\i t\hat\epsilon} \, {\cal D}(t) \, [\rho_\s(0)]_{12},\qquad (V=0)
\end{equation}
 where $\hat\epsilon$ is given in \eqref{epsilonhat} and 
\begin{equation}
\label{a66}
{\cal D}(t) =
\left\{
\begin{array}{l} \e^{-\frac{\i}{\pi}(\lambda_1^2-\lambda_2^2) Q_1(t)} \e^{-\frac{1}{\pi} (\lambda_1-\lambda_2)^2 Q_2(t)}\mbox{\quad (collective)}\\
	\mbox{\ \ and}\\
\e^{-\frac{i}{\pi} [ \lambda^2_1 Q_1^{(1)}(t) - \lambda^2_2 Q_1^{(2)}(t)]}    \e^{-\frac1\pi [\lambda^2_1 Q_2^{(1)}(t) +\lambda^2_2 Q_2^{(2)}(t)]} \mbox{\quad (local)}
\end{array}
\right.
\end{equation}
for the collective and the local resevoirs models, respectively, with $Q_{1,2}(t)$ and $Q_{1,2}^{(j)}(t)$ given in \eqref{b4} and \eqref{b3}. The following result is proven in Section \ref{subsectB2}, see  \eqref{a70}. 

\begin{prop}
\label{prop4}
We have 
\begin{equation}
\label{factor}
\lim_{t\rightarrow\infty}{\cal D}(t) = \e^{-\Gamma_\infty}
\end{equation}
where
\begin{equation}
\label{gammainfty}
\Gamma_\infty= 
\left\{
\begin{array}{l}
\frac1\pi (\lambda_1-\lambda_2)^2\int_0^\infty\frac{J(\omega)}{\omega^2}\coth(\beta\omega/2)\d\omega \mbox{\quad {\rm (collective)}}\\
\mbox{\rm \ \ and}\\
\frac1\pi \lambda_1^2\int_0^\infty\frac{J_1(\omega)}{\omega^2}\coth(\beta\omega/2)\d\omega + \frac1\pi \lambda_2^2\int_0^\infty\frac{J_2(\omega)}{\omega^2}\coth(\beta\omega/2)\d\omega \mbox{\quad {\rm (local)}}
\end{array}
\right.
\end{equation}
for the collective and the local reservoirs models, respectively. 
\end{prop}

{\em Remark.\ } For spectral functions $J(\omega)$ of the form \eqref{regj} we have 
\begin{equation}
\label{iff}
\Gamma_\infty =\infty \quad\Longleftrightarrow\quad p\le 0,
\end{equation}
provided $\lambda_1\neq\lambda_2$ (collective) and  $\lambda_1,\lambda_2\neq0$ (local), the divergence of the integrals in  \eqref{gammainfty} for $p\le 0$ stemming from a non-integrable singularity at low modes ($\omega\approx 0$).

We now introduce the {\em Lamb shift},
\begin{eqnarray}
\lefteqn{
	x_{\rm LS}=\frac{V^2}{2} \lim_{r\rightarrow 0_+}
	 	\label{lambshift}}\\
&&\left\{
\begin{array}{l}
	 \int_0^\infty
	 \e^{-r t} \sin(\hat\epsilon t) \cos\Big[\frac{(\lambda_1-\lambda_2)^2}{\pi} Q_1(t)\Big] \exp\Big[-\frac{(\lambda_1-\lambda_2)^2}{\pi}Q_2(t)\Big]\nonumber \d t \quad \mbox{(collective)}\\
	 \\
	 \int_0^\infty \e^{-r t} \sin(\hat\epsilon t) \cos\Big[\frac{\lambda^2_1}{\pi}  Q_1^{(1)}(t) + \frac{\lambda^2_2}{\pi}  Q_1^{(2)}(t)\Big] \exp \Big[ -\frac{\lambda_1^2}{\pi}Q_2^{(1)}(t) -\frac{\lambda_2^2}{\pi}Q_2^{(2)}(t) \Big] \d t\quad \mbox{(local)}\nonumber
\end{array}
\right.
\end{eqnarray}

In Section \ref{sectdynres}, we prove the following result.

\begin{thm}[Decoherence]
\label{thm2}
Let $\lambda_1$, $\lambda_2$ be arbitrary. There is a $V_0>0$ such that if  $0<|V|<V_0$, then 
\begin{equation}
\label{a64}
{}[\rho_\s(t)]_{12} = \e^{-\Gamma_\infty} \, \e^{-\gamma t/2}\, \e^{-\i t(\hat\epsilon+x_{\rm LS})}\  [\rho_\s(0)]_{12} +O(V) +R(t),
\end{equation}
where $\gamma$ is given in \eqref{b12} and  $x_{\rm LS}$ is the Lamb shift \eqref{lambshift}. The term $O(V)$ is independent of time $t$ and $R(t)$ satisfies
\begin{equation}
\label{a63}
|R(t)|\le C_1 \quad \mbox{and}\quad  |R(t)|\le C_2/t, 
\end{equation}
for some constants $C_1, C_2$.
\end{thm}

\subsection{Numerical simulations}
\label{subsnumsim}

The spectral functions of the collective and local environments are taken as in \eqref{e1}, with  $p_1=p_2=p$ in the case of local reservoirs (the results obtained below can be generalized in a straightforward way for the case $p_1 \neq p_2$). In the numerical simulations, we have chosen a weakly coupled dimer based on two chlorophylls, $Chl^*a-Chl^*b$, in their excited states. The following parameters of the dimer were chosen,
$$
\epsilon=150 ps^{-1}\approx 99 meV \quad\mbox{and} 
\quad V=25 ps^{-1}\approx 16.5 meV
$$
(so $|V/\epsilon|\approx 0.17$), and the room temperature: $T=25 meV\approx 37.88 ps^{-1}$ ($\beta\approx 0.026 ps$).

\subsubsection*{Relaxation}

It is convenient to define new dimensionless quantities
\begin{equation}
\label{newvar}
\varepsilon= \beta \epsilon,\quad 
\varepsilon^{c}_{i}= \beta\epsilon_{\c,i},\quad \varepsilon^{l}_{i}= 
\beta\epsilon_{\l, i} \quad \mbox{$i=1,2$\ \ \  and}\quad \tau =t/\beta.
\end{equation}
Using this notation, the relaxation rates \eqref{b12} become
\begin{align}
		\label{b12a}
	\gamma_\c & = {\beta}{V^2} \int_0^\infty 
	\cos\Big[\big(\varepsilon-\frac{\varepsilon^c_1-\varepsilon^c_2}{2}\big) 
	\tau\Big] \cos\left[\frac{\varepsilon^c_1+\varepsilon^c_2}{2}\mathcal 
	Q_1(\tau)\right] \exp\left[ 
	-\frac{\varepsilon^c_1+\varepsilon^c_2}{2}\mathcal Q_2(\tau)\right] \d 
	\tau,\\
		\gamma_\l &= \beta V^2 \int_0^\infty 
		\cos\Big[\big(\varepsilon-\frac{\varepsilon^l_1-\varepsilon^l_2}{2}\big)
		 \tau\Big] \, \cos\left[ \frac{\varepsilon^l_1}{2}{\mathcal 
		Q}_1^{(1)}(\tau) +\frac{\varepsilon^l_2}{2}{\mathcal 
		Q}_1^{(2)}(\tau)\right] \nonumber \\
		& \times \exp \left[-\frac{\varepsilon^l_1}{2}{\mathcal 
		Q}_2^{(1)}(\tau) -\frac{\varepsilon^l_2}{2}{\mathcal 
		Q}_2^{(2)}(\tau)\right] \d \tau,
	\label{b12b}
\end{align}
where we understand that a limit $r\rightarrow 0_+$ has to be performed (as in \eqref{b12}), and where we have set ${\mathcal Q}_{1,2}(\tau) = { Q}_{1,2}(\tau)/(\beta \nu)$ and 
${\mathcal Q}^{(i)}_{1,2}(\tau) = { Q}^{(i)}_{1,2}(\tau)/(\beta\nu_i)$ 
($i=1,2$). We have, 
\begin{align}
\nu &= \frac{A_p\eta^{2p+2}}{\beta^{2p+2}}\int_0^\infty   z^{ 2 p+1} 
e^{-z} \d 
z,\label{s1}\\
\mathcal Q_1(\tau) &=\frac{1}{\eta \Gamma(2p+2)}\int_0^\infty   z^{ 2 p} 
e^{-z}\sin((\eta\tau z ) \d z,\label{s2}\\
\mathcal Q_2(\tau) &=\frac{1}{\eta \Gamma(2p+2)} \int_0^\infty {z^{ 2 
p} 
e^{-z}(1-\cos((\eta\tau z))}\coth(\eta z/2) \d z,
\label{b4d}
\end{align}
where $
\eta = \beta \omega_c\quad\mbox{and}\quad z = \omega/\omega_c.
$

Performing the integrations in \eqref{s1}-\eqref{b4d}, we obtain, 
\begin{align}
\nu &= \frac{A_p\eta^{2p+2}\Gamma(2p+2)}{\beta^{2p+2}} ,\\
\mathcal Q_1(\tau) &=\frac{1}{( 2 p+1)\eta} \Im\Big(\frac{1}{(1- 
i\eta\tau)^{ 2 p+1}}\Big), \\
\mathcal Q_2(\tau) &=\frac{1}{( 2 p+1)\eta}  \Re(f(0) - f(\tau)),  \qquad\mbox{ $p>-1/2$, $p \neq 0$}.
\label{Zeta}
\end{align}
Here,
\begin{eqnarray}
f(\tau) =\frac{1}{\eta^{ 2 p+1}}\bigg(\zeta\Big( 2 p+1,\frac{1}{\eta}+ 
i\tau\Big ) +\zeta\Big( 2 p+1,\frac{1}{\eta} +1+ i\tau\Big )\bigg), 
\end{eqnarray}
and $\zeta(s,q) $ denotes the Hurwitz $\zeta$-function \cite{HZ}.

One can see that for $p>0$ the function ${\mathcal Q}_2(\tau)$ is bounded 
from above by 
\begin{eqnarray}\label{Q0}
{\mathcal Q}_0 = \lim_{\tau \rightarrow \infty} {\mathcal 
Q}_2(\tau)=\frac{1}{(2p+1)\eta^{ 2 p+2}}\bigg(\zeta\Big( 2 
p+1,\frac{1}{\eta}\Big ) +\zeta\Big( 2 p+1,\frac{1}{\eta} +1\Big )\bigg).
\end{eqnarray}
Using the asymptotic properties of the $\zeta$-function, we obtain 
\begin{align}
\mathcal Q_0 \approx	\left \{ \begin{array} {ll}
	\displaystyle \frac{1}{(2p+1)p\eta^2}, &  \eta \ll 1, \quad p>0\\
		\displaystyle \frac{1}{ (2 p+1)\eta} +\frac{2\zeta (2p+1)}{ (2 
		p+1)\eta^{2p+2}} - \frac{2\zeta (2p+2)}{ \eta^{2p+3}},&  \eta \gg 1, 
		\quad p>0
\end{array}
\right.
\end{align}
where $\zeta(s)$ is the Riemann  zeta function, $\zeta(s) = \zeta(s,1)$. One can estimate $\mathcal Q_2(\tau)$ as,
\begin{align}
\mathcal Q_2(\tau) \approx	\left \{ \begin{array} {ll}
\tau^2&  \eta\ll 1,  \eta \tau \ll 1 \,\, (p>-1/2)\\
	\displaystyle 
	\frac{(p+1)(2\zeta(2p+3,\frac{1}{\eta})-\eta^{2p+3})\tau^2}{\eta^{2p+2}},&
	  \eta \tau \ll 1\, (p>-1/2)\\
		\displaystyle \frac{1}{(2p+1)\eta}\Re\bigg( 1-  \frac{1}{(1+ i\eta 
		\tau)^{2p+1}}\bigg),&  \eta\gg 1,\,\eta \tau \gg 1\, (p>0)\\
		\displaystyle \frac{1}{(2p+1)p\eta^2}\Re\bigg( 1-  \frac{1}{(1+ i\eta 
		\tau)^{2p}}\bigg),&  {}\, \eta\tau \gg 1\, (-1/2< p <0) \\
		\displaystyle \frac{1}{(2p+1)p\eta^2}\Re\bigg( 1-  \frac{1}{(1+ i\eta 
		\tau)^{2p}}\bigg),&  \eta \ll 1,\, \tau >0\, (p>-1/2)
\end{array}
\right.
\end{align}

Similar considerations yield the asymptotic behavior for the function $\mathcal Q_1(\tau)$, 
\begin{align}
\mathcal Q_1(\tau) \approx	\left \{ \begin{array} {ll}
	\tau,&  \eta \tau \ll 1\\
		\displaystyle \frac{\tau\cos(\pi p)}{(2 p+1)(\eta \tau)^{2p+2}} +  
		\frac{\tau\sin(\pi p)}{(\eta \tau)^{2p+3}} ,&  \eta \tau \gg 1, \quad p > -1/2
\end{array}
\right.
\end{align}

\begin{figure}
{\scalebox{0.35}{\includegraphics{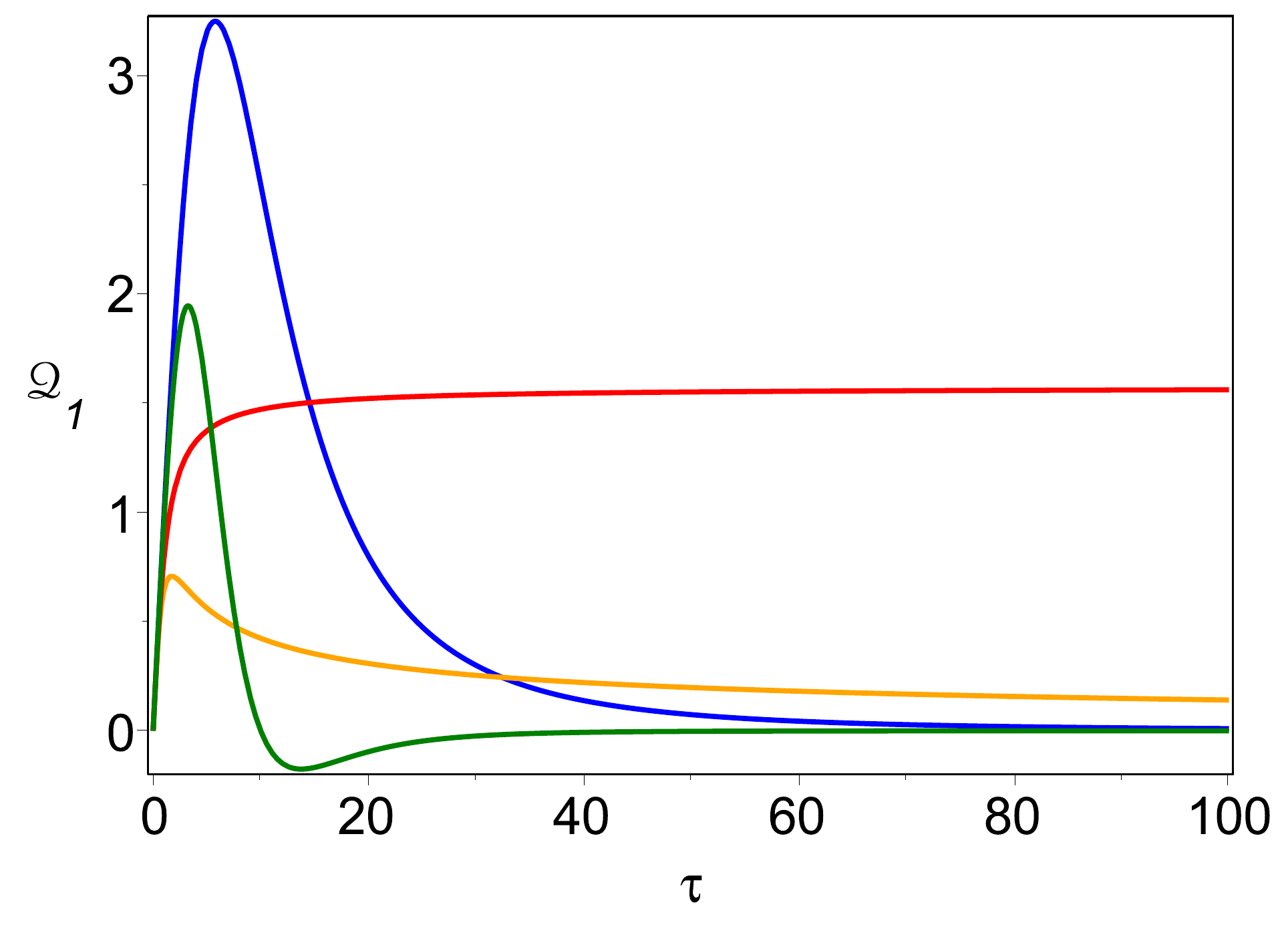}}	
(a)}
{\scalebox{0.355}{\includegraphics{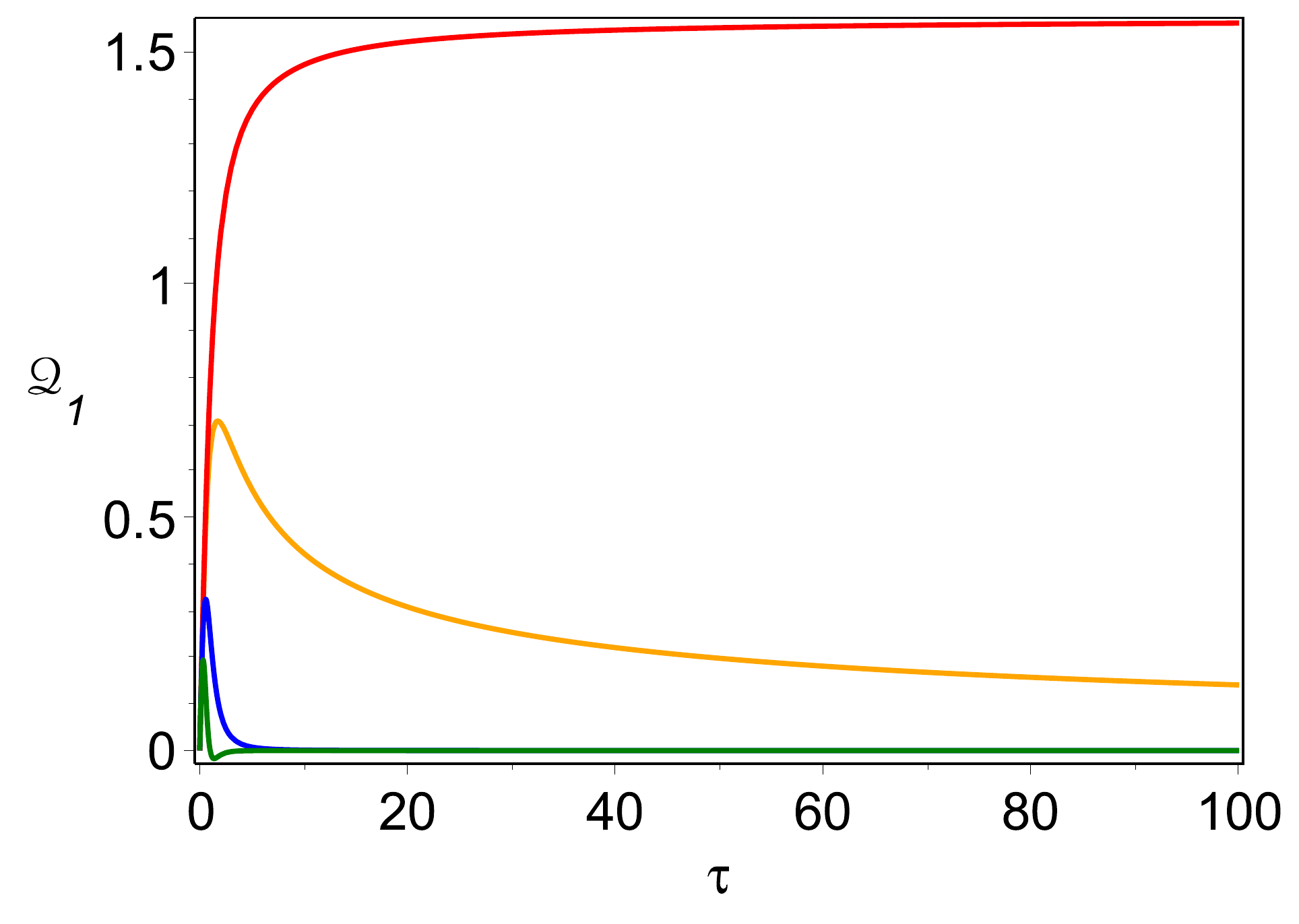}}	
(b)}
{\scalebox{0.375}{\includegraphics{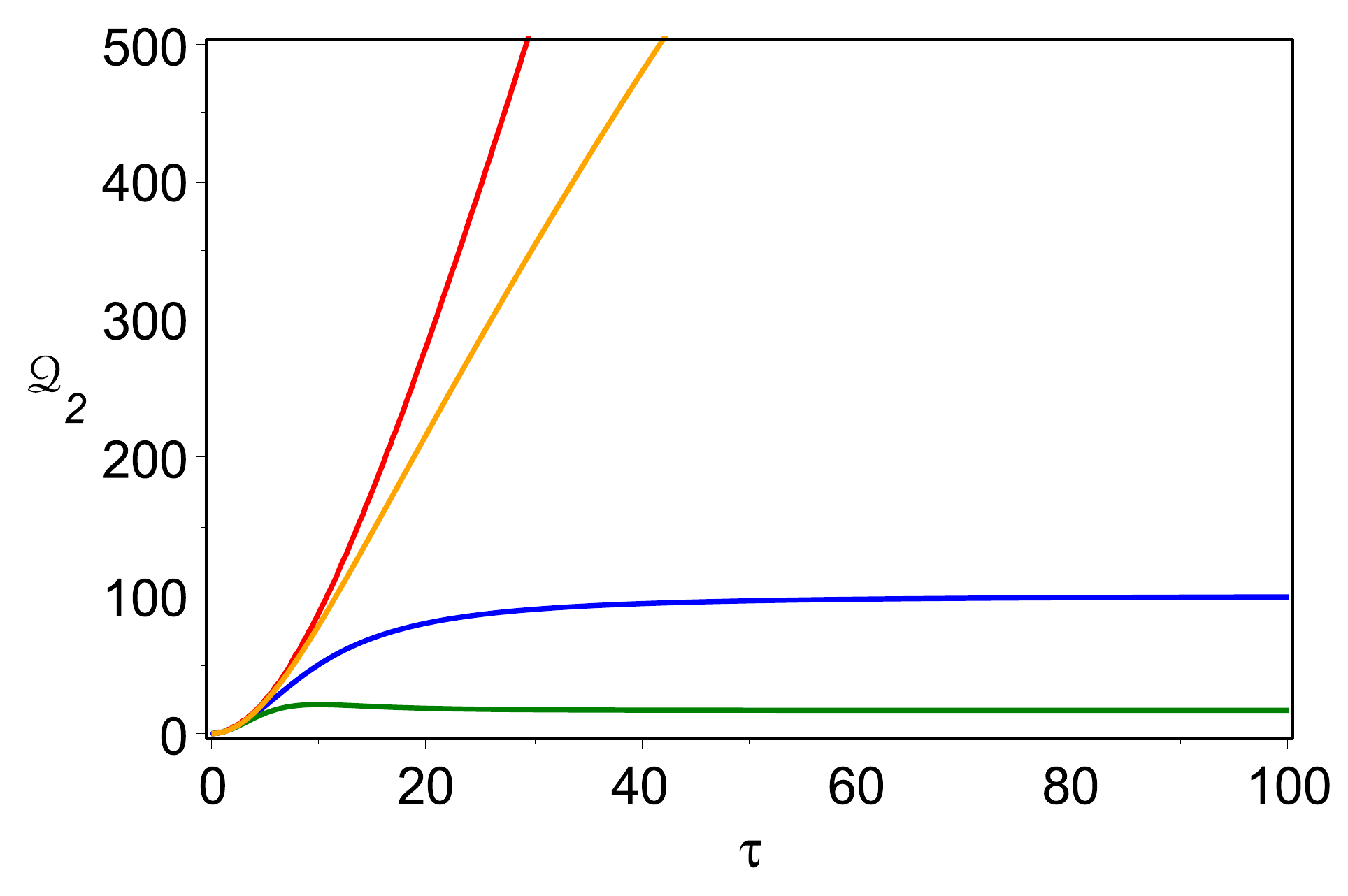}}	
(c)}
{\scalebox{0.365}{\includegraphics{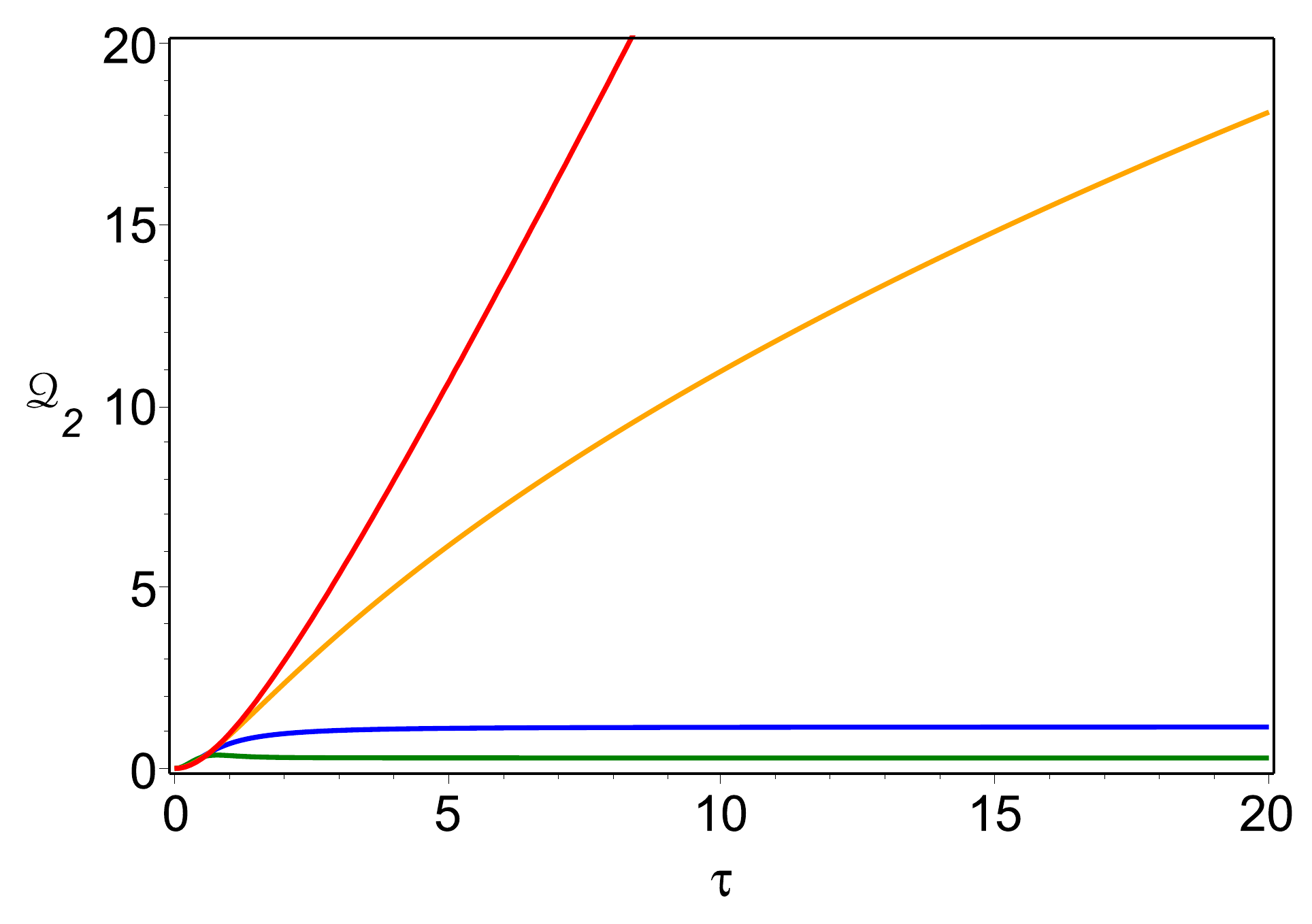}}	
(d)}
\caption{(Color online) The behavior of the functions ${\mathcal Q}_1(\tau)$, \eqref{s2}, in graphs 
(a,b) and ${\mathcal Q}_2(\tau)$, \eqref{b4d}, in graphs (c,d). In (a,c):  $\eta=0.1$, in (b,d):  $\eta =1 
$.   Parameters: $p=-1/4$  (orange curve), $p=-1/2+0$ (red curve), $ p =1/2$ (blue curve), $ 
p =3/2$ (green curve).
\label{R2}}
\end{figure}

In Fig. \ref{R2}, the functions, ${\mathcal Q}_1(\tau)$ and ${\mathcal 
Q}_2(\tau)$, are presented for various values of parameters, $p$ and 
$\eta$. As one can see, in the high-temperature regime, $\eta\ll 1$, the conditions \eqref{e14} and \eqref{e14'} are satisfied for $p\le0$, while for $p>0$ a saturation of ${\mathcal Q}_2(\tau)$ at finite values (${\mathcal Q}_2(\infty)<\infty$) takes place. The asymptotic behavior of the function ${\mathcal Q}_2(\tau)$ significantly influences the quantum coherence in the system.

\begin{figure}
{\scalebox{0.365}{\includegraphics{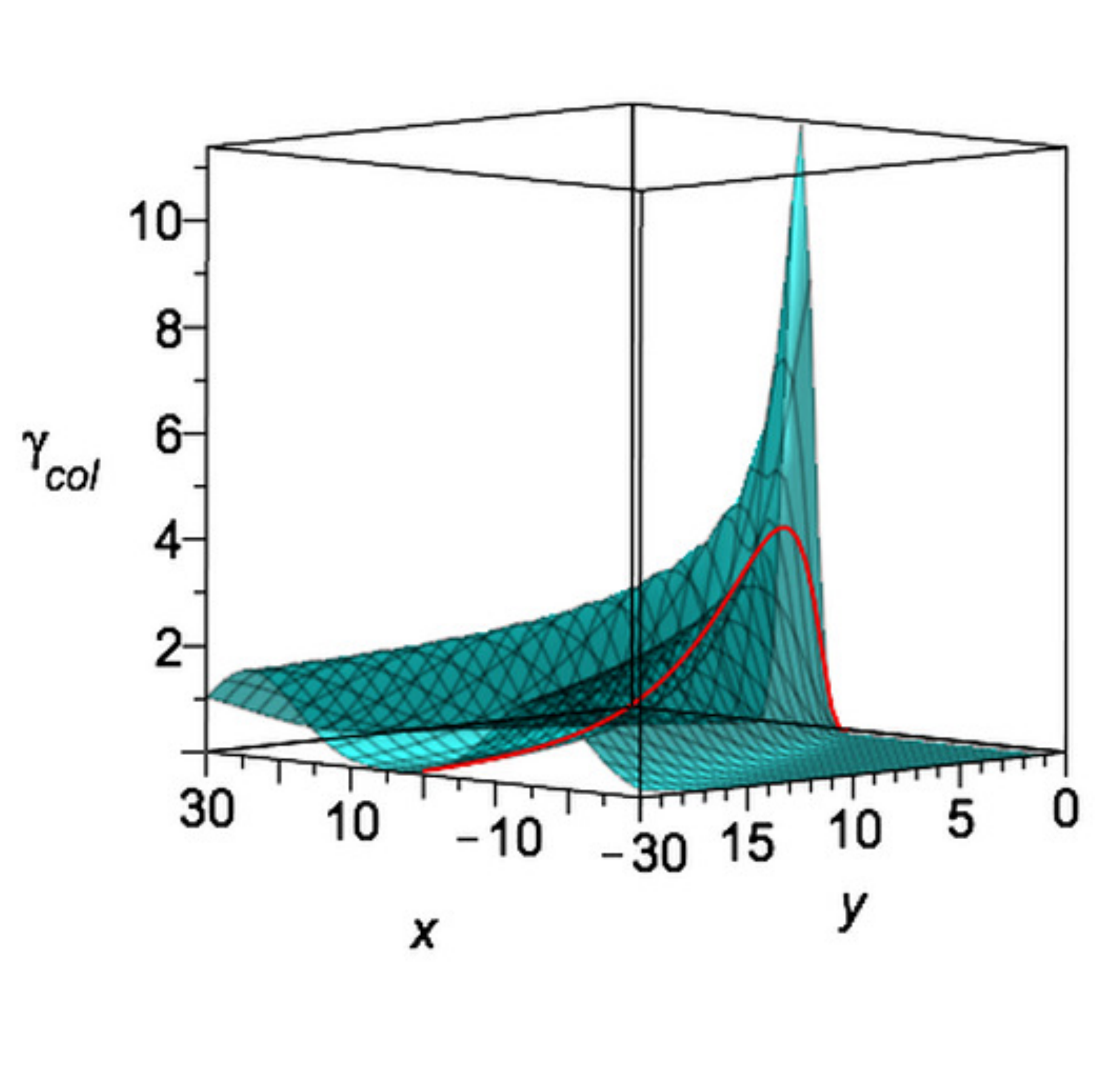}}	
(a)}
{\scalebox{0.35}{\includegraphics{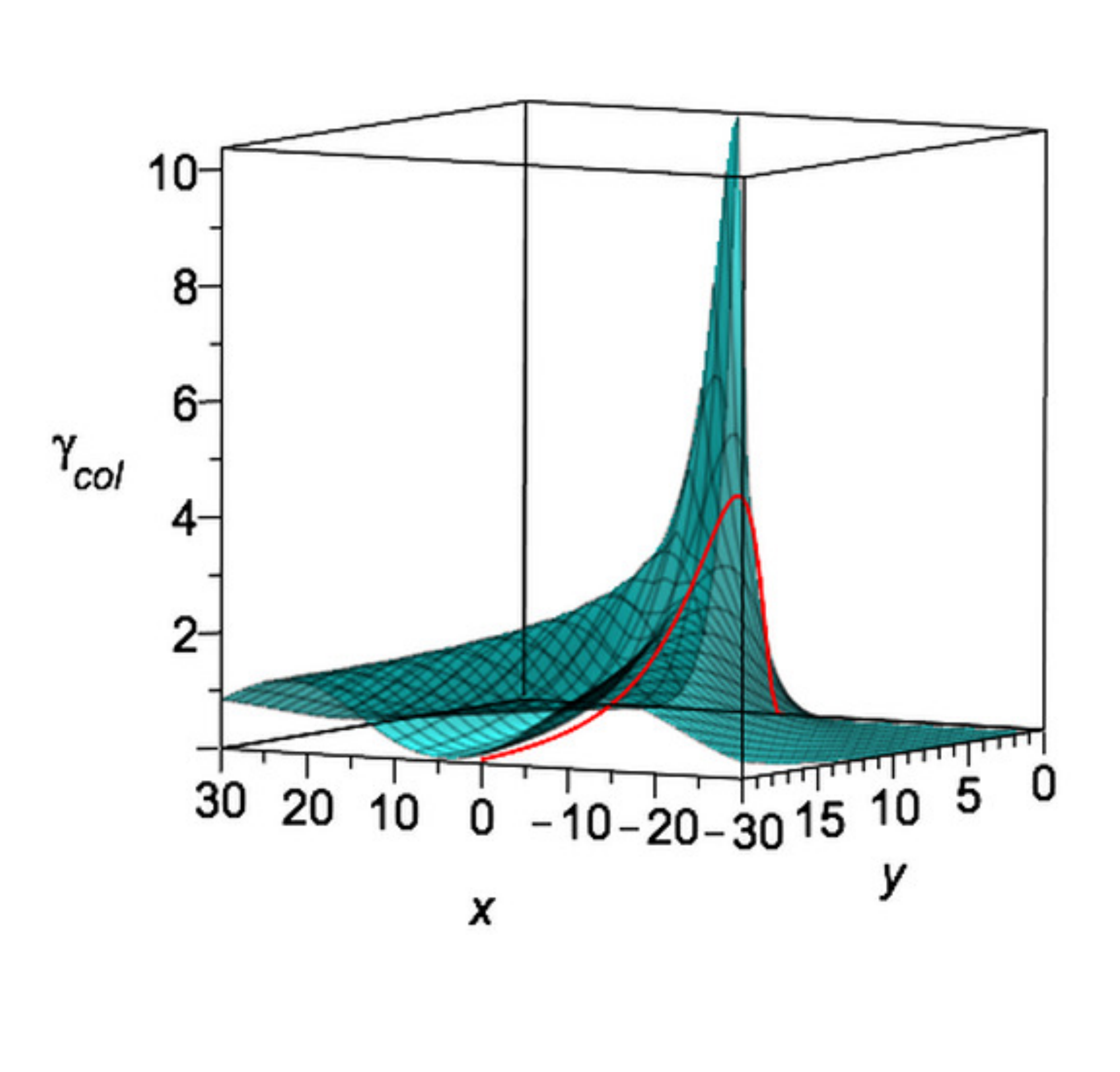}}	
(b)}
{\scalebox{0.35}{\includegraphics{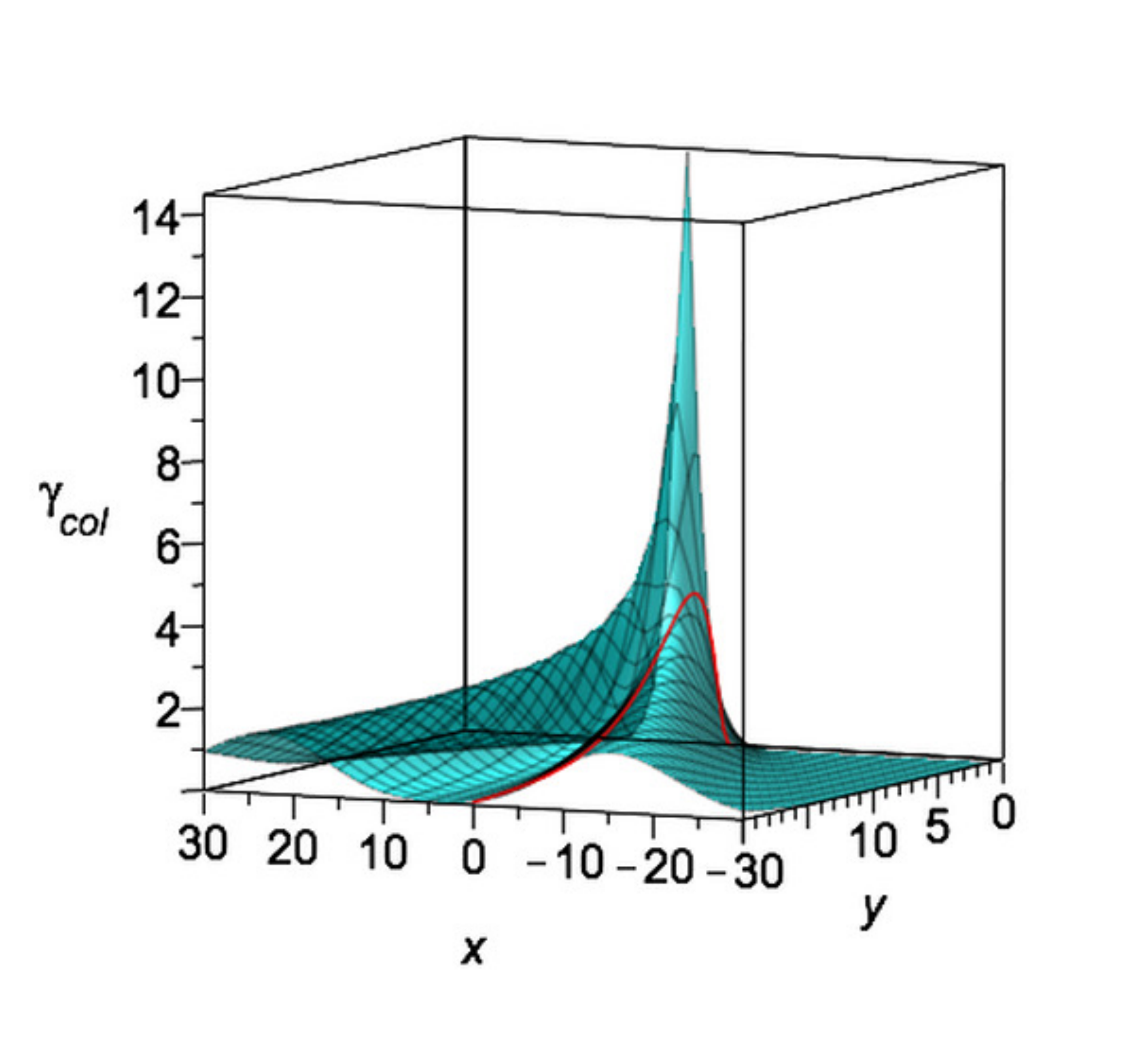}}	
(c)}
{\scalebox{0.365}{\includegraphics{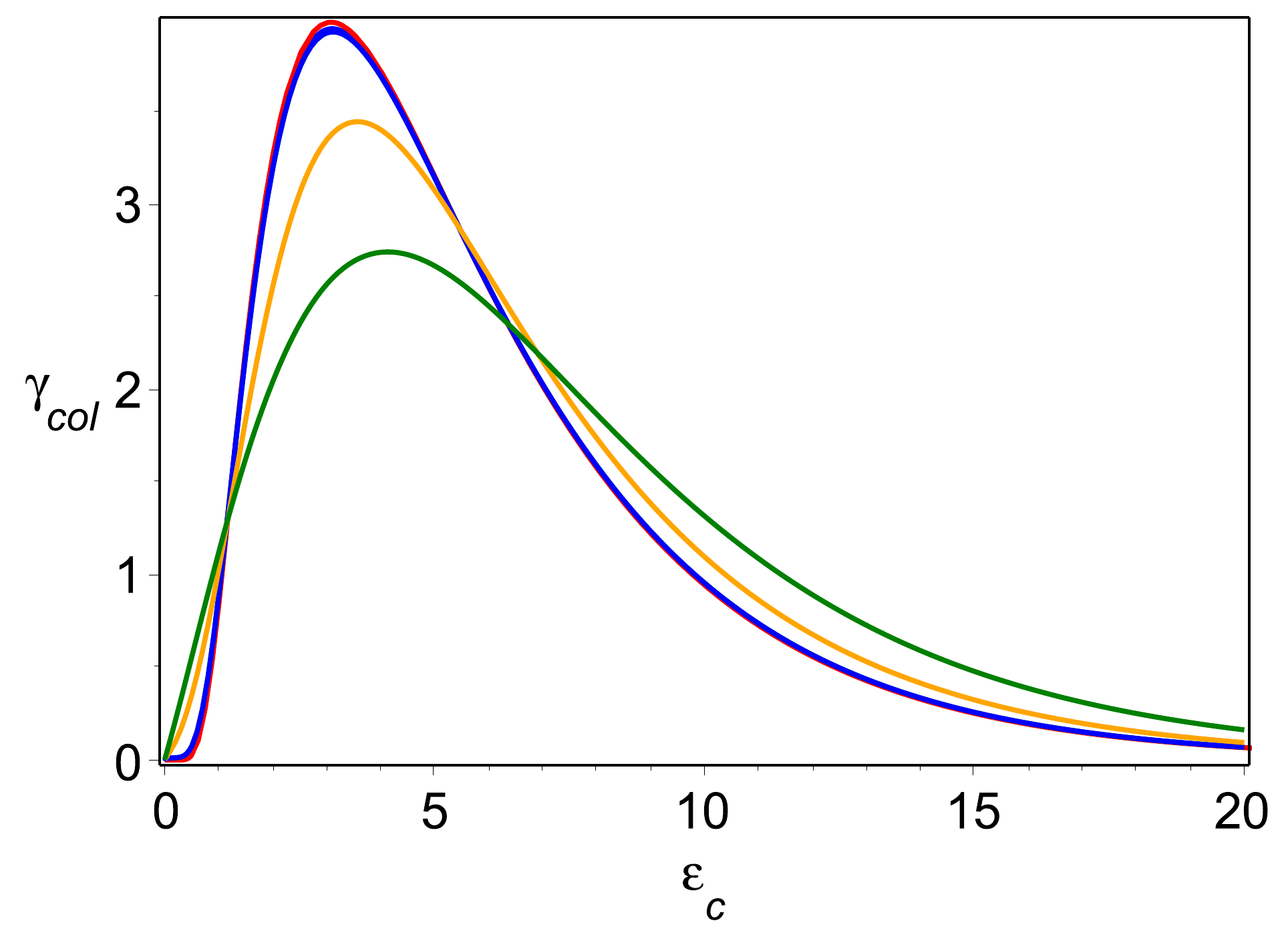}}	
(d)}
\caption{(Color online) The ET rates for the collective environment, $\gamma_\c$, given in \eqref{b12a}, as a 
function of $x=(\varepsilon^c_1- \varepsilon^c_2)/2$  and 
$y=(\varepsilon^c_1+ \varepsilon^c_2)/2$,  $\gamma_\c$ is measured in 
${\rm ps}^{-1}$.
The red curves correspond to the Marcus formula \eqref{XS}. (a) $ \eta =0.1$, $p=1/2$;  (b) $ \eta =1$, $p=1/2$. (c)  $\eta =1$, $p=-1/4$; (d)  $ \eta =0.1$, $p=1/2 $ (blue),  $\eta =1$, $p=1/2 $ (green), $ \eta =1$, $p=-1/4$ (orange). 
\label{Fig1}}
\end{figure}

\begin{figure}
{\scalebox{0.365}{\includegraphics{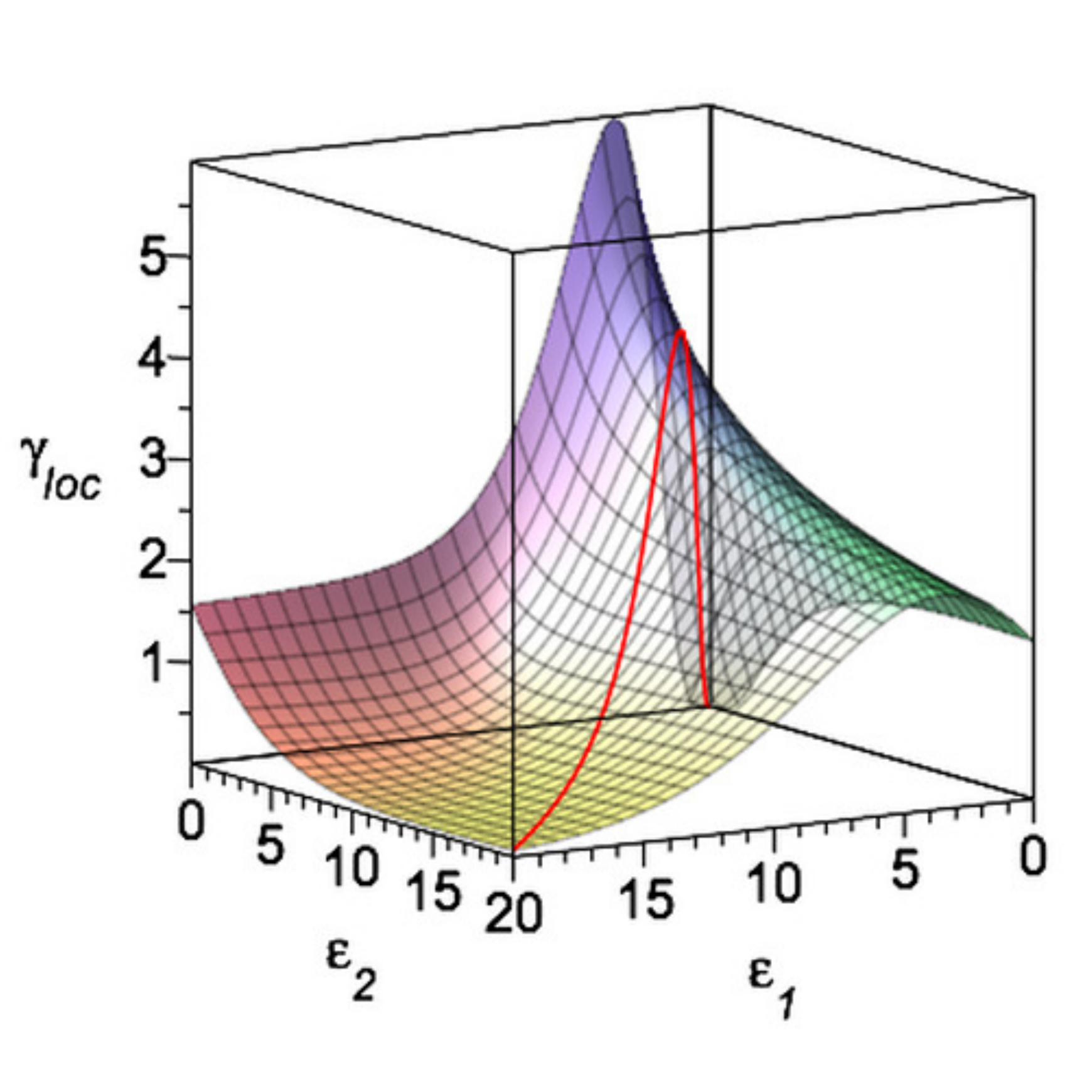}}	
(a)}
{\scalebox{0.365}{\includegraphics{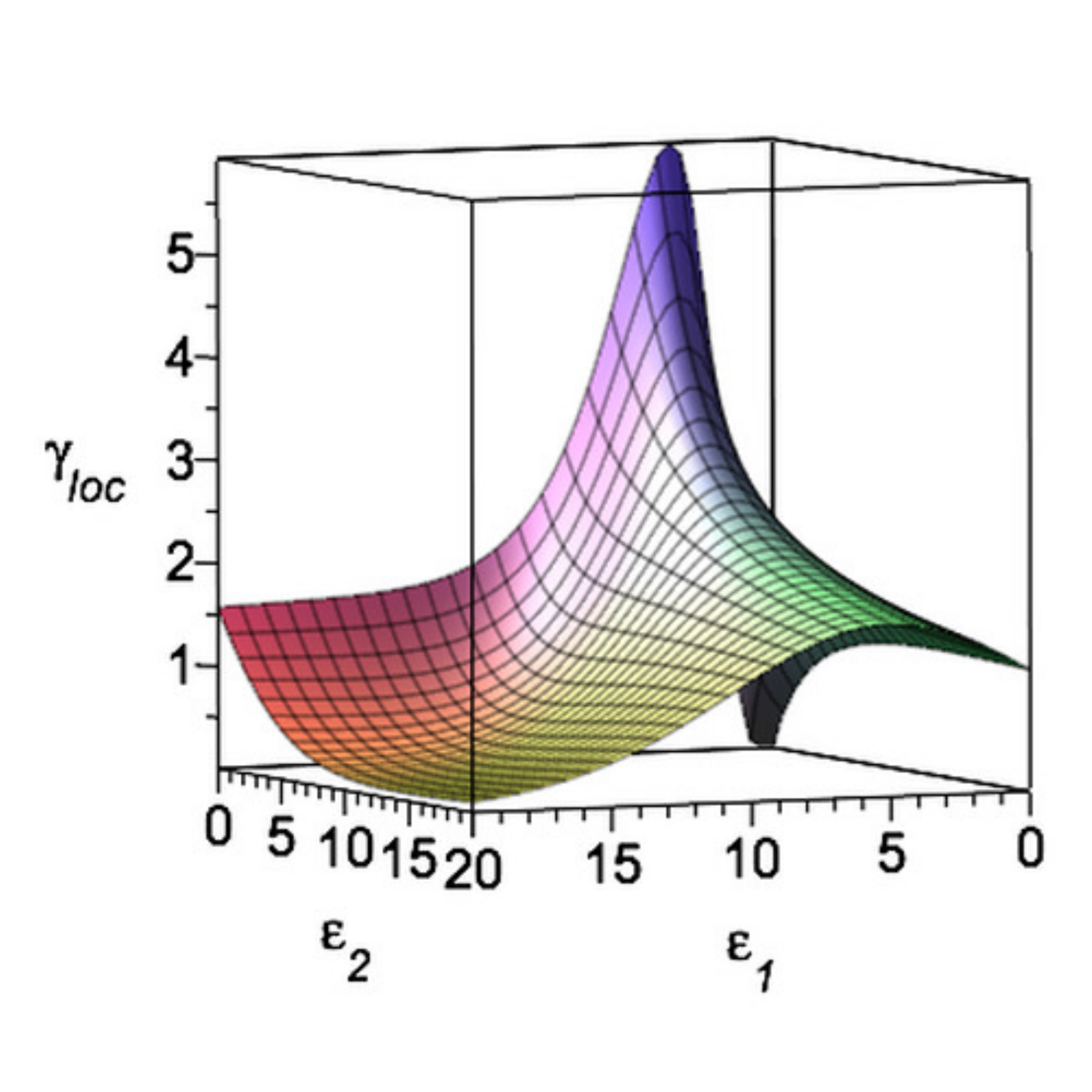}}	
(b)}
{\scalebox{0.365}{\includegraphics{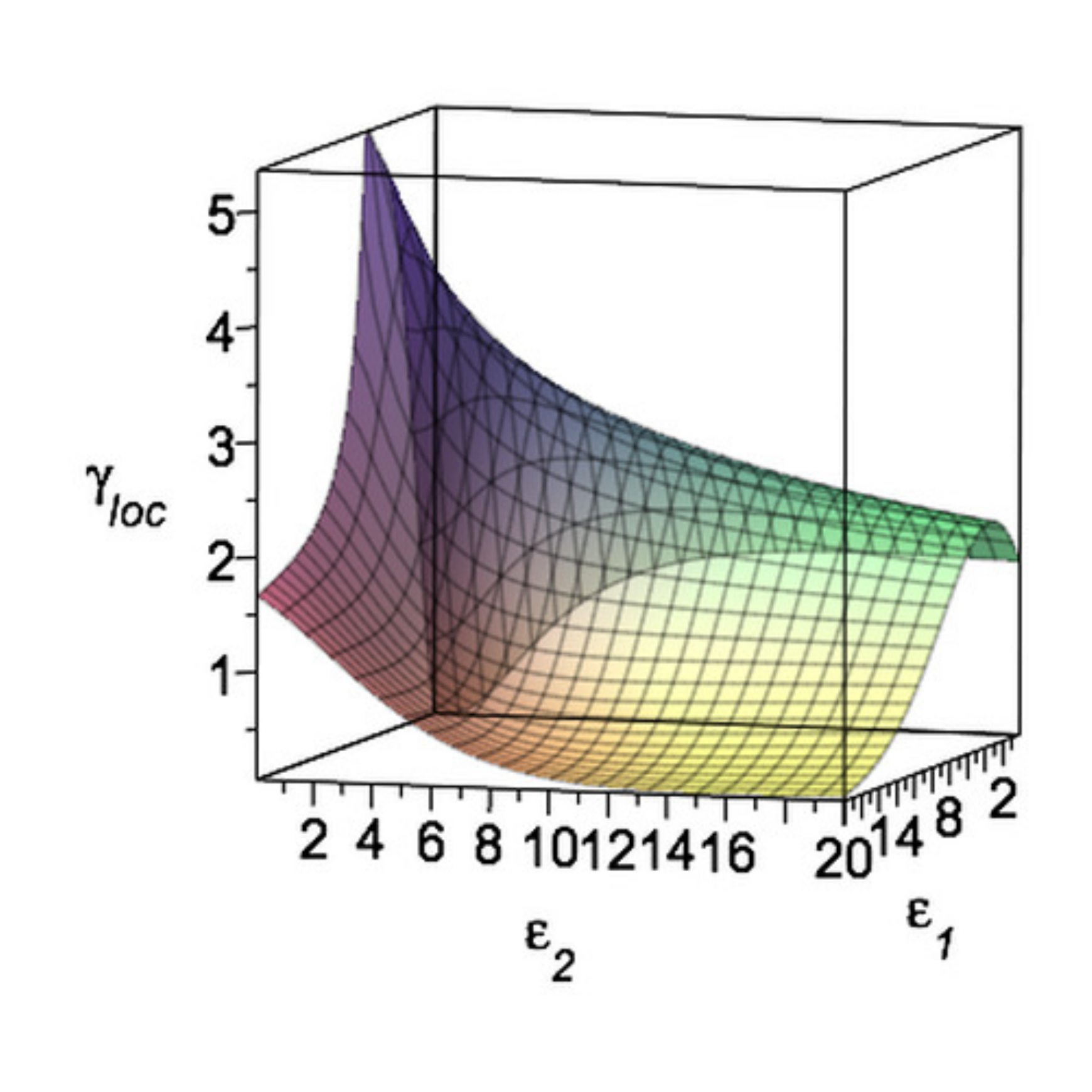}}	
(c)}
{\scalebox{0.365}{\includegraphics{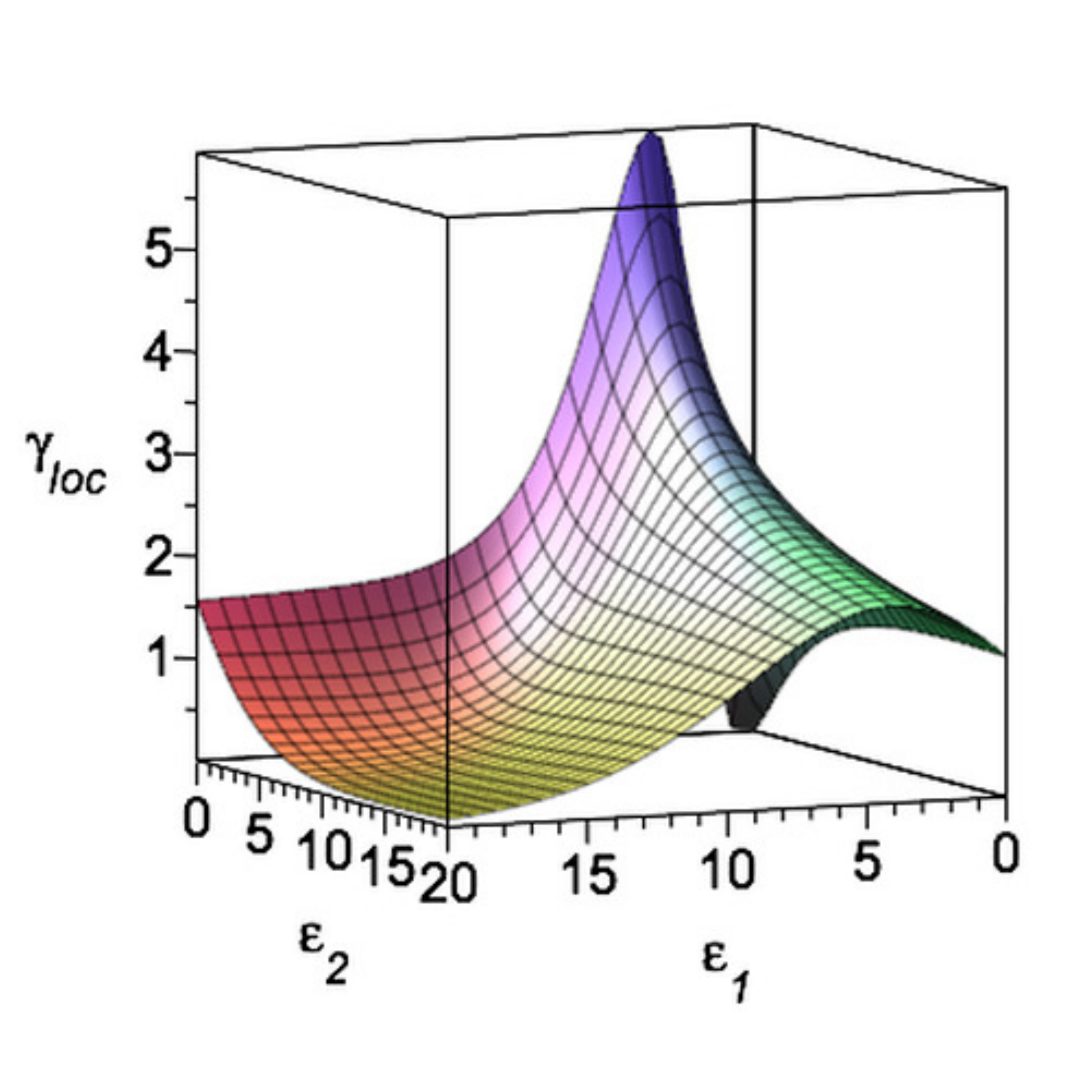}}	
(d)}
\caption{(Color online) The ET rates for local environments, $\gamma_\l$, given in \eqref{b12b}, as a 
function of $\varepsilon_1= \varepsilon^l_1$  and $\varepsilon_2 
=\varepsilon^l_2$, where $\varepsilon_1$ and $\varepsilon_2$ are  the 
reconstruction energies;  $\gamma_\l$ is measured in ${\rm ps}^{-1}$. The 
red curve corresponds to the Marcus formula (\ref{MRL1}).  (a) $p=1/2$, $ 
\eta_1 =\eta_2=0.1$;  (b) $p=1/2$, $ \eta_1 =0.1$,  $ \eta_2 =1$; (c)  
$p=1/2$, $ \eta_1 =1$,  $ 
\eta_2 =0.1$;  (d)  $p=-1/4$, $ \eta_1 =0.1$,  $ \eta_2 =1$. 
\label{Fig3}}
\end{figure}

\subsubsection*{Generalized Marcus limit}

The relaxation rates in the Generalized Marcus limit are given in \eqref{e18} and \eqref{MRL}. In the new variables, they are  
\begin{align}\label{MRC1}
\gamma_\c& = \beta\left(\frac{V}{2}\right)^2\sqrt{\frac{2\pi}{\varepsilon^{c}_{1}+\varepsilon^{c}_{2}}}\left\{ \exp\left[-\frac{(\varepsilon-\varepsilon^{c}_{1})^2}{2(\varepsilon^{c}_{1}+\varepsilon^{c}_{2})}\right] +  \exp\left[-\frac{(\varepsilon+\varepsilon^{c}_{2})^2}{2(\varepsilon^{c}_{1}+\varepsilon^{c}_{2})}\right]\right\}, \\
\gamma_\l & = \beta\left(\frac{V}{2}\right)^2\sqrt{\frac{2\pi}{\varepsilon^{l}_{1}+\varepsilon^{l}_{2}}}\left\{ \exp\left[-\frac{(\varepsilon-\varepsilon^{l}_{1})^2}{2(\varepsilon^{l}_{1}+\varepsilon^{l}_{2})}\right] +  \exp\left[-\frac{(\varepsilon+\varepsilon^{l}_{2})^2}{2(\varepsilon^{l}_{1}+\varepsilon^{l}_{2})}\right]\right\}.
\label{MRL1}
\end{align}
In the case $\varepsilon_c= \varepsilon^{c}_{1}=\varepsilon^{c}_{2}$, or $\varepsilon_l= \varepsilon^{l}_{1}=\varepsilon^{l}_{2}$, they take the standard Marcus form  (see also \cite{XuSch}), 
\begin{align}\label{XS}
\gamma_{\c,\l}& = \beta\left(\frac{V}{2}\right)^2\sqrt{\frac{\pi}{\varepsilon_{c,l}}}\left\{ \exp\left[-\frac{(\varepsilon-\varepsilon_{c,l})^2}{4\varepsilon_{c,l}}\right] +  \exp\left[-\frac{(\varepsilon+\varepsilon_{c,l})^2}{4\varepsilon_{c,l}}\right]\right\},
\end{align}
for the collective and local reservoirs models, respectively. 

In Figs.\,\ref{Fig1} and \ref{Fig3}, we compare the ET rates obtained by numerical integration of  (\ref{b12a}) and  (\ref{b12b}), with the generalized Marcus formulas given by \eqref{MRC1}-\eqref{XS}, corresponding to the high-temperature limit $\eta \ll 1$.
All two-dimensional surfaces shown in  Figs.\,\ref{Fig1} and \ref{Fig3} correspond to the exact Eqs. (\ref{b12a}) and  (\ref{b12b}). 
Fig. \ref{Fig1}(a) shows the high-temperature regime, $\eta=0.1$. In this case, the rate $\gamma_
\c(x,y)$, obtained from the generalized Marcus formula  \eqref{MRC1}, yields a surface that practically coincides with the surface defined $\gamma_
\c(x,y)$ obtained from the exact Eq. (\ref{b12a}), for all $x,y$. The large peak in Fig. \ref{Fig1}(a) lies inside the parameter region for the guaranteed domain of applicability of the theory, namely,  $\gamma_
\c <\!\!< 2\sqrt{2\pi}(V/2)^2/\omega_c = 2\sqrt{2\pi} (25/2)^2/3.7 {\rm ps}^{-1}\approx 206 {\rm ps}^{-1}$, see \eqref{upbnd2}. The values of $\gamma_\c$ for small values of $y$ lie outside the domain for which we can guarantee that the generalized Marcus formula is approximating correctly the rate (c.f. \eqref{e21}). So, in the numerical simulations of Fig. \ref{Fig1}(a) we take $y\geq 1$. The red curves in Fig. \ref{Fig1} correspond to the standard Marcus ET rate, given by Eq. (\ref{XS}). As expected,  the red curve in Fig. \ref{Fig1}(a) lies on the surface $\gamma_\c(x,y)$. The ET rates shown by the red curve in Fig. \ref{Fig1}(a) (Marcus high-temperature regime, $\eta=0.1$) are close to the experimental values reported in \cite{Fleming89} ($\gamma_{\rm exp}\approx 2 {\rm ps}^{-1}$). Still, the ET rates given by the surface but away from the red curve, corresponding to the generalized Marcus formula, can significantly exceed the values on the red curve.

In Fig. \ref{Fig1}(b) and Fig. \ref{Fig1}(c) we take $\eta=1$, which does not correspond to the high-temperature regime any longer. Our numerical simulations demonstrate that the generalized Marcus formula for $\gamma_\c$, given by Eq.  \eqref{MRC1}, does not work in this case. In particular, the red curves in Fig. \ref{Fig1}(b) and Fig. \ref{Fig1}(c) partly lie outside the two-dimensional surfaces corresponding to the exact formulas. The difference between the exact results for $\gamma_\c$ and the generalized Marcus formula is shown in Fig. \ref{Fig1}(d), for 
the dependence $\gamma_\c(\varepsilon_c)$, where $\varepsilon_c=\varepsilon_1^c=\varepsilon_2^c$. As one can see, only the blue curve, corresponding to the exact formula in the high-temperature regime ($\eta=0.1$), practically coincides with the Marcus formula  \eqref{XS}.

In Fig. \ref{Fig3}, similar results (to those presented in Fig. \ref{Fig1}) are 
shown for $\gamma_\l(\varepsilon_1,\varepsilon_2)$. The two-dimensional 
surfaces correspond in this case to the exact formulass of Eq. (\ref{b12b}). 
The high-temperature regime is shown in Fig. \ref{Fig3}(a) 
($\eta_1=\eta_2=0.1$).
In this case, the generalized Marcus formula, given by Eq. \eqref{MRL1}, 
practically describes the exact result in the whole region of the 
reconstruction energies, ($\varepsilon_1,\varepsilon_2$). In particular, the 
red curve in Fig. \ref{Fig3}(a) (corresponding to Eq. (\ref{XS})) lies on the 
two-dimensional surface, $\gamma_\l(\varepsilon_1,\varepsilon_2)$. Similar 
to the case of the collective environment, the ET rates, $\gamma_\l$, of the 
generalized Marcus formula \eqref{MRL1}, can exceed the ET rates 
corresponding to the standard Marcus formula given by Eq. (\ref{XS}). The 
results of our numerical simulations of the exact Eq. (\ref{b12b}) (shown in 
Figs. \ref{Fig3}(b,c,d)) demonstrate that outside the high-temperature regime 
($\eta\ll 1$), the generalized Marcus formula, given by Eq. \eqref{MRL1}, 
cannot be used. (However, we do have the correct expressions for 
$\gamma_\c$, $\gamma_\l$, even in this regime, c.f. 
\eqref{MRC1}-\eqref{XS}.)


\subsubsection*{Decoherence at $V=0$}

\begin{figure}
\begin{center}
\scalebox{0.4}{\includegraphics{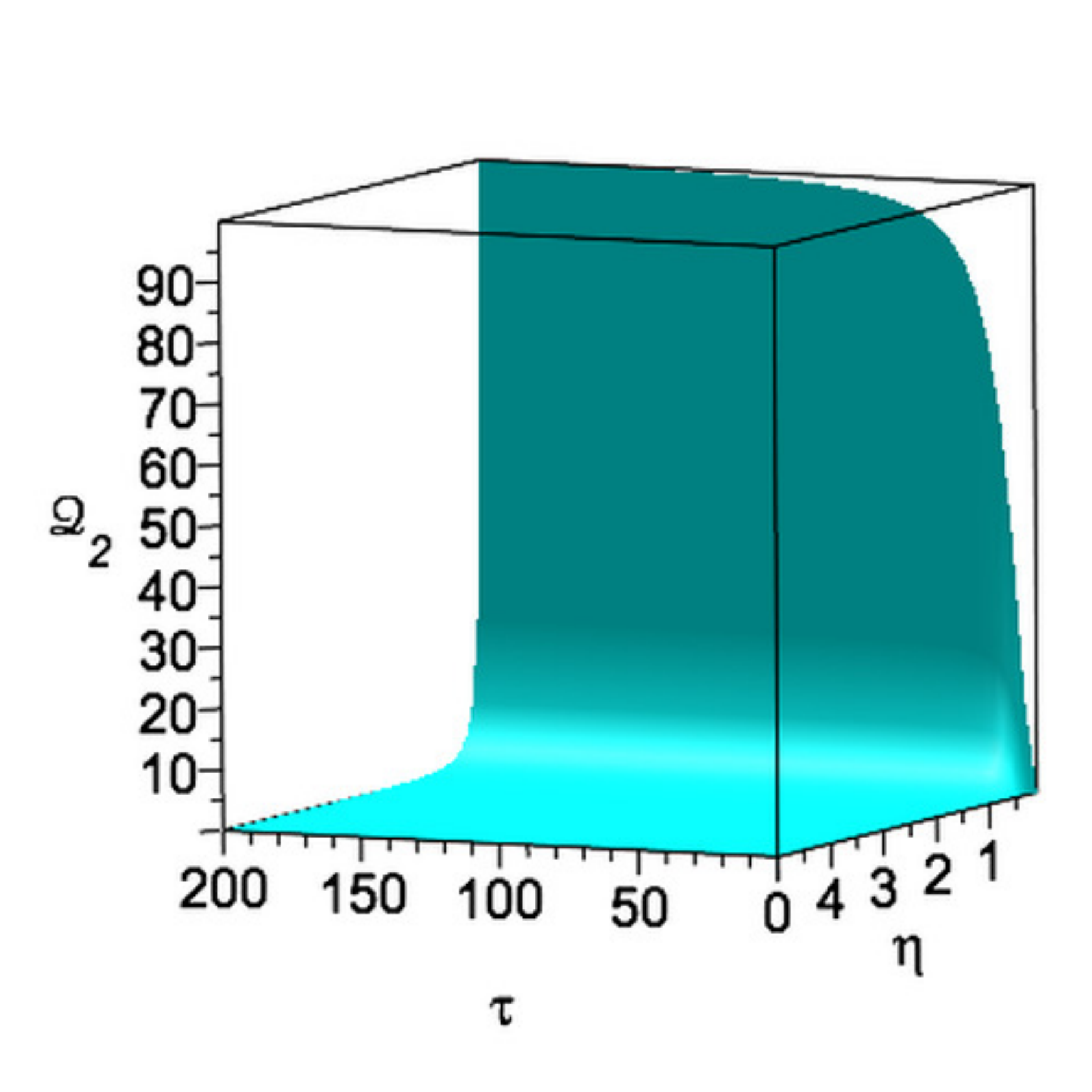}}	
\end{center}
\caption{(Color online)  The dependence $\mathcal Q_2$ on  $\tau$  and 
$\eta$ ($p=1/2$, $0.1\leq \eta \leq 5$).
\label{Fig4a}}
\end{figure}

For $V=0$ there is no relaxation, only decoherence in the energy basis occurs \cite{PSE}. The saturated value of the non-diagonal reduced density matrix elements is characterized  by the factor $\e^{-\Gamma_\infty}$ given in \eqref{gammainfty}. In the variables \eqref{newvar} we have
\begin{equation}
\label{gammatau}
\Gamma_\infty= 
\lim_{\tau \rightarrow \infty}\Gamma(\tau),\qquad \mbox{where}\qquad \Gamma(\tau) = \left\{
\begin{array}{l}
\displaystyle\frac{(\varepsilon^c_1+\varepsilon^c_2)}{2}\mathcal Q_2(\tau)  \mbox{\quad {\rm (collective)}}\\
{}\\
\displaystyle \frac{\varepsilon^l_1}{2}{\mathcal 
		Q}_2^{(1)}(\tau) +\frac{\varepsilon^l_2}{2}{\mathcal 
		Q}_2^{(2)}(\tau) \mbox{\quad {\rm (local)}}
\end{array}
\right.
\end{equation}
As one can see from Eq. (\ref{gammatau}), the asymptotic behavior of the function ${\mathcal Q_2}(\tau)$ significantly effects the quantum coherence in the system. 
We plot the function ${\mathcal Q_2}(\tau,\eta)$ in Fig. \ref{Fig4a} (see also Fig. \ref{R2}). One can see that in the high-temperature regime, $\eta \ll 1$,   $\mathcal Q_2(\tau)$ increases very fast when $\tau$ increases. Hence one expects strong decoherence in this case.
\begin{figure}
{\scalebox{0.325}{\includegraphics{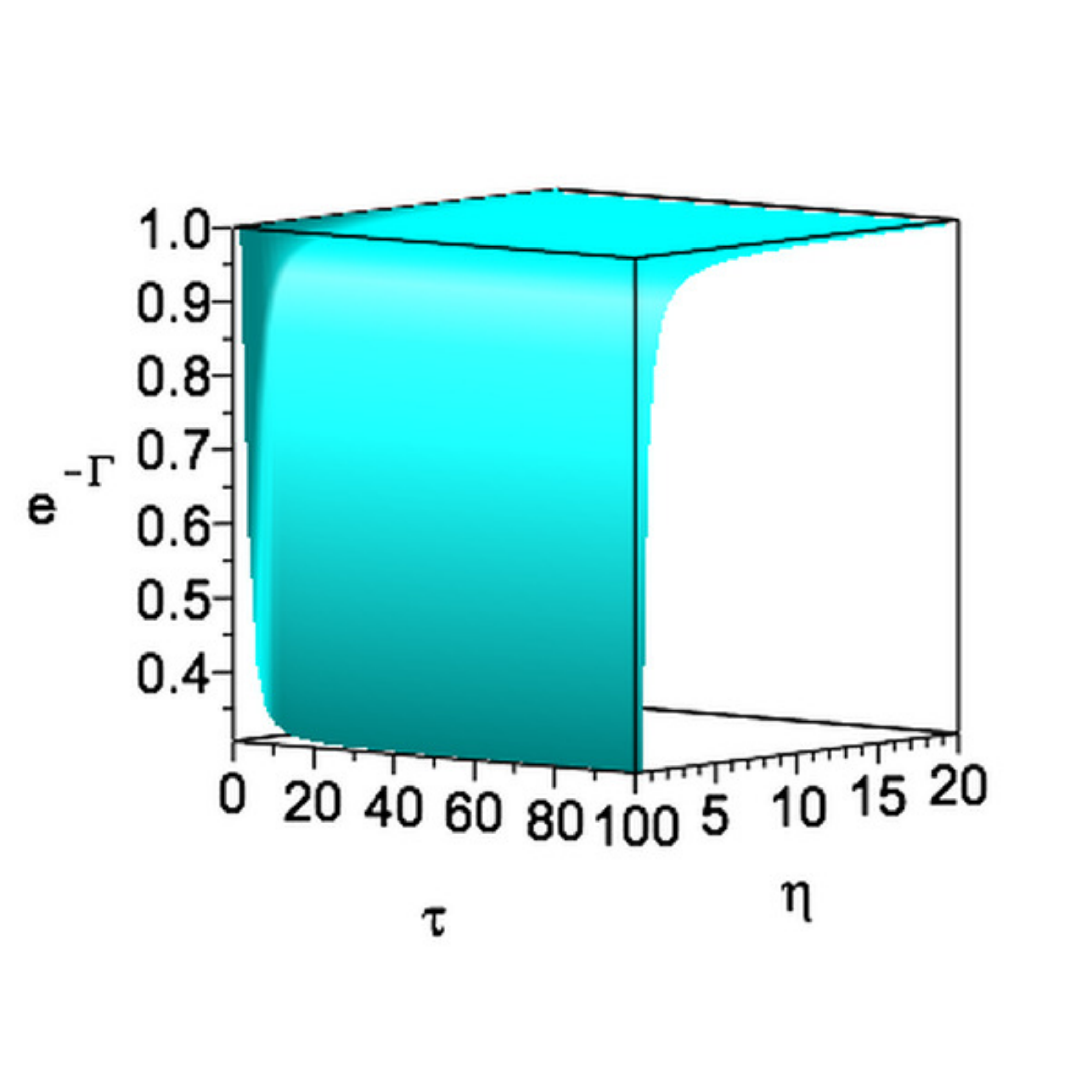}}	
(a)}
\vspace*{-3cm}
{\scalebox{0.375}{\includegraphics{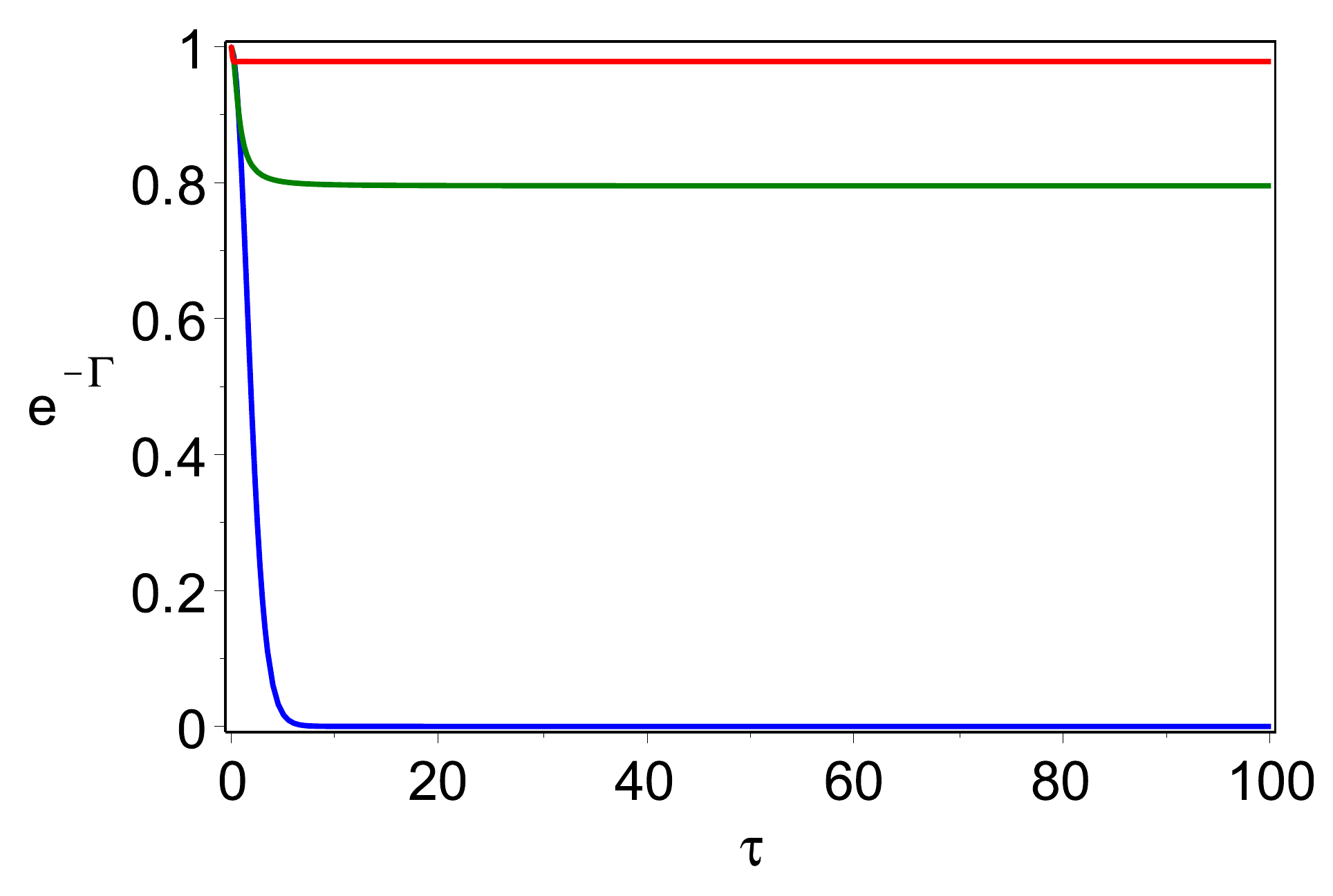}}	
(b)} 
\begin{center}
{\scalebox{0.35}{\includegraphics{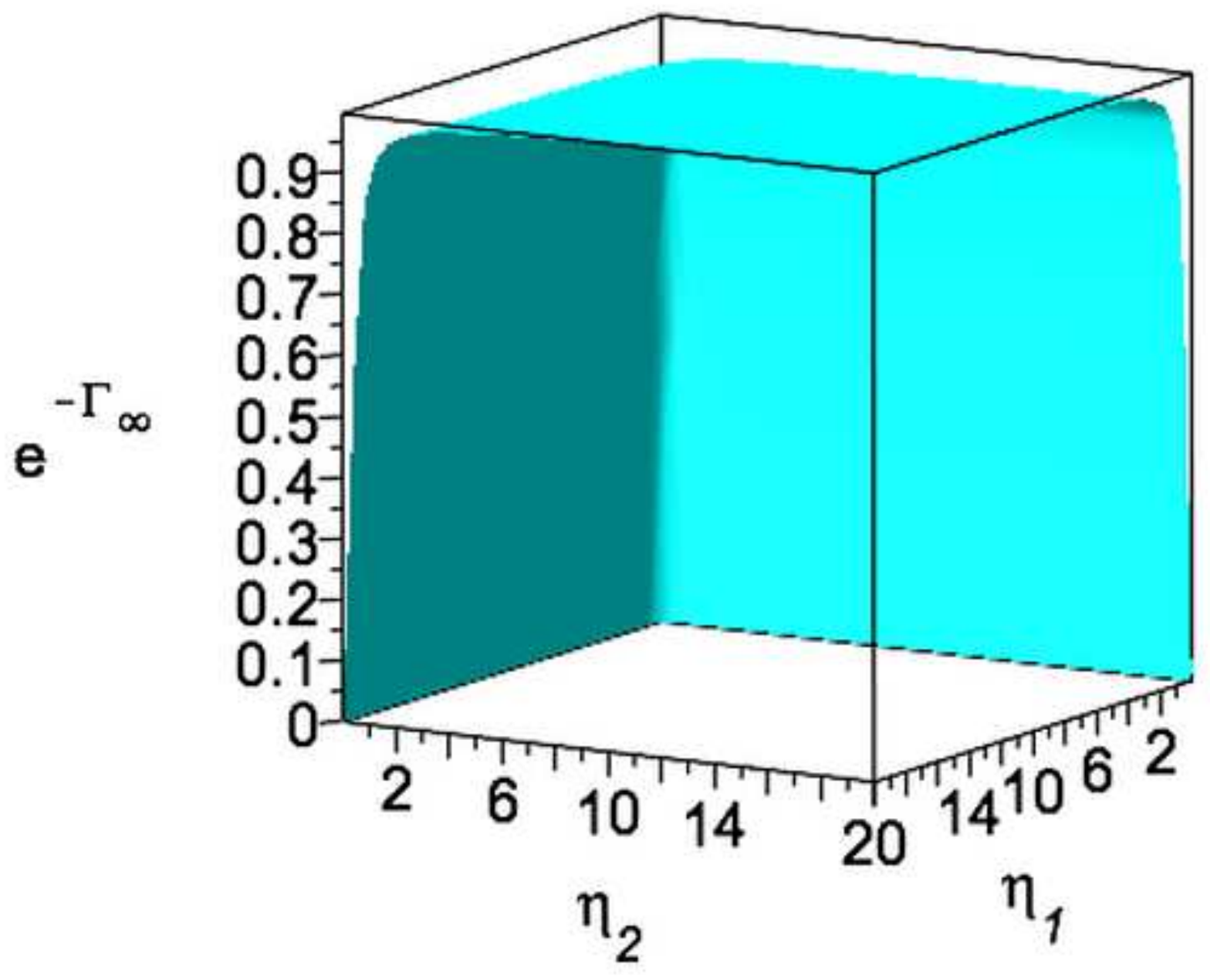}}(c)}
\end{center}
\caption{(Color online) The decoherence function $\e^{-\Gamma(\tau)}$, with $\Gamma(\tau)$ given in \eqref{gammatau}, as a function of 
$\tau$  and $\eta$. (a,b) collective environment;  (b) $ \eta =0.1$ (blue curve), $ \eta =1$ (green curve), $ \eta =5$ (red curve); $p=1/2$, $ \varepsilon =0.2$. (c) $\e^{-\Gamma_\infty}$ for local environments: $p=1/2$, $ \varepsilon_1 = \varepsilon_2 =0.1$.
\label{Fig4}}
\end{figure}
In the regime of partial phase decoherence, $p>0$, we can use the asymptotic formulas (\ref{Q0}) to obtain
\begin{equation}
\Gamma_\infty= 
\left\{
\begin{array}{l}
\displaystyle\frac{(\varepsilon^c_1+\varepsilon^c_2)}{2}\mathcal Q_0 \mbox{\quad {\rm (collective)}}\\
{}\\
\displaystyle \frac{\varepsilon^l_1}{2}{\mathcal 
		Q}_0^{(1)} +\frac{\varepsilon^l_2}{2}{\mathcal 
		Q}_0^{(2)}\mbox{\quad {\rm (local)}}
\end{array}
\right.
\end{equation}
Here,
\begin{align}
{\mathcal Q}_0 =\frac{1}{(2p+1)\eta^{ 2 p+2}}\bigg(\zeta\Big( 2 
p+1,\frac{1}{\eta}\Big ) +\zeta\Big( 2 p+1,\frac{1}{\eta} +1\Big )\bigg),
\end{align}
for the collective environment model and 
\begin{align}
{\mathcal Q}^{i}_0 =\frac{1}{(2p+1)\eta_i^{ 2 p+2}}\bigg(\zeta\Big( 2 
p+1,\frac{1}{\eta_i}\Big ) +\zeta\Big( 2 p+1,\frac{1}{\eta_i} +1\Big )\bigg), \quad i=1,2,
\end{align}
for the local environments model.

Fig. \ref{Fig4} shows the decoherence function $\e^{-\Gamma(\tau,\eta)}$ for both collective and local environments, and for $p>0$. Quantum coherence in this case ($p>0$) survives for large times (for all times if $V=0$). The level of quantum coherence strongly depends on the value of $\eta$ the for collective environment and on $\eta_1$ and $\eta_2$ for the local environments. Namely, quantum coherence is strong for $\eta$, $\eta_{1,2}$ not too small.
Indeed, for $\eta=0.1$ (blue curve in Fig. \ref{Fig4}b),  coherence is very small at large times.

\section{Dynamical resonance theory}
\label{sectdynres}

\subsection{The mathematical setup}

The description of a small system (dimer) coupled to infinitely extended Bose reservoirs, or continuous modes oscillator reservoirs, is phrased mathematically in terms of the language of $W^*$-dynamical systems. We note that we start off with a reservoir having continuous modes. Often in the physics literature, one first considers a reservoir of finitely many modes (quantized for instance because confined to a finite volume). Then one performs calculations, reaches expressions for, say, the relaxation rates, and finally takes the limit of continuous modes in the end. The arguments leading to expressions of relaxation rates in this way (the procedure being called the ``time-dependent perturbation theory"), are often based on taking large-time limits ($t\rightarrow\infty$). It is not clear if these arguments are really valid since in general, performing first the large time limit and then the continuous modes limit (or, infinite volume limit) is not the same as performing the limits in the opposite order (as taking first $t\rightarrow\infty$ at discrete modes or finite volume depends on `boundary effects'). Thus, for a rigorous analysis, we {\rm first} perform the infinite volume limit and get results which hold {\em for all times $t\geq 0$}. This approach has been fruitful in many situations recently \cite{MBSa,KoMe,KoMeSo,MBR,MeBeSo,MSB,MS}.

The total system is a $W^*$-dynamical system $({\mathcal H},{\frak M},\alpha)$, where ${\frak M}$ is a von Neumann algebra of observables acting on a Hilbert space $\mathcal H$ and where $\alpha^t$ is a group of $*$automorphisms of $\frak M$. The ``positive temperature Hilbert space''  is given by
\begin{equation}
	{\mathcal H} = {\mathbb C}^2\otimes {\mathbb C}^2\otimes{\mathcal F}_\beta \quad\mbox{or}\quad 
	{\mathcal H} = {\mathbb C}^2\otimes {\mathbb C}^2\otimes{\mathcal F}_\beta\otimes{\mathcal F}_\beta ,
	\label{2.1}
\end{equation}
depending on whether we have one or two reservoirs. Here, ${\mathcal F}_\beta$ is the Fock space 
\begin{equation}
	{\mathcal F}_\beta = \bigoplus_{n\geq 0} L^2_{\rm sym}(({\mathbb R}\times S^2)^{n},(d u\times d\Sigma)^{n}).
	\label{GluedFock}
\end{equation}
It differs from the `usual zero-temperature' Fock space $\oplus_{n\geq 0} L^2_{\rm sym}(({\mathbb R}^{3n},d^{3n} k))$  in that the single-particle space at positive temperature is the `glued' space $L^2({\mathbb R}\times S^2,d u\times d\Sigma)$ \cite{JP} ($d\Sigma$ is the uniform measure on $S^2$). ${\mathcal F}_\beta$ carries a representation of the CCR (canonical commutation relation) algebra. The represented Weyl operators are given by $W(f_\beta) = \e^{i\phi(f_\beta)}$, where $\phi(f_\beta)=\frac{1}{\sqrt{2}}(a^*(f_\beta)+a(f_\beta))$. Here, $a^*(f_\beta)$ and $a(f_\beta)$ denote creation and annihilation operators on ${\mathcal F}_\beta$, smoothed out with the function
\begin{equation}
	f_\beta(u,\Sigma) = \sqrt{\frac{u}{1-e^{-\beta u}}}\ |u|^{1/2} \left\{
	\begin{array}{ll}
		f(u,\Sigma), & u\geq 0\\
		-\overline{f}(-u,\Sigma), & u<0
	\end{array}
	\right.
	\label{2.3}
\end{equation}
belonging to $L^2({\mathbb R}\times S^2,d u\times d\Sigma)$. It is easy to see that the CCR are satisfied, namely,
\begin{equation}
	W(f_\beta)W(g_\beta) = e^{-\frac{i}{2}{\rm Im}\scalprod{f}{g}} W(f_\beta+g_\beta).
	\label{ccr}
\end{equation}
The vacuum vector $\Omega\in{\mathcal F}_\beta$ represents the infinite-volume equilibrium state of the free Bose field, determined by the formula 
\begin{equation}
	\label{thav}
	\scalprod{\Omega}{W(f_\beta)\Omega} = \exp\left\{ \textstyle-\frac14 \scalprod{f}{\coth(\beta|k|/2)f}\right\}.
\end{equation}
The CCR algebra is represented on \eqref{GluedFock} as $W(f)\mapsto W(f_\beta)$, for $f\in L^2({\mathbb R}^3)$ such that $ \scalprod{f}{\coth(\beta|k|/2)f}<\infty$. This representation was first derived by Araki and Woods \cite{AW}. We denote the von Neumann algebra of the represented Weyl operators by ${\mathcal W}_\beta$.

The doubled spin Hilbert space in \eqref{2.1} allows to represent any (pure or mixed) state of the two-level system by a vector, again by the GNS construction. This construction is also known as the {\em Liouville description} \cite{Muka} and goes as follows. Let $\rho$ be a density matrix on ${\mathbb C}^2$. When diagonalized it takes the form $\rho=\sum_i p_i |\chi_i\rangle\langle\chi_i|$, to which we associate the vector $\Psi_\rho = \sum_i \sqrt{p_i}\chi_i\otimes\overline\chi_i\in {\mathbb C}^2\otimes {\mathbb C}^2$ (complex conjugation in any fixed basis -- we will choose the eigenbasis of $H_\s$,  \eqref{eigenbasis}). Then ${\rm Tr}(\rho A)=\scalprod{\Psi_\rho}{(A\otimes\one_\S)\Psi_\rho}$ for all $A\in{\mathcal B}({\mathbb C}^2)$ and where $\one_\S$ is the identity in ${\mathbb C}^2$. This is the GNS representation of the state given by $\rho$ \cite{BRII,MSB}. 

The von Neumann algebra of observables of the total system is
\begin{equation}
	\label{vna}
	{\frak M}={\mathcal B}({\mathbb C}^2)\otimes\one_\S\otimes{\mathcal W}_\beta \subset {\mathcal B}({\mathcal H}).
\end{equation}
The modular conjugation $J$ is the antilinear operator defined by
\begin{equation}
	J( A\otimes\bbbone_\s\otimes W(f_\beta(u,\Sigma)) )J = \bbbone_\s\otimes\overline A\otimes W(\overline f_\beta(-u,\Sigma)),
	\label{2.6}
\end{equation}
where $\overline A$ is the matrix obtained from $A$ by taking entrywise complex conjugation (matrices are represented in the eigenbasis of $H_\s$). Note that by \eqref{2.3}, we have $\overline f_\beta(-u,\Sigma) = -e^{-\beta u/2}f_\beta (u,\Sigma)$. By the Tomita-Takesaki theorem \cite{BRII}, conjugation by $J$ maps the von Neumann algebra of observables \eqref{vna} into its commutant. That is, $V$ and $JVJ$ commute for any $V\in\frak M$.

The dynamics of the spin-boson system is given by
\begin{equation}
	\label{dyn}
	\alpha^t(A) =\e^{\i tL}A\e^{- \i tL}, \qquad A\in\frak M.
\end{equation}
It is generated by the self-adjoint Liouville operator acting on $\mathcal H$, associated to the Hamiltonian \eqref{1'}.  For the single reservoir model, 
\begin{equation}
	\label{a4}
	L_\c = L_\s +L_{\r} +I_\c -J I_\c J,
\end{equation}
where 
\begin{eqnarray}
	L_\S &=& \frac12
	\begin{pmatrix}
		\varepsilon  & V\\
		V & -\varepsilon 
	\end{pmatrix}
	\otimes \bbbone_\S 
	-\bbbone_\S\otimes
	\frac12 
	\begin{pmatrix}
		\varepsilon  & V\\
		V & -\varepsilon 
	\end{pmatrix}
	\label{a6}\\
	L_\r&=&\d\Gamma(u)
	\label{a7}\\
	I_{\c} &=& \big( \lambda_1 P_1+\lambda_2 P_2\big)\otimes\bbbone_\s\otimes \phi(g_\beta),
	\label{a8}
\end{eqnarray}
where $\d\Gamma(u)=\int ua^*(u,\Sigma) a(u,\Sigma)\d u\d\Sigma$ is the second quantization of the operator of multiplication by the radial variable $u$, $\phi(g_\beta)$ is the represented field operator and
\begin{equation}
	\label{p12}
	P_1 = |\varphi_1\rangle\langle\varphi_1| = \begin{pmatrix}
		1 & 0\\
		0 & 0
	\end{pmatrix},\qquad
	P_2= |\varphi_2\rangle\langle\varphi_2| = \begin{pmatrix}
		0 & 0\\
		0 & 1
	\end{pmatrix}.
\end{equation}	
Similarly, the Liouvillian associated to the Hamiltonian \eqref{1} is
\begin{equation}
	\label{a3}
	L_\l = L_\s+L_{\r_1}+L_{\r_2} +I_\l -JI_\l J,
\end{equation}
with $L_\s$ given as in \eqref{a6}, the $L_{{\r}_{1,2}}$ as in \eqref{a7} (on their individual reservoir spaces) and 
\begin{equation}
	\label{a10}
	I_\l = \lambda_1 P_1\otimes\bbbone_\s\otimes\phi_1(g_1) + \lambda_2 P_2\otimes\bbbone_\s\otimes\phi_2(g_2).
\end{equation}

It is convenient to transform the Liouvillians unitarily by a ``polaron transformation" \cite{Leggett,KoMeSo}.

\begin{prop}[Unitarily transformed Liouvillians]
\label{propunitrans} \  
\begin{itemize} 
\item[(1)]Define the unitaries $u=\e^{\i [P_1\otimes\bbbone_\s\otimes\phi(f_1) + P_2\otimes\bbbone_\s\otimes\phi(f_2)]}$ and $U=uJuJ$, where 
\begin{equation}
\label{f12}
f_{1,2}=(-\i \lambda_{1,2} \,g/\omega)_\beta.
\end{equation}
Then
\begin{equation}
\label{a13}
\wL_\c:=U L_\c U^* = L^{\rm ren}_\s +L_\r +\wI_\c,
\end{equation}
where
\begin{equation}
\label{a15}
L_\s^{\rm ren} = H_\s^{\rm ren}\otimes\bbbone_\s - \bbbone_\s\otimes H_\s^{\rm ren},\qquad 
H_\s^{\rm ren} = \frac12
\begin{pmatrix}
\epsilon -\alpha_1 & 0\\
 0& -\epsilon-\alpha_2
\end{pmatrix}
\end{equation}
with renormalization energies $\alpha_{1,2} = \lambda_{1,2}^2\|g/\sqrt\omega\|_2^2$. The transformed interaction is
\begin{equation}
\label{a14}
\wI_\c = \frac{V}{2}\big[ \sigma_+\otimes\bbbone_\s\otimes W\left(\big( \frac{\lambda_1-\lambda_2}{\i\omega}g\big)_\beta \right) - \bbbone_\s\otimes\sigma_+\otimes JW\left(\big( \frac{\lambda_1-\lambda_2}{\i\omega}g\big)_\beta\right)J +{\rm h.c.}\big],
\end{equation}
where
$$
\sigma_+ = 
\begin{pmatrix}
 0 & 1 \\
0 & 0
\end{pmatrix}.
$$

\item[(2)] Define the unitaries $u=\e^{\i[ P_1\otimes\bbbone_\s\otimes \phi_1(f_1) + P_2\otimes\bbbone_\s\otimes\phi_2(f_2)]}$ and  $U=uJuJ$, where 
\begin{equation}
\label{f1212}
f_{1,2} = (-\i\lambda_{1,2} g_{1,2}/\omega)_\beta.
\end{equation}
Then
\begin{equation}
\label{a11}
\wL_\l:=UL_\l U^*= L_\s^{\rm ren} +L_{\r_1}+L_{\r_2} +\wI_\l,
\end{equation}
where $L^{\rm ren}_\s$ is given as in \eqref{a15} with renormalization energies $\alpha_{1,2} = \lambda^2_{1,2}\|g_{1,2}/\sqrt\omega\|_2^2$. The transformed interaction is  
\begin{eqnarray}
\label{a12}
\wI_\l &=& \frac{V}{2} \big[ \sigma_+\otimes\bbbone_\s\otimes W_1\left( \big( \frac{\lambda_1 g_1}{\i\omega}\big)_\beta \right) \otimes W^*_2\left(\big( \frac{\lambda_2 g_2}{\i\omega}\big)_\beta \right) \\
&& - \bbbone_\s\otimes\sigma_+\otimes JW_1\left(\big( \frac{\lambda_1 g_1}{\i\omega}\big)_\beta \right) J\otimes JW^*_2\left(\big( \frac{\lambda_2 g_2}{\i\omega}\big)_\beta \right) J +{\rm h.c.}\big].
\nonumber
\end{eqnarray}
\end{itemize}
\end{prop}
{\em Remarks.\ } {\bf 1.} For a function $h(k)$, $k\in{\mathbb R}^3$, we have $\|h\|^2_2 = \int_{{\mathbb R}^3} |h(k)|^2\d^3k$.

{\bf 2.} If $\lambda_1=\lambda_2$ (and $g_1=g_2$ in the case of local reservoirs), then $L_\s^{\rm ren}=L_\s$.

{\bf 3.} Sometimes the following expressions for $u$ are useful. In the case of the single collective reservoir, $u=P_1\otimes\bbbone_\s\otimes W(f_1) +P_2\otimes\bbbone_\s\otimes W(f_2)$. In the case of the two local reservoirs, $u=P_1\otimes\bbbone_\s\otimes W_1(f_1) +P_2\otimes\bbbone_\s\otimes W_2(f_2)$.

\medskip

The proof of Proposition \ref{propunitrans} is simply a calculation, based on the relations
\begin{equation}
W(f) L_\r W(f)^* = L_\r -\varphi(\i \omega f) +\tfrac12 \|\sqrt{\omega}f\|_2^2
\end{equation}
and
\begin{equation}
W(f) \varphi(g)W(f)^* =\varphi(g) -{\rm Im}\scalprod{f}{g}.
\end{equation}

\subsection{Dynamics}
\label{sectdynamics}

\subsubsection{Resonance expansion of the propagator}

Let $\psi_0$ be the initial state (represented as a vector in the GNS Hilbert space). The expectation of a (represented) observable $A$ (of the dimer and/or the reservoir(s)) at time $t$ is 
\begin{equation}
\label{a18}
\av{A}_t = \scalprod{\psi_0}{\e^{\i tL} A\e^{-\i t L}\psi_0}.
\end{equation}

Since $\Omega_{\beta,\vlambda,V}$ is separating (see \cite{BRII}), there exists an operator $B$ from the commutant algebra (commuting with all observables)
, s.t.
\begin{equation}
\label{a20}
\psi_0 = B\Omega_{\beta,\vlambda,V}.
\end{equation}
Using relation \eqref{a20} together with the fact that $B$ commutes with $\e^{\i tL} A\e^{-\i tL}$ and the  invariance $\e^{\i tL}\Omega_{\beta,\vlambda,V} =\Omega_{\beta,\vlambda,V}$ gives
\begin{equation}
\label{a22}
\av{A}_t = \scalprod{\psi_0}{ B \e^{\i tL} A\Omega_{\beta,\vlambda,V}}.
\end{equation}
We now use the unitary transformation $U$ described in Proposition \ref{propunitrans},
\begin{equation}
\label{a22}
\av{A}_t = \scalprod{\psi_0}{ B U^* \e^{\i t\wL} UA\Omega_{\beta,\vlambda,V}},
\end{equation}
where $\wL$ is $\wL_\c$ or $\wL_\l$ (see \eqref{a13}, \eqref{a11}), depending on the model considered. \eqref{a22} has the following resonance expansion, which follows from a general resonance expansion of propagators proven in \cite{KoMe}. Set, for ease of notation, 
\begin{equation}
\Psi_1 = UB^*\psi_0 \quad\mbox{and}\quad \Psi_2=UA\Omega_{\beta,\vlambda,V}.
\label{psi12}
\end{equation} 
Then there is a $V_0>0$ s.t. for all $0<|V|<V_0$, and all $t>0$,
\begin{eqnarray}
\lefteqn{
\scalprod{\Psi_1}{\e^{\i t\wL}\Psi_2}= 
\omega_{\beta,\vlambda,V}(A)}\label{a23}\\
&+& \scalprod{\Psi_1}{\left[  \e^{-\gamma t} \Pi_0 + \e^{-\gamma t/2}\ \e^{\i t (\hat\epsilon +x_{\rm LS})} \Pi_+ +\e^{-\gamma t /2} \ \e^{-\i t (\hat\epsilon +x_{\rm LS})} \Pi_- \right] \Psi_2} +R(t).
\nonumber
\end{eqnarray}
Here, $\omega_{\beta,\vlambda,V}$ is the coupled dimer-reservoir(s) equilibrium state.
The ``decay rate'' is given by 
\begin{equation}
\label{a23'}
\gamma= \gamma_\c +O(V^4) \quad \mbox{or}\quad \gamma= \gamma_\l +O(V^4) 
\end{equation}
depending on whether we consider the model with the collective reservoir or the local ones, and where $\gamma_\c$, $\gamma_\l$ are given by \eqref{b12}. The quantity $x_{\rm LS}\in\mathbb R$ is the ``Lamb shift''. Moreover, the resonance projections are
\begin{eqnarray}
\Pi_0 &=& \frac{1}{1+\e^{-\beta\hat\epsilon}}\ | \varphi_{11} -\e^{-\beta\hat\epsilon/2}\varphi_{22}\rangle  \langle  \varphi_{11} -\e^{-\beta\hat\epsilon/2}\varphi_{22}|\otimes P_\r +O(V),\label{c1}\\ 
\Pi_+ &=& |\varphi_{12}\rangle\langle\varphi_{12}| \otimes P_\r  +O(V),\label{c2}\\
\Pi_- &=&  |\varphi_{21}\rangle\langle\varphi_{21}| \otimes P_\r +O(V),\label{c3}
\end{eqnarray}
where $\varphi_{ij}=\varphi_i\otimes\varphi_j$ (see also \eqref{eigenbasis}) and 
\begin{equation}
\label{pr}
P_\r =|\Omega_{\r_1}\rangle\langle \Omega_{\r_1}|\quad \mbox{or}\quad P_\r =|\Omega_{\r_1}\rangle\langle \Omega_{\r_1}|\otimes |\Omega_{\r_2}\rangle\langle \Omega_{\r_2}|,
\end{equation}
(projection onto the vacua of the reservoirs), depending on whether we consider the collective or local reservoirs models. The decay rates and Lamb shifts are (to lowest order in $V$) the real and imaginary parts of {\em level shift operators} which describe the perturbative movement of eigenvalues of $\wL$ for small $V$. We calculate them explicitly in Appendix \ref{LSOsect}. 

The remainder $R(t)$ in \eqref{a23} satisfies the following bounds: there are constants $C_1$ and $C_2$ (independent of $t$ and $V$), such that for all $t>0$, 
\begin{equation}
\label{a30}
|R(t)| \le \frac{C_1}{t},\quad |R(t)|\le C_2.
\end{equation}

{\em Remark.\ } To prove the expansion \eqref{a23} one needs some regularity of the form factors $g$, $g_{1,2}$ appearing in \eqref{1'} and \eqref{1}. We explain them in Section \ref{regularitysect}.

\subsubsection{Effective dynamics}

Define the operators $D_j$, $j=0,\pm$, belonging to the algebra of observables of the joint dimer-reservoir(s) system, as follows.
\begin{itemize}
\item[(1)] For the model with the single collective reservoir, 
\begin{equation}
	\label{a42}
	\begin{split}
		D_0 &= \ \ \e^{\beta\hat\epsilon/2} |\varphi_1\rangle\langle\varphi_1|\otimes \bbbone_\s \otimes\bbbone_\r- \e^{-\beta\hat\epsilon/2} |\varphi_2 \rangle\langle\varphi_2|\otimes \bbbone_\s\otimes\bbbone_\r \\ 
		D_+ &= \ \ \sqrt{1+\e^{-\beta\hat\epsilon}}\  |\varphi_1\rangle\langle\varphi_2|\otimes \bbbone_\s\otimes W^*(f_1)W(f_2)\\
		D_-&=\ \ \e^{\beta\hat\epsilon/2} \sqrt{1+\e^{-\beta\hat\epsilon}}\  |\varphi_2\rangle\langle\varphi_1|\otimes \bbbone_\s\otimes W^*(f_2)W(f_1).
	\end{split}
\end{equation}
\qquad \quad (Recall the definition \eqref{f12} of $f_{1,2}$.)

\item[(2)] For the model with the two local reservoirs, 
\begin{equation}
	\label{a43}
	\begin{split}
		D_0 &= \ \ \e^{\beta\hat\epsilon/2} |\varphi_1\rangle\langle\varphi_1|\otimes \bbbone_\s\otimes\bbbone_{\r_1}\otimes\bbbone_{\r_2} - \e^{-\beta\hat\epsilon/2} |\varphi_2 \rangle\langle\varphi_2|\otimes \bbbone_\s\otimes\bbbone_{\r_1}\otimes\bbbone_{\r_2} \\ 
		D_+ &= \ \ \sqrt{1+\e^{-\beta\hat\epsilon}}\  |\varphi_1\rangle\langle\varphi_2|\otimes \bbbone_\s\otimes W^*(f_1)\otimes W(f_2)\\
		D_-&=\ \ \e^{\beta\hat\epsilon/2} \sqrt{1+\e^{-\beta\hat\epsilon}}\  |\varphi_2\rangle\langle\varphi_1|\otimes \bbbone_\s\otimes W(f_1)\otimes W^*(f_2).
	\end{split}
\end{equation}
\qquad \quad (Recall the definition \eqref{f1212} of $f_{1,2}$.)
\end{itemize}

The main result of this section is the following representation for the dynamics.

\begin{thm}
\label{effdynprop}
Let $\omega_0$ be the initial dimer-reservoir(s) state and let $A$ be a dimer-reservoir(s) observable. Then
\begin{equation}
	\label{001}
\av{A}_t = \omega_{\beta,\vec\lambda,V}(A) + \sum_{j=0,\pm} \e^{\i t a_j}\, \omega_0(D_j)\,  \omega_{\beta,\vlambda,0}(D^*_jA) +R(t),
\end{equation}
where $\omega_{\beta,\vlambda,V}$ is the dimer-reservoir(s) equilibrium state, $R(t)$ satisfies \eqref{a30} and  
\begin{equation}
	\label{as}
a_j=\left\{
\begin{array}{ll}
\i\gamma & j=0,\\
\i\gamma/2\pm(\hat\epsilon +x_{\rm LS}) & j=\pm
\end{array}
\right.
\end{equation}
where $\gamma=\gamma_{\rm col}$ or $\gamma=\gamma_{\rm loc}$ given in \eqref{b12} and $x_{\rm LS}$ is the Lamb shift \eqref{lambshift}. 
\end{thm}

{\em Proof of Theorem \ref{effdynprop}.\ }

We use the expansion \eqref{a23}. 
The following result follows from an easy calculation combining \eqref{c1}-\eqref{c3} with the definition of the unitary $U$ given in Proposition \ref{propunitrans}. 
\begin{lem}
	\label{lem2}
	The projections $\Pi_j$, $j=0,\pm$, given in \eqref{c1}-\eqref{c3} have the form
	\begin{equation}
		U^*\Pi_jU = |X_j\rangle\langle X_j| +O(V),
		\label{a39}
	\end{equation}
	where the vectors $X_j$ are as follows.
\begin{itemize}
\item[(1)] For the model with the collective reservoir, we have 
	\begin{equation}
		\begin{split}
			X_0 &=\ \  \frac{1}{\sqrt{1+\e^{-\beta\hat\epsilon}}}\left( \varphi_{11}\otimes W^*(f_1)JW^*(f_1)J\Omega_\r -\e^{-\beta\hat\epsilon/2}\varphi_{22}\otimes W^*(f_2)JW^*(f_2)J\Omega_\r\right)\\
			X_+&=\ \ \varphi_{12}\otimes W^*(f_1)JW^*(f_2)J\Omega_\r \\
			X_- &= \ \  \varphi_{21}\otimes W^*(f_2)JW^*(f_1)J\Omega_\r 
		\end{split}
		\label{a41}
	\end{equation}
\item[(2)] For the model with the two local reservoirs, we have
	\begin{equation}
		\begin{split}
			X_0 &= \ \  \frac{1}{\sqrt{1+\e^{-\beta\hat\epsilon}}}\left( \varphi_{11}\otimes W^*_1(f_1)JW^*_1(f_1)J\Omega_{\r_1}\otimes\Omega_{\r_2}\right.\\
			& \ \ \  \left. -\e^{-\beta\hat\epsilon/2}\varphi_{22} \otimes\Omega_{\r_2}\otimes  W^*_2(f_2)JW^*_2(f_2)J\Omega_{\r_2}\right)\\
			X_+ &=\ \ \varphi_{12}\otimes W^*(f_1)\Omega_{\r_1}\otimes JW^*(f_2)J\Omega_{\r_2}\\
			X_- &=\ \ \varphi_{21}\otimes JW^*(f_1)J\Omega_{\r_1}\otimes W^*(f_2)\Omega_{\r_2}.
		\end{split}
		\label{a42}
	\end{equation}
\end{itemize}
\end{lem}
Lemma \ref{lem2} shows that
\begin{equation}
	\label{a38}
	\scalprod{\psi_0}{BU^*\Pi_j UA \Omega_{\beta,\vlambda,0}} = \scalprod{\psi_0}{BX_j}\scalprod{X_j}{A \Omega_{\beta,\vlambda,V}}.
\end{equation}

By Araki's perturbation theory of KMS states (\cite{BRII}, Theorem 5.4.4), the equilibrium states with $V=0$ and $V\neq 0$ satisfy
\begin{equation}
	\label{a19}
	\|\Omega_{\beta,\vlambda,V}-\Omega_{\beta,\vlambda,0}\| \le \e^{\beta V/2}-1 = O(V).
\end{equation}
Thus,
\begin{equation}
	\label{a038}
	\scalprod{\psi_0}{BU^*\Pi_j UA \Omega_{\beta,\vlambda,0}} = \scalprod{\psi_0}{BX_j}\scalprod{X_j}{A \Omega_{\beta,\vlambda,0}} +O(V).
\end{equation}
The first factor on the right side of \eqref{a038} is linked to the initial condition $\psi_0$, the second one to the equilibrium $\Omega_{\beta,\vlambda,0}$.

\begin{lem}
	\label{lem5}
The operators $D_j$ defined in \eqref{a42}, \eqref{a43} satisfy
	\begin{equation}
		\label{a40}
		X_j = D_j\Omega_{\beta,\vlambda,0}.
	\end{equation}
\end{lem} 
{\em Proof of Proposition \ref{lem5}.\ } The equilibrium states for the collective and local models are
\begin{equation}
	\label{a44}
	\Omega_{\beta,\vlambda,0} = 
	\tfrac{1}{\sqrt{1+\e^{-\beta\hat\epsilon}}}
	\left\{
	\begin{array}{l}
		\e^{-\beta\hat\epsilon/2} \varphi_{11}\otimes W^*(f_1)JW^*(f_1)J\Omega_\r + \varphi_{22}\otimes W^*(f_2)JW^*(f_2)J\Omega_\r\\
		\mbox{\qquad {and}} \ \\
		\e^{-\beta\hat\epsilon/2} \varphi_{11}\otimes W^*(f_1)JW^*(f_1)J\Omega_{\r_1}\otimes \Omega_{\r_2} +\\
		\qquad\qquad + \varphi_{22}\otimes \Omega_{\r_1}\otimes W^*(f_2)JW^*(f_2)J\Omega_{\r_2}
	\end{array}
	\right.
\end{equation}
The relations \eqref{a40}, \eqref{a42} and \eqref{a43} are then easily seen to hold.\hfill\qed

Since the operators $D_j$ commute with $B$, we have  
\begin{equation}
	\label{002}
	\scalprod{\psi_0}{BX_j} = \scalprod{\psi_0}{BD_j\Omega_{\beta,\vlambda,0}}= \scalprod{\psi_0}{D_j\psi_0}=\omega_0(D_j).
\end{equation}
 Moreover,  
\begin{equation}
	\label{003}
\scalprod{X_j}{A \Omega_{\beta,\vlambda,0}}= \omega_{\beta,\vlambda,0}(A)
\end{equation}
Combining \eqref{002} and \eqref{003} with \eqref{a38} and \eqref{a038}, the expansion \eqref{001} now follows from \eqref{a23}. This completes the proof of Theorem  \ref{effdynprop}. \hfill \qed

\subsubsection{Proofs of Theorems \ref{thm1} and \ref{thm2} }

{\bf Dynamics of populations and proof of Theorem \ref{thm1}.\ } Choosing $A=|\varphi_1\rangle\langle\varphi_1|\otimes\bbbone_\s\otimes \bbbone_\r$ in \eqref{001} yields the population of the level $1$, i.e., 
\begin{equation}
\label{a45}
\av{|\varphi_1\rangle\langle\varphi_1|\otimes\bbbone_\s\otimes \bbbone_\r}_t = {\rm Tr}\rho_\s(t) |\varphi_1\rangle\langle\varphi_1| = \scalprod{\varphi_1}{\rho_\s(t)\varphi_1} \equiv [\rho_\s(t)]_{11}.
\end{equation}
{}For this choice of $A$ and for both models, we have
\begin{equation}
\label{a46}
 \omega_{\beta,\vlambda,0}(D^*_\pm A)=0,\quad  \omega_{\beta,\vlambda,0}(D^*_0 A)= \frac{\e^{-\beta\hat\epsilon/2}}{1+\e^{-\beta\hat\epsilon}}
 \end{equation}
and
\begin{equation}
	\label{a046}
	\omega_0(D_0) = \e^{\beta\hat\epsilon/2}[\rho_\s(0)]_{11} -\e^{-\beta\hat\epsilon/2}[\rho_\s(0)]_{22} .
\end{equation}
Furthermore, 
\begin{equation}
\label{a47}
\omega_0(D_+) =\sqrt{1+\e^{-\beta\hat \epsilon}}\  [\rho_\s(0)]_{21}
\left\{
\begin{array}{l}
\av{W^*(f_1)W(f_2)}_{\r}\mbox{\quad (collective)}\\
\av{W^*(f_1)}_{\r_1} \av{W(f_2)}_{\r_2}  \mbox{\quad (local)},
\end{array}
\right. 
\end{equation}
and 
\begin{equation}
\label{a48}
\omega_0(D_-) =\e^{\beta\hat\epsilon/2}\  \overline{\omega_0({D_+})}.
\end{equation}
Combining \eqref{001} with \eqref{a46}-\eqref{a48} yields \eqref{a51}.

\medskip

\noindent
{\bf Dynamics of decoherence and proof of Theorem \ref{thm2}.\ } 
Choosing $A=|\varphi_2\rangle\langle\varphi_1|\otimes\bbbone_\s\otimes \bbbone_\r$ in \eqref{001} yields the off-diagonal dimer density matrix element,  
\begin{equation}
\label{a60}
\av{|\varphi_2\rangle\langle\varphi_1|\otimes\bbbone_\s\otimes \bbbone_\r}_t = {\rm Tr}\rho_\s(t) |\varphi_2\rangle\langle\varphi_1| = \scalprod{\varphi_1}{\rho_\s(t)\varphi_2} \equiv [\rho_\s(t)]_{12}.
\end{equation}
{}For this choice of $A$ and for both models, we have
\begin{equation}
\label{a61}
 \omega_{\beta,\vlambda,0}(D^*_0 A)=0=\omega_{\beta,\vlambda,0}(D^*_+ A)
\end{equation}
and
\begin{equation}
\label{a62}
\omega_{\beta,\vlambda,0}(D^*_- A) =
\frac{\e^{-\beta\hat\epsilon/2}}{\sqrt{1+\e^{-\beta\hat \epsilon}}}\ 
\left\{
\begin{array}{l}
\av{W(f_2)W^*(f_1)}_{\r} \mbox{\quad (collective)}\\
\av{W^*(f_1)}_{\r_1} \av{W(f_2)}_{\r_2} \mbox{\quad (local)}.
\end{array}
\right. 
\end{equation}
Furthermore,
\begin{equation}
	\label{0010}
	\omega_0(D_-) = \e^{\beta\hat\epsilon/2} \sqrt{1+\e^{-\beta\hat\epsilon}}\  [\rho_\s(0)]_{12}
\left\{
\begin{array}{l}
	\av{W^*(f_2)W(f_1)}_{\r} \mbox{\quad (collective)}\\
	\av{W(f_1)}_{\r_1} \av{W^*(f_2)}_{\r_2} \mbox{\quad (local)}.
\end{array}
\right. 	
\end{equation}
Taking into account \eqref{a61}-\eqref{0010}, the expression \eqref{a70} for $\Gamma_\infty$, and expansion \eqref{001}, we obtain \eqref{a64}.

\section*{Conclusion}

We give a theoretical analysis of a (chlorophyll-based) dimer interacting with collective (spatially correlated) and local (spatially uncorrelated) protein-solvent environments, as they appear in photosynthetic bio-complexes. We formulate the problem in the language of a spin-boson system, in which two excited electron energy levels of two light-sensitive molecules (such as chlorophylls or carotenoids) are described by an effective spin $1/2$. Both spin levels (excited electron states of donor and acceptor) interact with a thermal environment, modeled by bosonic degrees of freedom (quantum linear oscillators). In the case of a correlated thermal environment,  a single set of bosonic operators, with the characteristic frequencies of the environment, is introduced. In the case of an uncorrelated thermal environment, two sets of bosonic operators are introduced, one for the donor and one for the acceptor. In either situation, we introduce two independent constants of interaction between the donor and the acceptor and the environment(s).

We develop a mathematically rigorous perturbation theory in which the direct matrix element of the donor-acceptor interaction is a small parameter. This perturbation theory is different from the standard Bloch-Redfield theory where the interaction constant between the dimer and the environment(s) is used as a small parameter. Our approach allows us to consider, in a controlled way, the case of strong interaction constants and ambient temperatures. This is important for applications to real bio-systems.

We derive the explicit expressions for the thermal electron transfer rates, and demonstrate the differences with the standard Marcus expression for the electron transfer rates in the so-called high-temperature regime.
We analyze the dynamics of decoherence depending on the parameters of the system. In particular, we demonstrate how long-time coherences naturally occurs in the dimer. 

The results of the paper are important for a better understanding of the complicated quantum dynamics a chlorophyll-based dimer undergoes in photosynthetic complexes, for a wide region of parameters and under controlled approximations. Experiments could be used to verify our theoretical predictions.

\subsection*{Acknowledgments} This work was carried out under the auspices of the National Nuclear Security
Administration of the U.S. Department of Energy at Los Alamos National Laboratory
under Contract No. DE-AC52-06NA25396. M.M. and H.S. have been supported by NSERC through a Discovery
Grant and a Discovery Accelerator Supplement. M.M. is grateful for the hospitality and financial support of LANL, where part of this work was carried out. A.I.N. acknowledges support from the CONACyT, Grant No. 15349 and partial support during his visit from the Biology Division, B-11, at LANL.  G.P.B, S.G., and R.T.S. acknowledge  support from the LDRD program at LANL.

\appendix 

\section{The level shift operators}

\label{LSOsect}

Let $\chi_0$ and $\chi_{\pm}$ be the orthogonal projections onto the eigenspaces of 
\begin{equation}
\label{wlo}
\wL_0 = L_\s^{\rm ren} +L_\r \quad\mbox{or}\quad \wL_0 = L_\s^{\rm ren} +L_{\r_1}+L_{\r_2}
\end{equation}
associated to the eigenvalues zero and $\pm\hat\epsilon$, where
\begin{equation}
\label{hatepsilon}
\hat\epsilon = \epsilon -\frac{\alpha_1-\alpha_2}{2}.
\end{equation}
(See also Proposition \ref{propunitrans}.) 
Explicitly, $\chi_0 = |\varphi_{11}\rangle\langle\varphi_{11}|\otimes P_\r +|\varphi_{22}\rangle\langle\varphi_{22}|\otimes P_\r$ and $\chi_+=|\varphi_{12}\rangle\langle\varphi_{12}|\otimes P_\r$, $\chi_-=|\varphi_{21}\rangle\langle\varphi_{21}|\otimes P_\r$, where we recall that $P_\r$ is given in \eqref{pr}. To each unperturbed eigenvalue of $\wL_0$, we associate a level shift operator. The one associated to the eigenvalue zero is (the $2\times 2$ matrix)
\begin{equation}
\label{lso}
\Lambda_0 = \chi_0 \wI (\wL_0+\i 0_+)^{-1} \chi_0^\perp\wI\chi_0,
\end{equation}
where $\wI$ is the interaction operator, $\wI_\c$ or $\wI_\l$, see \eqref{a14}, \eqref{a12}. The level shift operators associated to the eigenvalues $\pm \hat\epsilon$ are both one-dimensional, 
\begin{equation}
\Lambda_+ = \chi_+\wI (\wL_0-\hat\epsilon +\i0_+)^{-1}\chi_+^\perp \wI \chi_+ \quad \mbox{and}\quad \Lambda_- = \chi_- \wI (\wL_0+\hat\epsilon +\i0_+)^{-1}\chi_-^\perp \wI \chi_-\ .
\end{equation}

\begin{prop}
\label{lso0prop}
{\rm \bf (1)} Consider the model with the collective environment and set 
\begin{equation}
x(\hat\epsilon) =-\frac{\i}{2}{\rm  Re} \int_0^\infty\e^{- \i\hat\epsilon t} \e^{\frac{\i  (\lambda_1-\lambda_2)^2}{\pi} Q_1(t) } \e^{-\frac{(\lambda_1-\lambda_2)^2}{\pi} Q_2(t)}  \d t,
\label{ma17'}
\end{equation}
where $Q_1(t)$, $Q_2(t)$ are given in \eqref{b4}. In the basis $\{\varphi_{11}, \varphi_{22}\}$, we have
\begin{equation}
\Lambda_0 =x(\hat\epsilon)
\begin{pmatrix}
 1& -\e^{-\beta\hat\epsilon/2} \\
-\e^{-\beta\hat\epsilon/2} & \e^{-\beta\hat\epsilon}
\end{pmatrix}.
\label{m16'}
\end{equation}
The eigenvalues of $V^2 \Lambda_0$ are $0$ and $i\gamma_\c$, where $\gamma_\c$ is given in \eqref{b12}.  Moreover, the one-dimensional level shift operators $\Lambda_\pm$ have the expressions
\begin{equation*}
V^2 \Lambda_+= \big(x_{\rm LS} +\tfrac{\i}{2} \gamma_\c\big)  \,|\varphi_{12}\rangle\langle\varphi_{12}|\otimes P_\r \quad\mbox{and}\quad  V^2 \Lambda_-= \big(-x_{\rm LS}+\tfrac{\i}{2} \gamma_\c \big) |\varphi_{21}\rangle\langle\varphi_{21}|\otimes P_\r.
\end{equation*}

{\rm \bf (2)} Consider the model with the two local reservoirs. The level shift operator $\Lambda_0$ has the form \eqref{m16'}, where now 
\begin{equation}
x(\hat\epsilon) =-\frac{\i}{2}{\rm  Re} \int_0^\infty\e^{- \i\hat\epsilon t} \e^{-\frac{\lambda_1^2}{\pi}\big[Q_2^{(1)}(t)-\i Q_1^{(1)} (t)\big]} \e^{-\frac{\lambda_2^2}{\pi} \big[Q_2^{(2)}(t)-\i Q_1^{(2)} (t)\big]}  \d t
\label{m17'}
\end{equation}
and $Q_1^{(j)}$, $Q_2^{(j)}$ are given in \eqref{b3}. 
The eigenvalues of $V^2\Lambda_0$ are $0$ and $i\gamma_\l$, where $\gamma_\l$ is given in \eqref{b12}.  Moreover, the one-dimensional level shift operators $\Lambda_\pm$ have the expression
\begin{equation*}
V^2\Lambda_+= \big(x_{\rm LS} +\tfrac{\i}{2} \gamma_\l\big)  \,|\varphi_{12}\rangle\langle\varphi_{12}|\otimes P_\r \quad\mbox{and}\quad V^2\Lambda_-= \big(-x_{\rm LS}+\tfrac{\i}{2} \gamma_\l \big) |\varphi_{21}\rangle\langle\varphi_{21}|\otimes P_\r.
\end{equation*}
\end{prop}

{\em Proof of Proposition \ref{lso0prop}.\ } {\rm \bf (1)} In the case of a single reservoir, the calculation of the level shift operator $\Lambda_0$, \eqref{m16'} is an easy transcription of that performed in Proposition 3.5 of \cite{KoMeSo}, where the case $\lambda_1=-\lambda_2$ was considered. We do not present the details. 

\noindent
{\rm \bf (2)} Once we have the form \eqref{m16'}, it is readily seen that the vector $\Psi=\varphi_{11} +\e^{\beta\hat\epsilon/2}\varphi_{22}$
is in the kernel of $\Lambda_0$. Therefore, the eigenvalues of $\Lambda_0$ are $0$ and ${\rm Tr}\Lambda_0 = (1+\e^{-\beta\hat\epsilon})x(\hat\epsilon)$. The relation $\e^{-\beta\hat\epsilon}x(\hat\epsilon) = x(-\hat\epsilon)$ (see \eqref{nm5}) together with \eqref{m17'} gives ${\rm spec} V^2 \Lambda_0 = \{ 0,\i \gamma_\l\}$.

Our task now is to show \eqref{m16'}. Let
$$
\LR =L_{R_1}+L_{R_2}
$$
and write $W_1$ instead of $W_1(f_1)\otimes\bbbone$ and $W_2$ for $\bbbone\otimes W_2(f_2)$, where the $f_j$ are given in \eqref{f1212}.  Then
\begin{equation}
\label{m10}
\begin{split}
4\Lambda_0\varphi_{11}= & \ \varphi_{11}\langle W_1 W_2^*(\LR-\hat\epsilon+\i
0^+)^{-1}W_1^* W_2\rangle\\
&+\varphi_{11}\langle JW_1 W_2^*J(\LR+\hat\epsilon+\i 0^+)^{-1}JW_1^* W_2J\rangle\\
&-\varphi_{22}\langle JW_1^*
W_2J(\LR-\hat\epsilon+\i
0^+)^{-1}W_1^* W_2\rangle\\
&-\varphi_{22}\langle W_1^* W_2(\LR+\hat\epsilon+\i 0^+)^{-1}JW_1^* W_2J\rangle.
\end{split}
\end{equation}
If we want to stress the dependence of $\Lambda_0$ on $\hat\epsilon$ and $\lambda_1,\lambda_2,$ we wirte $\Lambda_0(\hat\epsilon, \lambda_1,\lambda_2)$. Then \eqref{m10} gives
\begin{equation}
\label{m13}
\Lambda_0(\hat\epsilon,
\lambda_1,\lambda_2)\varphi_{11}=x(\hat\epsilon,
\lambda_1,\lambda_2)\varphi_{11}+ z(\hat\epsilon,
\lambda_1,\lambda_2)\varphi_{22}
\end{equation}
with
\begin{equation}
\begin{split}
\label{m11}
x(\hat\epsilon, \lambda_1,\lambda_2) = &\tfrac{\i}{2} {\rm Im} \langle W_1
W_2^*(\LR-\hat\epsilon+\i 0^+)^{-1}W_1^*W_2\rangle,\\
z(\hat\epsilon, \lambda_1,\lambda_2) = & -\tfrac{\i}{2} {\rm Im}\langle
W_1^* W_2(\LR+\hat\epsilon+\i 0^+)^{-1}JW_1^* W_2J\rangle.
\end{split}
\end{equation}
We use the relation $(\LR-\hat\epsilon+\i r)^{-1}=-\i \int_0^\infty 
\e^{\i t(\LR-\hat\epsilon+\i r )}\d t$ (for $r>0$) to obtain
\begin{equation}
\label{m12}
x(\hat\epsilon, \lambda_1,\lambda_2) = -\tfrac{\i}{2}  \lim_{r\rightarrow
0_+}{\rm Re} \int_0^\infty  \e^{\i(-\hat\epsilon+\i r )t}\langle
W_1(f_1)W_1(-\e^{\i t \omega}f_1)\rangle\,  \langle W_2(-f_2)
W_2(\e^{\i t \omega}f_2)\rangle \d t.
\end{equation}
With the canonical commutation relations $W(f)W(g)=\e^{-\frac{\i}{2} {\rm Im}\scalprod{f}{g}} W(f+g)$ and the thermal average $\langle W(f)\rangle=\exp {-\frac14}\scalprod{f}{\coth(\beta\omega/2)f}$  we arrive at
$$
\langle W(f)W(-\e^{\i t\omega} f)\rangle = \e^{\frac{\i}{2}\scalprod{f}{\sin(\omega t)f}} \e^{-\frac12\scalprod{f}{(1-\cos(\omega t))\coth(\beta\omega/2)f}}.
$$
Then, remembering the definition \eqref{f1212} of $f_j$, we arrive at 
\begin{eqnarray}
\lefteqn{
x(\hat\epsilon,\lambda_1,\lambda_2) }\nonumber\\
&=&-\tfrac{\i}{2} \lim_{r\rightarrow
0_+}{\rm Re } \int_0^\infty \e^{\i(-\hat\epsilon+\i r
)t}\e^{\frac{\i}{2}\lambda_1^2\scalprod{g_1}{\frac{\sin\omega t}{\omega^2} 
g_1} -\frac12\lambda^2_1\scalprod{g_1}{\frac{1-\cos\omega 
t}{\omega^2}\coth(\beta\omega/2)g_1}}\nonumber\\
&& \times \e^{\frac{\i}{2}\lambda_2^2\scalprod{g_2}{\frac{\sin\omega t}{\omega^2} g_2} -\frac12\lambda^2_2\scalprod{g_2}{\frac{1-\cos\omega t}{\omega^2}\coth(\beta\omega/2)g_2}} \d t,
\label{nm1}
\end{eqnarray}
which is the expression for $x(\hat\epsilon)$ given in \eqref{m17'}. Note that $x$ is invariant under a sign change of either $\lambda_1$ or $\lambda_2$. 

Next, we find an expression for $z(\hat \epsilon,\lambda_1,\lambda_2)$. Starting with the definition  \eqref{m11} and replacing the resolvent by an integral over the propagator, as above, yields the expression
\begin{equation}
\label{nm3}
z(\hat\epsilon,\lambda_1,\lambda_2) = \tfrac{\i}{2} \int_0^\infty \cos(\hat\epsilon t)\  \alpha(f_1,-\e^{\i t \omega}f_1) \alpha(f_2,-\e^{\i t \omega}f_2)   \d t,
\end{equation}
where 
\begin{equation}
\alpha(f,g)=\e^{\frac14\{\langle f,\e^{-\beta w/2}g\rangle-\langle g,\e^{\beta
w/2}f\rangle\}} \e^{-\frac{1}{4}\{\langle f,Cf\rangle+\langle
g,Cg\rangle-\langle g,C\e^{\beta w/2}f\rangle-\langle f,C\e^{-\beta
w/2}g\rangle\}},
\end{equation}
with $C=\coth(\beta\omega/2)$. Relation \eqref{nm3} shows that $z$ is invariant under changing signs of either of $\hat\epsilon$, $\lambda_1$ and $\lambda_2$.  Note also that $z$ is real. 

The symmetry
$$
(\sigma_x\otimes\sigma_x)\ \Lambda_0(\hat\epsilon,\lambda_1,\lambda_2) \
(\sigma_x\otimes\sigma_x) =
\Lambda_0(-\hat\epsilon,-\lambda_1,-\lambda_2)
$$
together with \eqref{m13} and the fact that $(\sigma_x\otimes\sigma_x) \varphi_{11} =\varphi_{22}$ yields  
\begin{equation}
\label{m12}
\Lambda_0(\hat\epsilon,\lambda_1,\lambda_2)  \varphi_{22} =
z(-\hat\epsilon,-\lambda_1,-\lambda_2)\varphi_{11}
+x(-\hat\epsilon,-\lambda_1,-\lambda_2)\varphi_{22}.
\end{equation}
We show below that 
\begin{equation}
z(\hat\epsilon,\lambda_1,\lambda_2) = -\e^{-\beta\hat\epsilon/2} x(\hat\epsilon,\lambda_1,\lambda_2). 
\label{m14}
\end{equation}
Together with the invariance of $z$ under flipping the sign of $\hat\epsilon$, this gives 
\begin{equation}
\label{nm5}
x(-\hat\epsilon) = \e^{-\beta\hat\epsilon} x(\hat\epsilon).
\end{equation}
This shows the form \eqref{m16'}, modulo showing relation \eqref{m14}. We 
prove \eqref{m14} now. Set $\chi=W_1^*\Omega_1\otimes W_2\Omega_2$ 
and consider, for $r>0$,
\begin{eqnarray}
\lefteqn{
{\rm Im}\scalprod{\chi}{(\LR-\hat\epsilon+\i r    )^{-1}\chi}  = 
-\scalprod{\chi}{\frac{r     }{(\LR-\hat\epsilon)^2+r^2}\chi}}\nonumber\\
&=& -\scalprod{\chi}{\frac{r}{(\LR-\hat\epsilon)^2+r^2}\ \big[ 
\e^{-\beta(\LR-\hat\epsilon)/2} + 1 -\e^{-\beta(\LR-\hat\epsilon)/2}\big] 
\chi}.
\label{m16}
\end{eqnarray}
Since 
\begin{equation}
\label{m17}
\e^{-\beta\LR/2}\chi= J\cdot J\e^{-\beta\LR/2} (W^*_1\Omega_1\otimes W_2\Omega_2) = J (W_1\Omega_1\otimes W^*_2\Omega_2) 
\end{equation}
is well defined, we can separate the terms in \eqref{m16}. We have used in the last step in \eqref{m17} that $J\e^{-\beta\LR/2}( A_1\Omega_1\otimes A_2\Omega) = A^*_1\Omega_1\otimes A^*_2\Omega_2$ (by the properties of the modular conjugation $J$ and the modular operator $\Delta=\e^{-\beta \LR}$). Then, by \eqref{m17},
\begin{eqnarray}
-\scalprod{\chi}{\frac{r}{(\LR-\hat\epsilon)^2+r^2}\e^{-\beta(\LR-\hat\epsilon)/2}
 \chi} &=& \e^{\beta \hat\epsilon/2} {\rm Im} 
\scalprod{\chi}{(\LR-\hat\epsilon+\i r )^{-1} \e^{-\beta \LR/2} 
\chi}\nonumber\\
&= &  \e^{\beta \hat\epsilon/2}{\rm Im}  \langle 
W_1W^*_2(\LR-\hat\epsilon+\i r )^{-1} J W_1 W^*_2\rangle.\ \ \qquad
\label{m18}
\end{eqnarray}
Next, by the functional calculus,
\begin{eqnarray}
\scalprod{\chi}{\frac{r}{(\LR-\hat\epsilon)^2+r^2}[1 -\e^{-\beta(\LR-\hat\epsilon)/2}] \chi} &=&\int_{\mathbb R}\frac{r\, (x-\hat\epsilon)}{(x-\hat\epsilon)^2+r^2}\frac{1 -\e^{-\beta(x-\hat\epsilon)/2}}{(x-\hat\epsilon)}\d\mu_\chi(x),\ \ \ \qquad 
\label{m19}
\end{eqnarray}
where $\d\mu_\chi(x)$ is the spectral measure of $\LR$ in the state $\chi$.  
One readily sees that the integrand satisfies the bound
$$
\left|\frac{r (x-\hat\epsilon)}{(x-\hat\epsilon)^2+r^2}\frac{1 -\e^{-\beta(x-\hat\epsilon)/2}}{(x-\hat\epsilon)}\right|\le C(1+\e^{-\beta(x-\hat\epsilon)/2}),
$$
independently of $r$. The right side is integrable w.r.t. $\d\mu_\chi(x)$ since $\chi$ is in the domain of definition of $\e^{-\beta \LR/2}$. Then, since the integrand in \eqref{m19} converges to zero as $r\rightarrow 0$, for all $x\in{\mathbb R}\backslash\{\hat\epsilon\}$, and since $\{\hat\epsilon\}$ has measure zero w.r.t. $\d\mu_\chi(x)$ (as $\hat\epsilon\neq 0$ is not an eigenvalue of $\LR$), we can apply the Dominated Convergence Theorem to conclude that 
\begin{eqnarray}
\lim_{r\rightarrow 0} \scalprod{\chi}{\frac{r}{(\LR-\hat\epsilon)^2+r^2}[1 -\e^{-\beta(\LR-\hat\epsilon)/2}] \chi} =0.
\label{m20}
\end{eqnarray}
It follows from \eqref{m16}, \eqref{m18} and \eqref{m20} that 
\begin{equation}
\label{m21}
{\rm Im} \langle W_1W^*_2( \LR-\hat\epsilon+\i0_+)^{-1} W^*_1W_2\rangle = \e^{\beta \hat\epsilon/2}{\rm Im}  \langle W_1W^*_2(\LR-\hat\epsilon+\i 0_+)^{-1} J W_1 W^*_2\rangle.
\end{equation}
Relation \eqref{nm1} shows that $x(\hat\epsilon)$ is invariant under a change of the sign of $\lambda_1$ and $\lambda_2$ (independently), so interchanging $W_1\leftrightarrow W_1^*$ and $W_2\leftrightarrow W_2^*$ in the expression \eqref{m11} for $x(\hat \epsilon)$ does not change the value of the expression. Thus we have 
$$
x(\hat\epsilon) = \tfrac{\i}{2} {\rm Im} \langle W_1^* W_2(\LR-\hat\epsilon+\i 0^+)^{-1} W_1 W^*_2\rangle = \tfrac{\i}{2}\e^{\beta\hat\epsilon/2}{\rm Im} \langle W^*_1W_2(\LR-\hat\epsilon+\i 0_+)^{-1} J W^*_1 W_2\rangle,
$$
where we have used \eqref{m21} in the last step (with $\lambda_{1,2}$ replaced by $-\lambda_{1,2}$). The definition \eqref{m11} for $z(\hat \epsilon)$ shows that the expression on the right side is $-\e^{\beta\hat\epsilon/2} z(-\hat\epsilon)$. Therefore we have proven the relation $z(-\hat\epsilon)=-\e^{-\beta\hat\epsilon/2}x(\hat\epsilon)$, which yields \eqref{m14} since $z(-\hat\epsilon)=z(\hat\epsilon)$. This completes the proof of \eqref{m16'}. 

Finally, the forms of $\Lambda_\pm$ given at the end of the proposition are checked directly by a simple calculation. This completes the proof of Proposition \ref{lso0prop}. \hfill\qed

\section{System for $V=0$, factor $\e^{-\Gamma_\infty}$}
\label{appB}

\subsection{Equilibrium for $V=0$}
\label{subsectB1}

In the setting of the collective reservoir, let 
\begin{equation}
\label{a56}
u=\e^{\i P_1\otimes \phi(f_1) + \i P_2\otimes \phi(f_2)} = P_1\otimes W(f_1) +P_2\otimes W(f_2),
\end{equation}
where $f_j=\i\lambda_j g/\omega$. A direct calculation shows that 
\begin{equation}
\label{a54}
u H_\c u^* = \tfrac12
\begin{pmatrix}
\epsilon-2\lambda^2_1\nu/\pi & 0\\
0 & -\epsilon-2\lambda^2_2\nu/\pi
\end{pmatrix}
+H_\r +\tfrac{V}{2}\left( \sigma_+\otimes W\big((\lambda_1-\lambda_2) g /(\i\omega)\big) +{\rm h.c.}\right),
\end{equation}
where $\nu$ is given in \eqref{b4} and $\sigma_+$ is the raising operator. To calculate the equilibrium state with $V=0$, we note that
\begin{equation}
\label{a55}
\e^{-\beta H_\c(V=0)} = u^*\e^{-\beta uH_\c(V=0) u^*}u = u^* \big( \e^{-\frac{\beta}{2}(\epsilon-2\lambda^2_1\nu/\pi)} P_1 + \e^{-\frac{\beta}{2}(-\epsilon-2\lambda^2_2\nu/\pi)} P_2\big) \otimes \e^{-\beta H_\r} \ u.
\end{equation}
Now \eqref{01'} follows readily from \eqref{a55} and \eqref{a56}. 

In the situation of the local reservoirs model, set $u=\e^{\i P_1\otimes \phi_1(f_1) + \i P_2\otimes \phi_2(f_2)}$, where $f_j=\i\lambda_j g_j/\omega$, and proceed as above to show \eqref{01}.

\subsection{Decoherence for $V=0$, proof of Proposition \ref{prop4}}
\label{subsectB2}

Using the same notation as in Section \ref{subsectB1} we have
\begin{equation}
\label{a68}
{}[\rho_\s(t)]_{12} = {\rm Tr}\left( \e^{-\i t uH_\c u^*} u (\rho_\s(0)\otimes\rho_\r ) u^* \e^{\i t uH_\c u^*} u( |\varphi_2\rangle \langle\varphi_1|\otimes\bbbone_\r) u^*\right).
\end{equation}
Using the form \eqref{a54} and that $u (|\varphi_2\rangle \langle\varphi_1|\otimes\bbbone_\r) u^*= |\varphi_2\rangle\langle\varphi_1| \otimes W(f_2)W^*(f_1)$ (and the suitable expressions for the local reservoirs model), we arrive at the following expression, for both the local and collective reservoirs models,
\begin{equation}
\label{a69}
{}[\rho_\s(t)]_{12} ={\cal D}(t) \ \e^{-\i t\hat \epsilon} [\rho_\s(0)]_{12},
\end{equation}
where
\begin{equation}
 {\cal D}(t)=
 \left\{
 \begin{array}{l} 
 	\label{be6}
 \av{W^*(f_2)W(f_2(t)) W^*(f_1(t))W(f_1)}_\r \mbox{\quad (collective)}\\
  \av{W^*(f_2)W(f_2(t))}_\r \av{W^*(f_1(t))W(f_1)}_\r  \mbox{\quad (local)}\\
 \end{array}
\right.
\end{equation}
Here, $W(f(t)) = \e^{\i tH_\r}W(f)\e^{-\i tH_\r}=W(\e^{\i t \omega}f)$. Then, using the canonical commutation relations $W(f)W(g)=\e^{-\frac\i2{\rm Im}\, \scalprod{f}{g}} W(f+g)$ and the thermal average $\av{W(f)}_\r = \e^{-\frac14\scalprod{f}{\coth(\beta\omega/2)f}}$, \eqref{a66} follows directly from \eqref{be6} and the definitions of the functions $Q$ (see \eqref{b4}, \eqref{b3}). 

\medskip

We now show Proposition \ref{prop4}. The reservoir alone has the property of `return to equilibrium', which can be expressed as follows.

\begin{lem}
\label{lemma1}
Let $A$, $B$ and $C$ be observables of the reservoir and set $B(t)=\e^{\i tL_\r} B\e^{-\i tL_\r}$. Then we have 
\begin{equation}
\label{lp0}
\lim_{t\rightarrow\infty} \av{A B(t) C}_\r = \av{AC}_\r\av{B}_\r.
\end{equation}
\end{lem}
{\em Proof of Lemma \ref{lemma1}.\ } For any $\epsilon>0$ there exists a $C'_\epsilon$ which commutes with all observables and which satisfies
\begin{equation}
\label{lp1}
\| C\Omega_\r - C'_\epsilon\Omega_\r\|\le \epsilon.
\end{equation}
This is simply the separability of $\Omega_\r$. Therefore,
\begin{equation}
\label{lp2}
\av{AB(t)C}_\r = \av{AB(t)C'_\epsilon}_\r + R_1(\epsilon),
\end{equation}
where $|R_1(\epsilon)|\le\epsilon \|A\|\, \|B\|$. Next, since $C'_\epsilon$ commutes with $B(t)$ and since $\e^{-\i tL_\r}\Omega_\r=\Omega_\r$, we have 
\begin{equation}
\label{lp3}
\av{AB(t)C'_\epsilon}_\r =  \av{AC'_\epsilon\e^{\i tL_\r} B}_\r= \av{AC'_\epsilon}_\r \av{B}_\r +R_2(\epsilon,t),
\end{equation}
with
\begin{equation}
\label{lp4}
\lim_{t\rightarrow 0} R_2(\epsilon,t)=0, \quad \forall \epsilon>0.
\end{equation}
Relation \eqref{lp4} follows from the fact that $\e^{\i tL_\r}\rightarrow |\Omega_\r\rangle\langle\Omega_\r|$ in the weak sense as $t\rightarrow\infty$ (which in turn is implied by the fact that $L_\r$ has absolutely continuous spectrum except for a simple eigenvalue at zero, with eigenvector $\Omega_\r$.) Also, the term $\av{AC'_\epsilon}_\r$ in \eqref{lp3} equals $\av{AC}+R_3(\epsilon)$, with $|R_3(\epsilon)|\le \epsilon \|A\|$, by \eqref{lp1}. We finally obtain
\begin{equation}
\label{lp5}
\av{AB(t)C}_\r - \av{AC}_\r\av{B}_\r = R_4(\epsilon) + R_2(\epsilon,t),
\end{equation}
with $|R_4(t)|\le 2\epsilon \|A\|\,\|B\|$ and $R_2(\epsilon,t)$ satisfying \eqref{lp4}. 

Relation \eqref{lp5} implies \eqref{lp0}. \hfill \qed

It follows from Lemma \ref{lemma1} that 
\begin{equation}
\label{a70}
\lim_{t\rightarrow\infty} {\cal D}(t) = \av{W^*(f_2)W(f_1)}_\r \av{W(f_2)W^*(f_1)}_\r=\e^{-\Gamma_\infty},
\end{equation}
where $\Gamma_\infty$ is given in \eqref{gammainfty}. 

For the model with the local reservoirs, the relations \eqref{a69} and $\lim_{t\rightarrow\infty}{\cal D}(t) = \e^{-\Gamma_\infty}$ are derived in the same way. This finishes the proof of Proposition \eqref{prop4} \hfill \qed

\subsection{Explanation of the factor $\e^{-\Gamma_\infty}$ in \eqref{a64}.} 
\label{explansect}

In the basic resonance expansion \eqref{a23}, for $A=|\varphi_2\rangle\langle\varphi_1|\otimes\bbbone_\s\otimes\bbbone_\r$,  all terms on the right side except the one with $\Pi_-$ are $O(V)$. Thus 
\begin{eqnarray}
{}[\rho_\s(t)]_{12} &=&\e^{-\gamma t/2}\e^{-\i t(\hat\epsilon +x_{\rm LS})} \scalprod{UB^*\psi_0}{\big( |\varphi_{21}\rangle\langle\varphi_{21}|\otimes|\Omega_\r\rangle\langle\Omega_\r| \big) UA\Omega_{\beta,\vlambda,V}}\nonumber\\
&& +O(V)+R(t).
\label{-01}
\end{eqnarray}
The projection $|\Omega_\r\rangle\langle\Omega_\r|$ is the (weak) limit $\lim_{t\rightarrow\infty}\e^{\i tL_\r}$ and so the scalar product term on the right side of \eqref{-01} is 
\begin{eqnarray}
\lefteqn{
 \lim_{t\rightarrow\infty}  \e^{\i t
 \hat\epsilon} \scalprod{UB^*\psi_0}{\e^{\i t\widetilde L_0}\big( |\varphi_{21}\rangle\langle\varphi_{21}|\otimes\bbbone_\r \big)  UA\Omega_{\beta,\vlambda,V}}}\nonumber\\
&& = \lim_{t\rightarrow\infty}  \e^{\i 
 	t \hat\epsilon} \scalprod{UB^*\psi_0}{\e^{\i t\widetilde L_0}\big( |\varphi_{21}\rangle\langle\varphi_{21}|\otimes\bbbone_\r \big)  UA\Omega_{\beta,\vlambda,0}} +O(V).
\label{-03}
\end{eqnarray}
Here, $\widetilde L_0$ is given in \eqref{wlo} and we have made use of the fact $L_\s^{\rm ren}\varphi_{21} =- \hat\epsilon\varphi_{21}$ and relation \eqref{a19}. One readily sees that
\begin{eqnarray}
\big( |\varphi_{21}\rangle\langle\varphi_{21}|\otimes\bbbone_\r \big)  UA\Omega_{\beta,\vlambda,0} &=& U\big(	|\varphi_{21}\rangle\langle\varphi_{21}|\otimes\bbbone_\r \big)  A\Omega_{\beta,\vlambda,0} \nonumber\\
&=& U \big(|\varphi_2\rangle\langle\varphi_1|\otimes |\varphi_1\rangle\langle\varphi_1| \otimes\bbbone_\r\big)\Omega_{\beta,\vlambda,0}\nonumber\\
&=& U A\Omega_{\beta,\vlambda,0}.
\label{-02}
\end{eqnarray}
The last equality follows from the explicit form \eqref{a44} of $\Omega_{\beta,\vlambda,0}$. Using \eqref{-02} in \eqref{-03} shows that
\begin{eqnarray}
	\lefteqn{
\scalprod{UB^*\psi_0}{\e^{\i t\widetilde L_0}\big( |\varphi_{21}\rangle\langle\varphi_{21}|\otimes\bbbone_\r \big)  UA\Omega_{\beta,\vlambda,0}} = \scalprod{\psi_0}{B \e^{\i t L|_{V=0}} A\Omega_{\beta,\vlambda,0}} }\nonumber\\
&=&\scalprod{\psi_0}{B \e^{\i t L|_{V=0}} A\e^{-\i t L|_{V=0}}\Omega_{\beta,\vlambda,0}} = \scalprod{\psi_0}{ \e^{\i t L|_{V=0}} A\e^{-\i t L|_{V=0}}B\Omega_{\beta,\vlambda,0}}\nonumber\\
&=&\scalprod{\psi_0}{ \e^{\i t L|_{V=0}} A\e^{-\i t L|_{V=0}}\psi_0} +O(V)\nonumber\\
&=& {\mathcal D}(t) \e^{-\i t\hat\epsilon} [\rho_\s(0)]_{12} +O(V).
\label{-05}
\end{eqnarray}
the last equality is obtained from \eqref{a69}, since $\scalprod{\psi_0}{ \e^{\i t L|_{V=0}} A\e^{-\i t L|_{V=0}}\psi_0}$ is exactly the $(1,2)$ matrix element of the dimer density matrix at time $t$, evolving according to the evolution with $V=0$. We combine \eqref{-05} with \eqref{-03} and  \eqref{-01} to reach 
\begin{equation}
	{}[\rho_\s(t)]_{12} = \e^{-\gamma t/2}\e^{-\i t(\hat\epsilon +x_{\rm LS})}
	\big\{ \lim_{t\rightarrow\infty}  {\mathcal D}(t)\big\} \,  [\rho_\s(0)]_{12} +O(V)+R(t).
	\label{-04}
\end{equation}

{\bf Upshot.\ }
The rigorous derivation the formula \eqref{-04} given above can be put into heuristic words, uncovering the mechanism making $\e^{-\Gamma_\infty} =\lim_{t\rightarrow\infty}  {\mathcal D}(t)$ appear in \eqref{a64}, as follows. The resonance approximation (c.f. \eqref{a23}) consists in replacing the propagator $\e^{\i tL}\sim \sum_j\e^{\i t a_j} \Pi_j +R(t)$, where $a_j\in\mathbb C$ are complex resonance energies and $\Pi_j$ are projections. While the $a_j$ depend on $V$, the projections $\Pi_j$ are lowest oder approximations (i.e. $O(V^0)$) and hence independent of $V$. The observable $A=|\varphi_2\rangle\langle\varphi_1|\otimes\bbbone_\s\otimes\bbbone_\r$ selects a single projection, $\Pi_- = |\varphi_2\rangle\langle\varphi_2|\otimes |\varphi_1\rangle \langle\varphi_1|\otimes P_\r$, in that all other terms are $O(V)$,
$$
\e^{\i tL} A\sim \sum_j \e^{\i ta_j}\Pi_j A  +R(t) =   \e^{\i ta_-} \Pi_- A +O(V)+R(t). 
$$
Due to the dispersiveness of the reservoir(s), we have $P_\r=\lim_{t\rightarrow\infty} \e^{\i tL_\r}$, so that $\Pi_-$ is the long-time limit of the dynamics with $V=0$, i.e., 
 $\Pi_j\sim\lim_{t\rightarrow\infty} 
\e^{\i tL|_{V=0}}$. Thus, 
$$
\e^{\i tL} A \sim  \e^{\i ta_-} \lim_{t\rightarrow\infty}\e^{\i t L|_{V=0}} A +O(V) +R(t).
$$
This is the form \eqref{-04}, which can be rewritten as (see \eqref{a69})
\begin{equation}
\label{-08}
[\rho_\s(t)]_{21} = \e^{-\gamma t/2} \e^{-\i t(\hat\epsilon +x_{\rm LS})} \lim_{t\rightarrow\infty} \e^{\i t\hat\epsilon} \big[\rho_\s(t,V=0)\big]_{12} +O(V)+R(t),
\end{equation}
where $\rho_\s(t,V=0)$ is the dimer density matrix at time $t$, evolving under the dynamics coupled with the reservoir(s) with $V=0$. In \eqref{-08}, $\lim_{t\rightarrow\infty} \e^{\i t\hat\epsilon} \big[\rho_\s(t,V=0)\big]_{12}$ can be considered as a `shifted initial condition' for the off-diagonal dimer density matrix. 

The analogous analysis holds for the populations, e.g. for $[\rho_\s(t)]_{1,1}$. However, since the populations are stationary under the coupled dimer-reservoir(s) evolution with $V=0$, we have $[\rho_\s(t,V=0)]_{11}=[\rho_\s(0)]_{11}$ for all times. Thus the shifted initial condition coincides with the true one and the analogue of the factor $\e^{-\Gamma_\infty}$ is just the factor $1$ for the populations.

\subsection{Origin of full and partial phase decoherence}
\label{decoorigin}

We limit our discussion here to the situation of a collective reservoir and symmetric coupling, i.e., $\lambda_1=-\lambda_2$.

\subsubsection{Quantum noise} Consider the dimer coupled to the collective reservoir with $V=0$, and where $\lambda_1=-\lambda_2=\lambda$. The Hamiltonian is (c.f. \eqref{1'})
\begin{equation}
\label{b20}
	H=\frac\epsilon2\sigma_z +H_R+ \lambda\sigma_z\otimes\phi(g).
\end{equation}
One readily verifies that the exact solution for the off-diagonal dimer density matrix element is\footnote{One way to do this is to pass to the interaction picture, $\widetilde \rho(t) = \e^{\i tH_0}\widetilde\rho(0)\e^{-\i tH_0}$, where $H_0$ is \eqref{b20} with $\lambda=0$, and then write the evolution for $\widetilde\rho(t)$ using a time-dependent generator for the propagator.}
\begin{equation}
	\label{rho12}
	[\rho_\S(t)]_{12} = \e^{-\i\epsilon t} [\rho_\S(0)]_{12}\, {\rm Tr}_\R (\rho_\R\,  \e^{-2\i\lambda\int_0^t\phi(\e^{\i\omega s}g)}).
\end{equation}
Let 
\begin{equation}
	\phi_s = \phi(e^{i\omega s}g)
\end{equation}
and denote
\begin{equation}
	\langle \cdot\rangle = {\rm Tr}_\R (\rho_\R\, \cdot).
\end{equation}
Then
\begin{equation}
	\label{av}
	\langle \e^{-2\i\lambda\int_0^t \phi_s ds}\rangle = \sum_{n\ge 0} \frac{(-2\i\lambda)^n}{n!}\int_0^tds_1\cdots\int_0^t ds_n \, \langle \phi_{s_1}\cdots \phi_{s_n}\rangle.
\end{equation}
The process $\phi_t$ is (non-commutative) Gaussian. Namely, the $n$-point correlations functions are expressed solely using two-point correlations according to ``Wick's theorem" (c.f. \cite{BRII} for example),
\begin{equation}
	\label{pairings}
	\langle \phi_{s_1}\cdots \phi_{s_{2k}}\rangle = \sum_{{\rm pairings}} \langle\phi_{s_{i_1}}\phi_{s_{j_1}}\rangle \cdots \langle\phi_{s_{i_k}}\phi_{s_{j_k}}\rangle,
\end{equation}
and the correlation functions of odd order vanish. Here, the sum in \eqref{pairings} is over the 
\begin{equation}
	(k-1)!! = (k-1)(k-3)\cdots 5\cdot3\cdot1 =2^{-k+1}\frac{(2k-1)!}{(k-1)!}
\end{equation}
pairings of the indices. Now
\begin{equation}
	\label{ints}
	\int_0^tds_1\cdots\int_0^t ds_{2k} \, \langle \phi_{s_1}\cdots \phi_{s_{2k}}\rangle
	=\sum_{\rm pairings} \big(\alpha_Q(t)\big)^{k} = 2^{-k+1}\frac{(2k-1)!}{(k-1)!}(\alpha_Q(t)\big)^{k},
\end{equation}
where
\begin{equation}
	\alpha_Q(t) =\int_0^t ds\int_0^t dr\  \langle\phi_s\phi_r\rangle.
\end{equation}
We use \eqref{ints} in \eqref{av} and get
\begin{equation}
	\label{f1}
	\langle \e^{-2\i\lambda\int_0^t \phi_s ds}\rangle = \sum_{k\ge 0} \frac{(-2\i\lambda)^{2k}}{{(2k)}!}2^{-k+1}\frac{(2k-1)!}{(k-1)!}(\alpha_Q(t)\big)^{k} = \e^{-2\lambda^2\alpha_Q(t)}.
\end{equation}
A direct calculation shows that
\begin{equation}
	\label{f2}
	\alpha_Q(t) = \frac2\pi Q_2(t),
\end{equation}
where $Q_2(t)$ is given in \eqref{b4}. Relation \eqref{rho12} is thus the same as \eqref{introa65} and ${\cal D}(t)$ is expressed via $\alpha_Q$,  the double integral over the correlation function, as
\begin{equation}
\label{dcor}
{\cal D}(t) = \e^{-2\lambda^2\alpha_Q(t)} = \e^{-\frac{4}{\pi}Q_2(t)}.
\end{equation}

\subsubsection{Comparison with classical noise} Instead of considering a quantum mechanical noise as in the previous section, one may introduce a time-dependent Hamiltonian (including an "external noise") as
\begin{equation}
	H(t) = \frac\epsilon2\sigma_z +\lambda\sigma_z\xi_t,
\end{equation}
where $\xi_t$ is a commutative stochastic process. The exact solution for the off-diagonal density matrix element for each realization of the noise $\xi_t$ is
$[\rho_\S(t)]_{12} = \e^{-\i\epsilon t} [\rho_\S(0)]_{12}\,  \e^{-2\i\lambda\int_0^t\xi_s ds}$ and taking the average $\langle\cdot\rangle$ over the noise gives
\begin{equation}
	[\rho_\S(t)]_{12} = \e^{-\i\epsilon t} [\rho_\S(0)]_{12}\,  \langle \e^{-2\i\lambda\int_0^t\xi_s ds}\rangle.
\end{equation}
We have again
\begin{equation}
	\label{avc}
	\langle e^{-2i\lambda\int_0^t \xi_s ds}\rangle = \sum_{n\ge 0} \frac{(-2i\lambda)^n}{n!}\int_0^tds_1\cdots\int_0^t ds_n \, \langle \xi_{s_1}\cdots \xi_{s_n}\rangle.
\end{equation}
and if $\xi_t$ is a Gaussian process, then by the Gaussian Moment Theorem (Wick's Theorem), just as in the quantum case,
\begin{equation}
	\label{pairingsc}
	\langle \xi_{s_1}\cdots \xi_{s_{2k}}\rangle = \sum_{{\rm pairings}} \langle\xi_{s_{i_1}}\xi_{s_{j_1}}\rangle \cdots \langle\xi_{s_{i_k}}\xi_{s_{j_k}}\rangle,
\end{equation}
and the correlation functions of odd order vanish. We then get in the same way as for the quantum case
\begin{equation}
	\label{clde}
	\langle \e^{-2\i\lambda\int_0^t \xi_s ds}\rangle =  \e^{-2\lambda^2\alpha_C(t)},\quad\mbox{
where}\quad 
	\alpha_C(t) = \int_0^t ds \int_0^t dr \ \langle\xi_s\xi_r\rangle.
\end{equation}

Comparing \eqref{f1} and \eqref{clde} shows that decoherence in the classical and the quantum noise case is {\em exactly the same} if the noises have the same characteristics, i.e., if $\xi_t$ is Gaussian with zero average and two-point function
$\langle\xi_s\xi_r\rangle = \langle\phi_s\phi_r\rangle$.

\subsubsection{Full versus partial decoherence} For either the classical or quantum noise, consider the correlation function
\begin{equation}
	C(t,s) = \frac12\big(\langle \phi_t\phi_s\rangle +\langle\phi_s \phi_t\rangle\big) \quad\mbox{or}\quad C(t,s) = \frac12\big(\langle \xi_t\xi_s\rangle +\langle\xi_s \xi_t\rangle\big).
\end{equation}
For stationary processes, $C(t,s)=C(t-s)$ and so $C(t,s)=C(|t-s|)$. Then
\begin{equation}
	\alpha(t)=\int_0^tds \int_0^t dr \, C(|s-r|) =2 \int_0^t ds\int_0^s d\tau\,  C(\tau).
\end{equation}
We have
\begin{equation}
	\mbox{Full decoherence} \quad\Longleftrightarrow \quad \lim_{t\rightarrow\infty}\alpha(t) = \infty.
\end{equation}
Only the decay asymptotics of the correlation function plays a role to determine if $\alpha(t)\rightarrow\infty$ or not. Assume 
\begin{equation}
	\label{dis}
	C(\tau) \sim \tau^{-\delta},\quad \mbox{for some $\delta\ge 0$ and for $\tau$ large}.
\end{equation}
Then $\alpha(t)\sim t^{2-\delta}$ for large $t$, so we have full decoherence exactly if $\delta\le 2$ (for $\delta=2$, we have $\alpha(t)\sim \ln t$). This holds equally well for quantum and classical noises.

\bigskip

{\bf Discussion for the quantum thermal noise.} One has
\begin{equation}
	C(\tau) = \frac12\, {\rm Re}\, \int_0^\infty e^{i\omega\tau} J(\omega) \coth(\beta\omega/2) d\omega,
\end{equation}
where $J(\omega)$ is the spectral density of noise \eqref{a57}. To find the decay of the correlation function, write
\begin{equation}
	\label{find}
	(i\tau)^n C(\tau) = \frac12\, {\rm Re}\, \int_0^\infty \big(\tfrac{d^n}{d\omega^n}e^{i\omega \tau}\big) J(\omega) \coth(\beta\omega/2) d\omega.
\end{equation}
Assume the form
\begin{equation}
	J(\omega) \sim\omega^\gamma\quad\mbox{for $\omega$ small}
\end{equation}
and that $J(\omega)$ decays at least as $\omega^{-n}$ for large $\omega$.
By integrating \eqref{find} by parts to transfer the action of the $\omega$-derivatives onto the function $J(\omega)\coth(\beta\omega/2)$, one easily finds that 
\begin{equation}
	\limsup_{\tau\rightarrow\infty}\big| \tau^n C(\tau)| <\infty \quad \Longleftrightarrow \quad \gamma > n.
\end{equation}
According to the discussion after \eqref{dis}, the critical value for full decoherence is $n=2$, so if $J(\omega)$ vanishes more quickly than $\omega^2$ for small $\omega$, then we do not have full decoherence. In this way, the decay speed of correlations is governed by the low frequency behavior of the spectral density of noise and therefore, this behavior determines whether we have or do not have full decoherence. We can sum this up:

\bigskip
{\em The low frequency modes of the reservoir are responsible for full decoherence. We have full decoherence if and only if the low frequency modes are well coupled to the dimer. More precisely, we have full decoherence if and only if $J(\omega)\sim\omega^\gamma$ with $\gamma\le 2$ as $\omega\sim 0$.}

\bigskip

[Note: In \cite{PSE}, (p. 577), it is mentioned that the effect of non-full decoherence is due to ``...the suppressed influence of low-frequency fluctuations...", which coincides with the picture we uncover here.]

\section{Mathematical regularity requirements}
\label{regularitysect}

We specify the precise regularity requirements which lead to \eqref{regj} and \eqref{1.7j}. The spectral function of the reservoir is linked to the form factor by $J(\omega) = \frac\pi2 \omega^2\int_{S^2}|g(\omega,\Sigma)\d\Sigma$ and the form \eqref{regj} corresponds to
\begin{equation}
\label{exformfact}
g(k) = \frac{|k|^p}{(1+|k|)^\sigma}\widetilde g(k), 
\end{equation}
where $\widetilde g(k)$ is a (real valued) function which satisfies $\sup_{k\in{\mathbb R}^3} |\partial^j_\omega \widetilde g(\omega,\Sigma)|<\infty$ for $j=0,\ldots,4$.
Here, $k=(\omega,\Sigma)\in{\mathbb R}_+\times S^2$ is the spherical representation of vectors $k$. There are two origins of the infrared regularity needed here. One is the application of a `polaron transformation' (c.f. Proposition \ref{propunitrans}), which is indispensable in order to deal with arbitrary values of $\lambda_1$ and $\lambda_2$ and which requires only the milder regularity $p>-1/2$. The other origin is that we use a `dynamical resonance theory' developed in \cite{KoMe}, which gives a proof of \eqref{a23} under the assumptions that 
\begin{equation}
	\label{a32}
	u h, \ \partial^j_u h \in L^2({\mathbb R}\times S^2,\d u\times\d\Sigma), \qquad j=0,1,\ldots 4,
\end{equation}
where $h=(-\i g/\omega)_\beta$ in the case of the collective reservoir model, and $h=(-\i g_1/\omega)_\beta$ and $h= (-\i g_2/\omega)_\beta$ for the local reservoirs model. We recall \eqref{2.3} where $(\cdot)_\beta$ is defined. The square root factor $(u/(1-\e^{-\beta u}))^{1/2}$ in \eqref{2.3} is infinitely many times differentiable for $u\in
\mathbb R$ and bounded above by (a constant times) $|u|^{1/2}$. A form factor $g$ such that $h=(-\i g/\omega)_\beta$ satisfies \eqref{a32} needs to obey $g(0)=0$ (recall that $\omega(k)=|k|$) and it needs to decay at infinity.  Let $p>0$ be the strength of the zero and $\sigma$ the speed of decay, namely, $g(k)=|k|^p (1+|k|)^{-\sigma}\,\widetilde g(k)$, for some function $\widetilde g(k)$ which is four times differentiable w.r.t. its radial component ($r=|k|)$ and satisfies  $\sup_{k\in{\mathbb R}^3} |\partial_{|k|}^j\widetilde g(k)|<\infty$ for $j=0,\ldots,4$. Assume that $g$ is real valued (this is not necessary, but it slightly simplifies the exposition). Close to $u=0$, the singularity structure of $h$ is $\propto |u|^{-1/2+p}$. To have local square integrability of $\partial^j_uh$ at $u=0$ we thus need that either $-1/2+p$ is an integer $n\ge0$, or that $p>j$. Hence $p=(2\ell+1)/2$, $\ell=0,1,2,\ldots$ or $p>4$ (as $j=0,\ldots,4$). The integrability at $u\rightarrow\infty$ is guaranteed for $\sigma>3/2$.  

Note that the above infrared condition is stronger than the infrared condition necessary to apply the polaron transformation, which would only ask for $p >-1/2$ (so that $g_{1,2}/\omega\in L^2({\mathbb R}^3,\d^3k)$, which is the condition used to be able to deal with arbitrary sizes of $\lambda_1$, $\lambda_2$ with the help of the polaron transformation, see Proposition \ref{propunitrans}).

We note that in \cite{KoMe}, only a dynamical resonance theory for model with a single reservoir is considered. One must slightly adapt the arguments for the case of two reservoirs to be able to describe our local reservoirs model. An easy way to do this is to use a well-known property of Fock space, the isometric isometry
$$
{\cal F}(L^2) \otimes {\cal F}(L^2) = {\cal F}(L^2\oplus L^2),\qquad \mbox{where}\quad L^2 \equiv L^2({\mathbb R}\times S^2). 
$$
Under this mapping, the formalism of two reservoirs turns into one of a single reservoir (albeit on a different one-particle space). The arguments of \cite{KoMe} leading to the resonance expansion of the propagator (c.f. \eqref{a23}) can then be adapted in a straightforward way to this new single-reservoir setting. We do not present the details of the analysis in the present work.

\section{Parameter constraints}
\label{parcon}

The large coupling results of \cite{KoMeSo, KoMe} our resonance approach in the current work is based on are stated as follows. Given arbitrary $\lambda_1, \lambda_2\in\mathbb R$, there is a constant $V_0>0$ such that if $|V|<V_0$, then the results hold. The upper bound $V_0$ depends on the fixed $\lambda_1, \lambda_2$. This dependence can be found by tracing the parameters $\lambda_1,\lambda_2$ through the arguments of \cite{KoMeSo,KoMe}. 
We set
\begin{equation}
	\label{mu}
\widetilde\gamma = \gamma/V^2,\quad \theta=\min\{|\hat\epsilon|,\widetilde\gamma\},
\end{equation}
where $\gamma$ and $\hat\epsilon$ are given in \eqref{b12} and \eqref{epsilonhat}, respectively. Note that $\widetilde \gamma$ and $\hat\epsilon$ depend on $\lambda_1$ and $\lambda_2$. The following are  constraints used to prove the validity of the dynamical resonance theory.

\medskip
\noindent
{\bf Collective reservoir model.}\ 
The interaction \eqref{a14} depends only on the difference $\lambda_1-\lambda_2$ of the coupling constants. Define
$$
\xi=|\lambda_1-\lambda_2|.
$$
Let $C$ be a constant which does not depend on $V,\lambda_2,\lambda_2$. 
\begin{itemize}
\item Validity of the {\em Born approximation},
\begin{equation}
	\label{need1}
|V|\le C(1+\xi^2)^{-1}.
\end{equation}
This constraint is used to isolate the main dynamics in which the reservoir can be considered in its equilibrium state at all times, i.e., to justify the  Born approximation. [Technical remark: the condition is needed to show that $\bar L(\eta)$ has spectrum in the lower complex half plane, c.f. Lemma 4.3 of \cite{KoMeSo}; the main dynamics can then be expressed on ${\rm Ran}P_\Omega$.]

\item Separation of resonances, 
\begin{equation}
	\label{need2}
|V|\le C \min\{ \widetilde\gamma\, (1+\xi^6)^{-1}, \sqrt{\widetilde\gamma}\, \xi^{-5}\}.
\end{equation}
 This condition ensures that the effective complex dimer energy differences, i.e., the resonances are well separated. Then they determine three decaying directions (in the dimer Liouville space) and one equilibrium direction. If this condition is not satisfied, then one must develop a theory of overlapping resonances \cite{MS, MeBeSo}.\\
 The resonances are the complex $a_0$, $a_\pm$, \eqref{as} and $a'_0=0$. Rescaled by $V^{-2}$, their separation (smallest distance among them) is $\propto\widetilde\gamma = \widetilde \gamma(\lambda_1,\lambda_2)$. Even if the real part $\hat\epsilon \pm x_{\rm LS}$ of $a_\pm$ is large, shifting those two resonances horizontally far away from the origin, one resonance ($a_0$) stays at a distance $\widetilde\gamma$ from $a'_0=0$. The separation of resonances (rescaled by $V^{-2}$) must persist for perturbations small in $V$, hence the above upper bound on $V$. [Technical remark: This condition is used to guarantee that $A_z$ is close to the level shift operator in Lemma 3.1 of \cite{KoMe}.]

\item No backreaction on the dimer, 
\begin{equation}
	\label{need3}
|V|\le C\min\{\theta,  \sqrt\theta(1+\xi^6)^{-1}\}.
\end{equation}
The stationary states of the uncoupled dimer-reservoir system are (superpositions of) states with the dimer in a dimer-energy eigenstate and the reservoir in equilibrium. The above condition ensures that away from these states, the coupled dynamics is dispersive (local observables decay sufficiently quickly in time). The dispersiveness estimates are obtained separately for different stationary dimer Bohr energies, so they involve their separation, $\hat\epsilon$ (on which $\theta$ depends). [Technical remark: This condition is used to prove the Limiting Absorption Principles of the interacting  resolvent, in particular when reduced to ${\rm Ran}P_e^\perp$, Theorem A.1 in \cite{KoMe}.]
\end{itemize}

{\bf Discussion.\ }  Consider spectral functions of the form \eqref{e1}, $J(\omega) = A\omega^{2p+2} \e^{-\omega/\omega_c}$, with $p$ as in \eqref{1.7j}.

\begin{itemize}
\item[(1)] Consider $\xi <\!\!<1$.  We now show that 
\begin{equation}
	\label{sh1}
\gamma_\c\le C
\left\{
\begin{array}{ll}
\xi^6(\hat\epsilon)^{2p} \min\{ \e^{-|\hat\epsilon|/\omega_c}, 
|\hat\epsilon|^{-1}\} & \mbox{if $|\hat\epsilon|\ge C\xi^2$},\\
\xi^2|\hat\epsilon|^{2p+1} & \mbox{if  $|\hat\epsilon|\le C\xi^2$}.
\end{array}
\right.
\end{equation} 
For small $\xi$, we have $\widetilde\gamma\propto \frac{\xi^2}{\hat\epsilon^2}J(|\hat\epsilon|)\coth(\beta|\hat\epsilon|/2)$, see Section 5 of \cite{MBSa}. Taking into account the form of the spectral density \eqref{e1}, we obtain
\begin{equation}
	\label{est}
\gamma_\c \ =\ V^2\widetilde\gamma\  \propto \ V^2 \xi^2  
(\hat\epsilon)^{2p}\e^{-|\hat\epsilon|/\omega_c}\coth(\beta|\hat\epsilon| 
/2).
\end{equation}
Consider two cases: 

(a) $|\hat\epsilon|\ge C\xi^2$ hence $\theta=\xi^2$. The conditions \eqref{need1} -- \eqref{need3} become $|V|\le C\xi^2$ and from \eqref{est} we obtain the first estimate in \eqref{sh1}.

(b) $|\hat\epsilon|\le C\xi^2$, so that $\theta=|\hat\epsilon|$. The conditions \eqref{need1} -- \eqref{need3} become $|V|\le C|\hat\epsilon|$ so  $\gamma_\c\le C \xi^2 |\hat\epsilon|^{2p+1}$. 

\item[(2)] Consider $\xi\approx 1$. The conditions \eqref{need1} -- \eqref{need3} become $|V|\le\min\{ \widetilde\gamma,\sqrt{\widetilde\gamma}, |\hat\epsilon|, \sqrt{|\hat\epsilon|}\}$.  

\item[(3)] Consider $\xi>\!\!>1$. From \eqref{b12} we expect that $\widetilde\gamma\propto \e^{-c\xi^2}$ for some $c>0$. The conditions \eqref{need1} -- \eqref{need3} become $|V|\le C\e^{-c\xi^2}$ and therefore $\gamma_\c\le C\e^{-c\xi^2}$ is extremely tiny.
\end{itemize}

{\bf Conclusion.\ } Sizeable values of $\gamma_\c$ will be obtained in the regime $\xi\approx 1$ only, that is, when the difference of $\lambda_1$ and $\lambda_2$ is not too large nor too small. For $\lambda_1\approx \lambda_2$, the dynamical process is suppressed (weakly, $\gamma_\c$ is proportional to a power of $|\lambda_1-\lambda_2|$) because for $\lambda_1=\lambda_2$ there is no relaxation or decoherence at all. For $|\lambda_1-\lambda_2|>\!\!>1$ the dynamical process is is strongly suppressed ($\gamma_\c\propto\e^{-c|\lambda_1-\lambda_2|^2})$, which is an effect analogous to  Anderson localization.

\medskip
{\em Remark.\ } If we use the expression \eqref{e18} for $\gamma_\c$ in the Marcus regime, we have $\gamma_\c\le C V^2/\xi$.  The above constraints \eqref{need1} -- \eqref{need3} on $V$ give an upper bound on $\gamma_\c$,
\begin{equation}
\gamma_\c \le C
\left\{
\begin{array}{ll}
\xi & \mbox{for $\xi <\!\!<1$},\\
\min\{ \widetilde\gamma,\sqrt{\widetilde\gamma}, |\hat\epsilon|, \sqrt{|\hat\epsilon|}\} & \mbox{for $\xi\approx 1$},\\
\e^{-c\xi^2} & \mbox{for $\xi>\!\!>1$}.
\end{array}	
\right.
	\qquad \mbox{(Marcus regime)}
\end{equation}
These upper bounds are qualitatively correct (they coincide qualitatively with the ones obtained from the true expression \eqref{b12} for $\gamma_\c$, even though for small $\xi$, the power of the upper bound is different).

\medskip
\noindent
{\bf Local reservoirs model.}\ Here the bounds \eqref{need1}-\eqref{need3} need to be verified with $\xi=\lambda_1$ and $\xi=\lambda_2$.

\section{Reservoirs and continuous mode limits}
\label{reservoirssect}

A reservoir of vibrational modes is typically modeled by a collection of (independent) quantum oscillators with Hamiltonian ($\hbar=1$)
\begin{equation}
\label{fin1}
H_N = \sum_{k=1}^N \omega_k a^\dagger_k a_k, 
\end{equation}
with $a_k a^\dagger_\ell-a^\dagger_\ell a_k=\delta_{k,\ell}$ (Kronecker symbol). In the context of a {\em `molecular reservoir'}, one has $N$ (protein, solvent...) atoms in the environment, each one being modeled by three degrees of freedom, i.e., by a three-dimensional linear quantum oscillator with frequencies $\omega_{n,d}$, $n=1,\ldots,N$, $d=1,2,3$. The Hamiltonian of the (finite mode) reservoir could then be written as in \eqref{fin1} with either $N$ replaced by $3N$ or it could be written as a sum over $\vec k =(k_1,k_2,k_3)$. In a {\em `structured environment'}, say, given by a periodic structure, by phonons or photons, the index $\vec k$ is the wave vector and has associated frequency $\omega_{\vec k}$. This fixes a relation between $k$ and $\omega$, the dispersion relation (e.g. $\omega=|k|$ for acoustic phonons or photons). However, the frequencies present in a molecular reservoir may be more `random'. One can list them increasingly as $\omega_1\le\omega_2\le\cdots\le\omega_n$, but there is no inherent relation between the index $j$ and the value $\omega_j$. The defining property for a molecular reservoir is the {\em fraction among all oscillators having a given frequency}. It is determined by a function $\varrho(\omega)$  such that for any set $I\subset {\mathbb R}_+$, the quantity $\int_I \varrho(\omega)\d\omega$ is the proportion of all oscillators having frequencies lying in $I$. In particular, $\int_0^\infty \varrho(\omega)\d\omega=1$.  The function $\varrho$ is the {\em  frequency density of the molecular environment}. 

Mathematically, the treatment of molecular and structured reservoirs is the same. Our method does not rely on the presence or absence of a dispersion relation. The only quantity that matters is the spectral function $J(\omega)$, which is defined via the correlation function of the reservoir, c.f. \eqref{mr39}. The {\em explicit relation} between the spectral function $J$ and the form factor $g$ will depend on the nature of the reservoir (dispersion relation or frequency density). For instance, for three dimensional phonons or photons with $\omega_k=|k|$, we have 
\begin{equation}
\label{twores1}
J(\omega) = \frac\pi2\omega^2 \int_{S^2} |g(\omega,\Sigma)|^2 \d\Sigma
\end{equation}
(where $g$ is represented in spherical coordinates of $k\in{\mathbb R}^3$). For a molecular reservoir with frequency density $\varrho(k)$, 
\begin{equation}
\label{twores2}
J(\omega) = \pi \varrho(\omega) |g(\omega)|^2. 
\end{equation}
However, in either case, the relevant physical characteristics of the reservoir are condensed into the function $J(\omega)$, which is typically of the form \eqref{regj} or \eqref{e1}.


\begin{thebibliography}{99}

\bibitem{Vib4}
M. Aghtar, J. Str\"{u}mpfer, C. Olbrich, K. Schulten, and U. Kleinekath\"{o}fer: {\it Different Types of Vibrations Interacting with Electronic Excitations in Phycoerythrin 545 and Fenna-Matthews-Olson Antenna Systems}, J. Phys. Chem. Lett., {\bf  5}, 3131-3137 (2014)

\bibitem{AW}
 H. Araki, E. Woods: {\em Representations  of  the  canonical  commutation  relations  describing  a  nonrelativistic
 infinite free bose gas}, J. Math. Phys. {\bf 4}, 637-662 (1963)



\bibitem{Dimer}
G.P. Berman, A.I. Nesterov, S. Gurvitz, and R.T. Sayre: {\it Possible Role of Interference and Sink Effects in Nonphotochemical Quenching in Photosynthetic Complexes}, arXiv:1412.3499v1 [physics.bio-ph]


\bibitem{BRII}
O. Bratteli, D.W. Robinson: Operator algebras and quantum
statistical mechanics  II, Springer Verlag, 1987


\bibitem{BP}
H.-P. Breuer, F. Petruccione: The theory of open quantum systems, Oxford University Press, 2006

\bibitem{1f}
T.G. Dewey and J.G. Bann, {\em Protein dynamics and 1 /f noise}, Biophys. J., {\bf 63}, 594-598 (1992) 



\bibitem{Vib2}
J. Du, T. Teramoto,  K. Nakata, E. Tokunaga, and T. Kobayashi: {\it Real-Time Vibrational Dynamics in Chlorophyll a Studied with a Few-Cycle Pulse Laser}, Biophysical Journal, {\bf 101},  995-1003 (2011)

\bibitem{Fleming89}

D.D. Eads, E.W. Castner Jr., R.S. Alberte, L. Mets, and G.R. Fleming: {\em Direct observation of energy transfer in a photosynthetic membrane: chlorophyll b to chlorophyll a transfer in LHC}, J. Phys. Chem., {\bf 93}, 8271-8275 (1989)

\bibitem{JP}
V. Jaksic, C.-A. Pillet: {\em On a model for quantum friction II. Fermi's golden rule and
	dynamics at positive temperature}, Comm. Math. Phys. {\bf 176},  619-644 (1996)



\bibitem{Forster}
R.S. Knox, {\em F\"{o}rster's resonance excitation transfer theory:
	not just a formula}, J. Biomedical Optics, {\bf 17}, 011003-6 (2012)


\bibitem{KoMe}
M. K\"onenberg, M. Merkli: {\em  On the irreversible dynamics emerging from quantum resonances}, submitted (2015), preprint archive:  arXiv:1503.02972v2


\bibitem{KoMeSo}
M. K\"onenberg, M. Merkli, H. Song: {\em Ergodicity of the Spin-Boson Model for Arbitrary Coupling Strength}, Comm. Math. Phys. {\bf 336}, 261-285 (2014)

\bibitem{Leggett}
A.J. Leggett, S. Chakravarty, A.T. Dorsey, M.P. A. Fisher, A. Garg, W. Zwerger: {\em Dynamics of the dissipative two-state system}, Rev. Mod. Phys. {\bf 59}(1), 1-85 (1987)


\bibitem{Marcus}
R.A. Marcus: {\em On the Theory of Oxidation-Reduction Reactions Involving Electron Transfer. I}, J. Chem. Phys. {\bf 24}, no.5, 966-978 (1956)

\bibitem{MarcusNobel}
http://www.nobelprize.org/nobel$\underline{\  }\,$prizes/chemistry/laureates/1992/marcus-lecture.pdf

\bibitem{MBSa}
M. Merkli, G.P. Berman, and R. Sayre, {\em Electron Transfer Reactions: Generalized Spin-Boson Approach}, Journal of Mathematical Chemistry, {\bf 51}, Issue 3, 890-913 (2013)

\bibitem{MBR}
M. Merkli, G.P. Berman, A. Redondo: {\em  Application of Resonance Perturbation Theory to Dynamics of Magnetization in Spin Systems Interacting with Local and Collective Bosonic Reservoirs}, J. Phys. A: Math. Theor. {\bf 44}, 305306-305330 (2011)

\bibitem{MeBeSo}
M. Merkli, G.P. Berman, H. Song: {\em Multiscale dynamics of open three-level quantum systems with two quasi-degenerate levels}, J. Phys. A: Math. Theor. {\bf 48}, 275304 (2015) 


\bibitem{MSB} M. Merkli, I.M. Sigal, G.P. Berman: {\em Resonance theory of decoherence and thermalization}, Ann. Phys. {\bf 323}, 373-412 (2008)

\bibitem{MS} M. Merkli, H. Song: {\em Overlapping Resonances in Open Quantum Systems}, Ann. Henri Poincar\'e {\bf 16}, Issue 6, 1397-1427 (2015) 



\bibitem{QEB}
M. Mohseni, Y. Omar, G.S. Engel, and M.B. Plenio (Eds):  Quantum Effects in Biology, Cambridge University Press, 2014

\bibitem{CP29}
F. M\"{u}h, D. Lindorfer, M. S. am Busch, and T. Renger: {\it Towards a structure-based exciton Hamiltonian for the CP29 antenna of photosystem II}, Phys. Chem. Chem. Phys.,
{\bf 16}, 11848-11863 (2014)

\bibitem{Muka}
S. Mukamel: Principles of Nonlinear Spectroscopy. Oxford Series in Optical and Imaging Sciences, Oxford University Press, 1995

\bibitem{NB}
A.I. Nesterov and G.P. Berman: {\it The Role of Protein Fluctuation Correlations in Electron Transfer in Photosynthetic Complexes}, Phys. Rev. E., {\bf 91}, 
042702-8 (2015)


\bibitem{CF2}
C. Olbrich, J. Str\"{u}mpfer, K. Schulten,  and U. Kleinekath\"{o}fer, {\it Theory and Simulation of the Environmental Effects on FMO Electronic Transitions}, J. Phys. Chem. Lett., {\bf  2}, 1771-1776 (2011)


\bibitem{HZ}
Frank W. J. Olver, Daniel W. Lozier, Ronald F. Boisvert,  Charles W. Clark, eds.
{ NIST Handbook of Mathematical Functions}. Cambridge University Press, 2010.

\bibitem{PSE}
G.M. Palma, K.-A. Suominen, A.K. Ekert, A.K.: {\em  Quantum Computers and Dissipation}, Proc. R. Soc. Lond. A {\bf 452}, 567-584 (1996)


\bibitem{Lloyd1}
P. Rebentrost, M. Mohseni, I. Kassal, S. Lloyd, and A. Aspuru-Guzik: {\it Environment assisted
	quantum transport}, New J. Phys., 11, 033003-12 (2009).



\bibitem{CF1}
A. Shabani, M. Mohseni, H. Rabitz, and S. Lloyd: {\it Numerical evidence for robustness of environment-assisted quantum transport}, Phys. Rev. E, {\bf 89}, 042706-7 (2014)

 \bibitem{T1}
 Y. Tanimura and R. Kubo, Time evolution of a quantum system in contact with a nearly Gaussian-Markoffian noise bath,  J. Phys. Soc. Japan {\bf 58}, 101-114 (1989) 
 
 \bibitem{T2}
 Y. Tanimura, {\em Reduced hierarchical equations of motion in real and imaginary time: Correlated initial states and thermodynamic quantities}, J. Chem. Phys. {\bf 141}, 044114-13 (2014)




\bibitem{Vegte}
C. P. van der Vegte,   J.D. Prajapati,  U. Kleinekath\"ofer, J. Knoester, and T. L. C. Jansen: {\it Atomistic Modeling of Two-Dimensional Electronic Spectra and Excited-State Dynamics for a Light Harvesting 2 Complex}, J. Phys. Chem. B, {\bf 119} 1302-1313 (2015)


\bibitem{Vib1}
R. Wang, S. Parameswaran, G. Hastings: {\it Density functional theory based calculations of the vibrational properties of chlorophyll-a}, Vibrational Spectroscopy, {\bf 44}, 357-368 (2007)


\bibitem{XuSch}
D. Xu, K. Schulten: {\em Coupling of protein motion to electron transfer in a photosynthetic reaction center: investigating the low temperature behavior in the framework of the spin-boson model}, Chem. Phys. {\bf 182}, 91-117 (1994)
                     

\bibitem{Vib3}
W. Zhuang, T. Hayashi, and S, Mukamel: {\it Coherent Multidimensional Vibrational Spectroscopy of Biomolecules: Concepts, Simulations, and Challenges}, Angew. Chem. Int. Ed., {\bf 48}, 3750-3781 (2009)









\end{thebibliography}
\end{document}